\documentclass[journal,comsoc]{IEEEtran}
\usepackage{hyperref}
\usepackage{cite}
\usepackage{tikz}
\usepackage{amssymb,amsfonts}
\usepackage{algorithmic}
\usepackage{graphicx}
\usepackage{xcolor}
\usepackage{array}
\usepackage{xurl}
\usepackage{ragged2e}
\usepackage{acronym}
\usepackage{amsmath}
\ifCLASSOPTIONcompsoc
    \usepackage[caption=false, font=normalsize, labelfont=sf, textfont=sf]{subfig}
\else
\usepackage[caption=false, font=footnotesize]{subfig}
\fi
\usepackage{multirow}
\usepackage{url}
\usepackage{lipsum}
\usepackage{enumitem}
\usepackage{balance}
\usepackage{tabularx}
\usepackage{rotating}
\usepackage{graphicx}
\usepackage{amsmath}
\usepackage{rotating}
\usepackage{wasysym}
\usepackage{color, colortbl}
\usepackage{array}
\usepackage{booktabs}
\usepackage{soul}
\usepackage{multicol}
\begin{document}

\title{\LARGE Intelligent Control in 6G Open RAN: Security Risk or Opportunity? }
\author{\IEEEauthorblockN{Sanaz Soltani\IEEEauthorrefmark{1}, Mohammad Shojafar\IEEEauthorrefmark{1}, Ali Amanlou\IEEEauthorrefmark{1}, and Rahim Tafazolli\IEEEauthorrefmark{1}}

\IEEEauthorblockA{\IEEEauthorrefmark{1} 5GIC \& 6GIC, University of Surrey, Guildford, UK\\ \{s.soltani, m.shojafar, aliamanlou, r.tafazolli\}@surrey.ac.uk}

	}

	\maketitle

\begin{abstract}
The Open Radio Access Network (Open RAN) framework, emerging as the cornerstone for Artificial Intelligence (AI)-enabled Sixth-Generation (6G) mobile networks, heralds a transformative shift in radio access network architecture.  As the adoption of Open RAN accelerates, ensuring its security becomes critical. The RAN Intelligent Controller (RIC) plays a central role in Open RAN by improving network efficiency and flexibility. Nevertheless, it also brings about potential security risks that need careful scrutiny. Therefore, it is imperative to evaluate the current state of RIC security comprehensively. This assessment is essential to gain a profound understanding of the security considerations associated with RIC. This survey combines a comprehensive analysis of RAN security, tracing its evolution from 2G to 5G, with an in-depth exploration of RIC security, marking the first comprehensive examination of its kind in the literature. Real-world security incidents involving RIC are vividly illustrated, providing practical insights. The study evaluates the security implications of the RIC within the 6G Open RAN context, addressing security vulnerabilities, mitigation strategies, and potential enhancements. It aims to guide stakeholders in the telecom industry toward a secure and dependable telecommunications infrastructure. The article serves as a valuable reference, shedding light on the RIC's crucial role within the broader network infrastructure and emphasizing security's paramount importance. This survey also explores the promising security opportunities that the RIC presents for enhancing network security and resilience in the context of 6G mobile networks. It outlines open issues, lessons learned, and future research directions in the domain of intelligent control in 6G open RAN, facilitating a comprehensive understanding of this dynamic landscape.
\end{abstract}

\section{Introduction}\label{sec:Intro}
\IEEEPARstart{O}{pen} RAN framework has gained significant attention as a potential basis for AI-enabled 6G mobile networks. The Open RAN is an architecture that aims to disaggregate, virtualize, and intelligently control various components of the traditional RAN, allowing for more flexibility, interoperability, and innovation in the design and deployment of cellular networks.  In the transition from 5G to 6G, specific requirements emerge that 5G RAN~\cite{3GPPTS38.401} struggles to fulfill. 6G necessitates seamless integration of diverse technologies, low latency with real-time edge processing, and intelligent management of highly heterogeneous networks, all of which surpass the capabilities of current RAN due to its constrained flexibility and openness. As a result, the Open RAN is being explored as the leading option for the future generation of RAN, as it can address these diverse requirements effectively~\cite{singh2020evolution}. By leveraging Open RAN along with AI technologies, 6G networks could potentially offer enhanced performance, improved resource allocation, and better user experiences~\cite{rahman2021network},~\cite{wang2023road}. 
The existing RAN architectures are tightly woven together, obtained from a singular provider as a closed, exclusive package of combined hardware and software. The RAN hardware is specifically designed for that supplier and cannot accommodate software from other sources. Open RAN represents a fresh approach to constructing the RAN. Its goal is to provide mobile operators with the capability to implement hardware and software solutions from a variety of suppliers within the same geographic region. Open RAN strives to construct RAN solutions on universally applicable hardware not tied to any specific vendor, employing open interfaces and software~\cite{Intelligent2023innovation, Intelligent2023ecosystems}.

Forecasts from the Body of European Regulators for Electronic Communications (BEREC)~\cite{ORANBEREC} and market projections indicate that Open RAN technology is poised to constitute a substantial share of the global RAN market, exceeding 15 to 20\% by 2027 with a projected market value exceeding \$1 billion~\cite{oranmarket}. This trajectory underscores the imperative need to safeguard the security of Open RAN networks, as potential security breaches could have far-reaching consequences. Moreover, the Network and Information Systems (NIS) Cooperation Group emphasizes Open RAN's capacity to introduce novel security risks, particularly in 6G networks~\cite{nis2022euoran}. Failing to comprehensively address these risks could imperil the security of extensive 6G deployments relying on Open RAN technology, underscoring the pressing need for robust security measures within this domain. Given this trajectory, safeguarding the security of Open RAN networks holds paramount importance for many enterprises, as potential threats could have widespread implications~\cite{Intelligent2023security}.

In this context, one of the pivotal components of Open RAN that demands careful attention is the RAN Intelligent Controller or RIC. This controller is instrumental in managing and controlling network operations in near-real-time, with response times ranging from 10 milliseconds to 1 second. The near-real-time RIC plays a central role in orchestrating and optimizing the RAN due to its remarkable capabilities~\cite{balasubramanian2021ric}. The RIC also introduces intelligence, agility, and programmability into RANs, fostering multivendor interoperability, seamless integration of third-party applications, and the resolution of critical challenges, such as cost optimization, network performance enhancement, and revenue diversification. Its distinctive role in enhancing network efficiency, flexibility, and overall performance underscores the imperative of the RIC~\cite{polese2023understanding}. Although RIC plays a transformative role in navigating the shift from traditional monolithic network architectures to the disaggregated and software-defined landscape of Open RAN, this transition reveals novel security challenges and opportunities. Given its paramount importance in Open RAN architecture, the security of the RIC is of utmost significance, necessitating dedicated attention and thorough investigation.

The RIC, positioned as a critical interface linking diverse network elements, becomes a focal point for potential security vulnerabilities. For instance, centralizing control in the RIC could make it more appealing for Denial of Service (DoS) attacks, and if key Application Programming Interfaces (APIs) are exposed to unauthorised software, it risks taking down the entire network. Any compromise in RIC security could cascade across the network, demanding vigilant attention~\cite{O-RAN.SFGRICXAPP}. Furthermore, as Open RAN technology gains momentum and market projections soar, the importance of RIC security magnifies. Security breaches within the RIC not only imperil network stability and performance but also erode trust within the Open RAN ecosystem, potentially impeding its widespread adoption. It's essential to recognize that RIC's origins trace back to the concept of a centralized control plane, akin to a Software-Defined Network (SDN)  controller, to manage and program nodes in the data plane~\cite{schmidt2021flexric,balasubramanian2021ric}. Consequently, RIC naturally inherits many of the security vulnerabilities associated with SDN controllers, necessitating consideration in securing the RIC. Therefore, the emphasis on RIC security complements broader discussions on Open RAN security, ensuring a comprehensive approach to safeguarding the integrity, confidentiality, and availability of Open RAN networks. From a different perspective, the open-source nature of Open RAN, combined with the RIC's capabilities, presents an opportunity to bolster network security. The RIC's real-time device identification and anomaly detection, augmented by external data, can effectively mitigate threats. This proactive approach plays a vital role in addressing security challenges and enhancing network security within the dynamic Open RAN framework, striking a balance between risks and opportunities.

This paper does not seek to determine whether Open RAN networks are inherently "more secure" compared to older architectures. Instead, its focus is on examining RIC vulnerabilities, attacks, and relevant solutions, as well as the opportunities that RIC gives us to make Open RAN more secure. This survey presents valuable insights, aiming to propel the Open RAN ecosystem forward. 
We perform a comprehensive examination of the security consequences associated with the intelligent controller in RAN. This investigation covers potential security benefits as well as drawbacks, given that this concept is still in its initial phases and detailed technical specifications are under formulation. The primary objective of this analysis is to ascertain whether additional suggestions or measures are necessary for the effective implementation of Open RAN networks. To this end, the study aims to answer the fundamental questions: Q1. How does Open RAN adoption and the introduction of 6G's RIC influence network security dynamics? Q2. What specific security vulnerabilities are linked to the RIC, and what effective mitigation strategies exist? Q3. In the context of 6G, what security enhancements can be harnessed through the capabilities of the RIC?

\subsection{Scope and Contributions}
This paper offers a thorough and comprehensive exploration of security aspects related to RIC within the context of 6G Open RAN. The primary contributions of this article are delineated as follows:
\begin{itemize}[leftmargin=*]
    \item Our survey combines an in-depth tutorial, tracing the evolution of RAN from 2G to 5G, with a comprehensive analysis of RAN security. This dual perspective offers readers valuable insights into the path leading to secure Open RAN in 6G.
    \item We offer an in-depth analysis and technical discussion on the vulnerabilities, threats, and countermeasures associated with RIC. This represents the first comprehensive exploration of RIC security in the literature.
    \item Our paper presents a compelling exploration of real-world security incidents involving RIC, offering detailed examples and scenarios that vividly illustrate the practical implications of these vulnerabilities.
    \item We conduct a comprehensive review of up-to-date published works and ongoing research endeavors pertaining to the security aspects of Open RAN, particularly RIC.
    \item  We identify potential security opportunities that stem from using RIC solutions in 6G. 
    \item Furthermore, we contribute by delineating several open issues, lessons learned, and future directions for secure RIC integration with emerging technologies, such as Satellite, Reconfigurable Intelligent Surfaces (RIS), Self-Organizing Networks (SON), and Digital Twin (DT) in 6G Open RAN domain.
\end{itemize}

In this study, our objective is to present a thorough analysis centered around the RIC as the 6G Open RAN intelligent controller, emphasizing its security aspects and potential vulnerabilities. The purpose is to offer an all-encompassing resource tailored to individuals seeking insight into this domain. The focus remains dedicated to comprehending the intricate landscape of the controller's functions within the realm of 6G, with a keen emphasis on safeguarding its operations amidst emerging threats. This work aims to serve as a singular point of reference, catering not only to those embarking on an exploration of this subject but also to those keen on understanding the pivotal role the RIC plays within the broader framework of network infrastructure.

\begin{figure}[!htbp]
\centering
\includegraphics[width=0.46\textwidth]{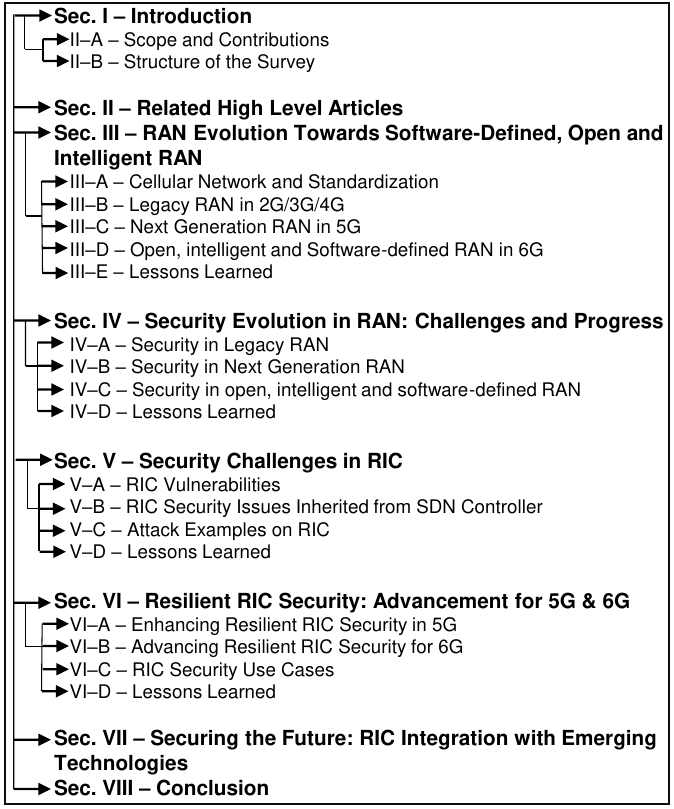}
\caption{\small Overview of the organization of the survey.}
\label{fig:SECTION}
\end{figure} 

\subsection{Structure of the Survey}

The remaining sections of this paper follow the structure outlined in Fig.~\ref{fig:SECTION}. Section~\ref{sec:Intro} serves as an introduction and delineates the precise boundaries and objectives that define the scope of the present survey. Advancing to the subsequent section, Section~\ref{sec:RelatedWorks} revolves around an analysis of related high-level articles. Notably, these articles have taken the form of surveys and reviews, specifically addressing the Software-defined Open RAN security landscape. 
Section~\ref{sec:RANEvolution}, as a natural progression, is dedicated to developing a comprehensive understanding. To achieve this, it lays a crucial foundation through an exploration of the RAN's evolution across 2G, 3G, 4G, and 5G generations. By establishing this context, the section seamlessly transitions into an examination of the intricate security aspects within the Open RAN framework. The spotlight remains on the transformation of RAN, emphasizing the progression from Legacy RAN to Next Generation RAN, ultimately culminating in the software-defined open and intelligent RAN paradigm. Section~\ref{sec:ORAN-SecurityRisks} comprehensively covers various dimensions of security, including security in Legacy RAN, security in Next Generation RAN, as well as security considerations in the context of Intelligent and Open RAN architectures. The insights garnered from the first four sections of this survey are instrumental in addressing our initial inquiry (Q1) concerning the impact of Open RAN adoption and 6G's RIC on network security dynamics. Section~\ref{sec:RIC-SecurityRisks} addresses security challenges in RIC, directly relating to the second question (Q2). This section provides insights into a range of topics, encompassing RIC vulnerabilities, instances of attacks on RIC, and the security issues carried over from SDN Controllers. In Section~\ref{sec:RICSecurity5G6G}, our emphasis turns to bolstering resilient RIC security, with a specific focus on advancements tailored to the 5G and 6G contexts. Within this section, we explore security solutions and practices designed to bolster the resilience of RIC security in the presence of these evolving threats, offering answers to Q3. Section~\ref{sec:lessonlearn} is dedicated the future research directions. This part reflects on insights gained and outlines potential paths for future research, effectively synthesizing the discussions relevant to all three questions (Q1, Q2, and Q3) and setting the stage for ongoing inquiry. For reader convenience, a table labeled Table~\ref{tbl:Acronym} contains the list of utilized acronyms.


\begin{table}[ht]
	\caption{Summary of Main Acronyms used in this survey}
 \label{tbl:Acronym}
    \centering
    \scriptsize
	\begin{tabular}{|l m{5.5cm}|}
		\hline
		\textbf{Acronym} 
		&\textbf{Definition}
		\\\hline \hline
        AI & Artificial Intelligence \\
        API & Application Programming Interface \\
        BBU & Based-Band Unit \\
        BS & Base Station \\
        BSC & Base Station Controller\\
        BSS & Base Station Subsystem \\
        C-RAN & Cloud-Radio Access Network \\
        CN & Core Network \\
        CP & Control Plane\\
        DDoS & Distributed Denial of Service \\
        DTLS & Datagram Transport Layer Security\\
        eNB & Evolved NodeB \\
        GERAN & GSM/EDGE Radio Access Network \\
        GUTI & Global Unique Temporary Identifier \\
        IETF & The Internet Engineering Task Force \\
        IMSI & International Mobile Subscriber Identity \\
        IPsec  & Internet Protocol Security\\
        KPI & Key Performance Indicator \\
        MEC & Multi-Access Edge Computing \\
        ME & Mobile Equipment \\
        ML & Machine Learning \\
        MNO & Mobile Network Operator \\
        MITM & Man-in-the-Middle \\
        NFV & Network Function Virtualization \\
        NGRAN & Next Generation Radio Access Network \\
        NIB & Network Information Base\\
        NR & New Radio \\
        nRT RIC & Near-Real Time RAN Intelligent Network \\
        non-RT RIC & Non-Real Time RAN Intelligent Network \\
        O-CU-CP & O-RAN Central Unit – Control Plane \\
        O-CU-UP & O-RAN Central Unit – User Plane \\
        O-DU & O-RAN Distributed Unit \\
        O-RU & O-RAN Radio Unit \\
        QoS & Quality of Service \\
        RAN & Radio Access Network \\
        RIC & RAN Intelligent Controller \\
        RRM & Radio Resource Management\\
        RRU & Remote Radio Unit \\
        SD-RAN & Software-Defined Radio Access Network \\
        SDN & Software-Defined Networking \\
        SDR & Software Defined Radio \\
        SDL & Shared Data Layer \\
        SMO & Service Management and Orchestration \\
        SDAP & Service Data Adaptation Protocol \\
        SDK & Software Development Kit \\
        SDO & Standard Development Organization \\
        SFG & Security Focus Group \\
        TMSI & Temporary Mobile Subscriber Identity \\
        TN & Transport Network \\
        OAM & Operations, Administration, and Maintenance \\
        ONF & Open Networking Foundation \\
        ONOS & Open Networking Operating System \\
        OSC & O-RAN Software Community \\
        PDCP & Packet Data Convergence Protocol \\
        R-NIB & RAN NIB \\
        RLC & Radio Link Control \\
        RRC & Radio Resource Control \\
        TIP & Telecom Infra Project \\
        TLS & Transport Layer Security \\
        UE & User Equipment \\
        UMTS & Universal Mobile Telecommunications System \\
        UP & User Plane \\
        ZTA& Zero Trust Architecture\\
\hline
	\end{tabular}
\end{table}

\section{Related High-Level Articles}\label{sec:RelatedWorks}

\begin{table*}[htbp]
    \caption{RIC Security Features in Surveys. Notations: $\CIRCLE$: Detailed Discussion. $\LEFTcircle$: Partial Discussion (i.e. At least one specialized section is presented but lacks a thorough examination). $\Circle$: Not Supported}
    \label{tbl:ProsandCons}
    \centering
    \scriptsize
    \begin{tabular}{l*{13}{>{\centering\arraybackslash}m{0.45cm}}}
         \hline
         \rule{0pt}{4ex}
        Covered topics & \shortstack{\cite{polese2023understanding}\\(2023)} & \shortstack{\cite{liyanage2022open}\\(2023)} & \shortstack{\cite{mao2023security}\\(2023)} & \shortstack{\cite{abdalla2022toward}\\(2022)} & \shortstack{\cite{zhao2022survey}\\(2022)} & \shortstack{\cite{mimran2022evaluating}\\(2022)} & \shortstack{\cite{nguyen2021security}\\(2021)} & \shortstack{\cite{boswell2020security}\\(2020)} & \shortstack{\cite{habibi2019comprehensive}\\(2019)} & \shortstack{\cite{abdou2018comparative}\\(2018)} & \shortstack{\cite{tian2017survey}\\(2017)} & \shortstack{\cite{khan2016topology}\\(2016)} & \shortstack{\textbf{Our}\\\textbf{survey}}\\
        \hline
        RAN architecture (2G/3G/4G) & $\Circle$ & $\LEFTcircle$ & $\Circle$ & $\Circle$ & $\Circle$ & $\Circle$ & $\Circle$ & $\Circle$ & $\CIRCLE$ & $\Circle$ & $\CIRCLE$ & $\Circle$ & $\CIRCLE$ \\
        RAN security (2G/3G/4G) & $\Circle$ & $\Circle$ & $\Circle$ & $\Circle$ & $\Circle$ & $\Circle$ & $\Circle$ & $\Circle$ & $\CIRCLE$ & $\Circle$ & $\CIRCLE$ & $\Circle$ & $\CIRCLE$ \\
        5G RAN architecture and security & $\LEFTcircle$ & $\Circle$ & $\Circle$ & $\LEFTcircle$ & $\CIRCLE$ & $\LEFTcircle$ & $\Circle$ & $\Circle$ & $\CIRCLE$ & $\Circle$ & $\CIRCLE$ & $\Circle$ & $\CIRCLE$ \\
        6G RAN architecture and security & $\CIRCLE$ & $\LEFTcircle$ & $\CIRCLE$ & $\CIRCLE$ & $\LEFTcircle$ & $\Circle$ & $\CIRCLE$ & $\Circle$ & $\Circle$ & $\Circle$ & $\Circle$ & $\Circle$ & $\CIRCLE$ \\
        SD-RAN security & $\Circle$ & $\Circle$ & $\LEFTcircle$ & $\LEFTcircle$ & $\CIRCLE$ & $\Circle$ & $\Circle$ & $\Circle$ & $\CIRCLE$ & $\LEFTcircle$ & $\Circle$ & $\LEFTcircle$ & $\CIRCLE$ \\
        6G Open RAN architecture & $\CIRCLE$ & $\CIRCLE$ & $\LEFTcircle$ & $\LEFTcircle$ & $\CIRCLE$ & $\CIRCLE$ & $\Circle$ & $\LEFTcircle$ & $\Circle$ & $\Circle$ & $\Circle$ & $\Circle$ & $\CIRCLE$ \\
        6G Open RAN security & $\CIRCLE$ & $\CIRCLE$ & $\Circle$ & $\Circle$ & $\CIRCLE$ & $\CIRCLE$ & $\Circle$ & $\LEFTcircle$ & $\Circle$ & $\Circle$ & $\Circle$ & $\Circle$ & $\CIRCLE$ \\
        RIC architecture & $\CIRCLE$ & $\Circle$ & $\Circle$ & $\LEFTcircle$ & $\Circle$ & $\Circle$ & $\Circle$ & $\Circle$ & $\Circle$ & $\Circle$ & $\Circle$ & $\Circle$ & $\CIRCLE$ \\
        RIC-specific vulnerabilities & $\Circle$ & $\Circle$ & $\Circle$ & $\Circle$ & $\Circle$ & $\LEFTcircle$ & $\Circle$ & $\Circle$ & $\Circle$ & $\Circle$ & $\Circle$ & $\Circle$ & $\CIRCLE$ \\
        RIC's inheritance of SDN vulnerabilities & $\Circle$ & $\Circle$ & $\Circle$ & $\Circle$ & $\Circle$ & $\CIRCLE$ & $\Circle$ & $\Circle$ & $\Circle$ & $\Circle$ & $\Circle$ & $\Circle$ & $\CIRCLE$ \\
        \textbf{RIC-specific security solutions in 5G \& 6G} & $\Circle$ & $\Circle$ & $\Circle$ & $\Circle$ & $\Circle$ & $\Circle$ & $\Circle$ & $\Circle$ & $\Circle$ & $\Circle$ & $\Circle$ & $\Circle$ & $\CIRCLE$ \\
        \textbf{Real-World threat scenarios for RIC} & $\Circle$ & $\Circle$ & $\Circle$ & $\Circle$ & $\Circle$ & $\Circle$ & $\Circle$ & $\Circle$ & $\Circle$ & $\Circle$ & $\Circle$ & $\Circle$ & $\CIRCLE$ \\
        \textbf{Practical scenario of RIC capability in 6G security} & $\Circle$ & $\Circle$ & $\Circle$ & $\Circle$ & $\Circle$ & $\Circle$ & $\Circle$ & $\Circle$ & $\Circle$ & $\Circle$ & $\Circle$ & $\Circle$ & $\CIRCLE$ \\
        \textbf{Ensuring secure RIC integration with future technologies} & $\Circle$ & $\Circle$ & $\Circle$ & $\Circle$ & $\Circle$ & $\Circle$ & $\Circle$ & $\Circle$ & $\Circle$ & $\Circle$ & $\Circle$ & $\Circle$ & $\CIRCLE$ \\
        \hline
    \end{tabular}
\end{table*}

In recent years, several notable surveys and tutorials related to RAN security have emerged in academic literature, as referenced in~\cite{polese2023understanding,liyanage2022open,mao2023security,abdalla2022toward,zhao2022survey,mimran2022evaluating,nguyen2021security,boswell2020security,habibi2019comprehensive,abdou2018comparative,tian2017survey,khan2016topology}. These works have aimed to document and analyze the technical advancements and issues within the realm of RAN security. In the subsequent sections, we will concisely overview these pertinent surveys' research focus and contributions. To elucidate the distinction and originality of our survey, we provide a comprehensive comparison between our survey and closely related ones in Table~\ref{tbl:ProsandCons}. It becomes evident that our survey stands out as the sole one that offers a wide spectrum of critical aspects in the realm of RIC security. We clearly define RIC-specific vulnerabilities, investigate how RIC inherits vulnerabilities from SDN, outline real-world threat scenarios specific to RIC, and offer practical insights into RIC's role in 6G security. Furthermore, our survey uniquely explores the security risks of RIC's integration with future technologies such as satellite communication systems and SON. We also scrutinize the security implications of integrating RIC with RIS, a distinctive approach to optimizing wireless communication via passive surface elements. Finally, we delve into the integration of RIC with DT technology, enabling real-time network monitoring and predictive analysis, setting our survey apart from others in the field. Additionally, the limitations of other surveys are evident through this comparison. While various existing surveys focus on Open RAN, 5G RAN, 6G RAN, or Software Defined RAN (SD-RAN), our survey distinguishes itself by its specific areas of emphasis. It conducts an in-depth review that centers on the correlation between the security aspects of the RIC. Notably, our survey addresses a critical need within the field. A comprehensive survey that delves deeply into the realm of security vulnerabilities and solutions concerning the RIC is notably absent. This need for an extensive exploration is underscored by the tabulated comparison in Table~\ref{tbl:ProsandCons}, which outlines and contrasts the findings of these prior surveys on Open RAN, SDN, and security aspects. 
\\\indent The proposed survey by Polese et al.~\cite{polese2023understanding} presents an examination of how Open RAN is poised to transform upcoming cellular networks. The authors achieve this by evaluating the technical specifications and architectural elements of Open RAN, the interconnections between these elements, and the Machine Learning (ML) and closed-loop control processes enabled and standardized by Open RAN. Conclusively, the authors review recent findings concerning Open RAN's design and optimization while delving into the unresolved matters that must be dealt with to fully materialize the Open RAN vision. The primary aim is to provide interested readers with a lucid portrayal of the current status of Open RAN, along with a profound comprehension of the opportunities introduced by Open RAN within the realm of cellular networks~\cite{polese2023understanding}. However, it's worth noting that the specific focus on the RIC and its security risks and challenges is extremely limited in the paper. In fact, the paper does not provide explicit solutions or countermeasures to address potential attacks or vulnerabilities related to RIC.

A classification system identifying security risks inherent in the Open RAN framework is furnished by Liyanage et al.~\cite{liyanage2022open}. Each of these risks is expounded upon, outlining the potential impact. Distinctive remedies aimed at addressing security vulnerabilities within Open RAN are introduced, leveraging technologies such as blockchain, the physical layer, and artificial intelligence. A succinct overview of prevalent design flaws related to Open RAN, along with their repercussions and potential mitigation approaches, is provided. Finally, a compilation of security advantages unique to Open RAN and those already observed in virtual RAN and 5G networks is presented for discussion~\cite{liyanage2022open}. It's important to highlight that the security aspects of the RIC were not a primary focus of their discussion.

Mao et al.~\cite{mao2023security} present the service prerequisites for 6G and concentrate on critical methods pertaining to the 6G network's edge. They elucidate the interconnections between edge computing, edge caching, and edge intelligence. Furthermore, they delve into the security needs and potential risks associated with edge computing, edge caching, and edge intelligence. Their proposed article comprehensively examines research on strategies for safeguarding security and privacy. It also addresses the incorporation of Open RAN, Federated Learning (FL), and blockchain in the context of how these approaches can be employed to formulate security and privacy strategies. The article concludes by summarizing the limitations of current research in meeting 6G requirements and highlights possible future directions for investigation~\cite{mao2023security}. It is important to highlight that in their work, the authors also did not delve into any discussions regarding RANs, including the Open RAN architecture, RIC architecture, and their associated security aspects and vulnerabilities. This omission is notable as these aspects are critical components of the evolving network landscape and play a pivotal role in shaping the security and functionality of next-generation networks. 

The paper proposed by Abdalla et al.~\cite{abdalla2022toward}, characterized by a limited bibliography, focuses on discerning the constraints within the present architecture and technologies of Open RAN. Additionally, it outlines opportunities for research and development to overcome these limitations. It is important to note that the paper does not encompass any discussions related to the security aspects of RIC~\cite{abdalla2022toward}.

Zhao et al.~\cite{zhao2022survey} address challenges in open-source-defined wireless networks including Open RANs and propose solutions. They detail integrating open-source software, hardware, projects, and frameworks into the core, RAN, and edge networks. They analyze upcoming evolution trends in open-source wireless networks and discuss 5G's industrial security view. They highlight research efforts from academia and industry on the impact of these networks. They also present a conceptual federated SDN controller-based open-source-defined wireless network. They share lessons learned, ongoing research, and future directions for adopting open-source-defined networks in communications~\cite{zhao2022survey}.

Mimran et al.~\cite{mimran2022evaluating} formulated a comprehensive procedure for assessing the security of the Open RAN architecture. This procedure encompasses various components, including a purpose-built ontology to evaluate security vulnerabilities inherent in Open RAN's structure. Leveraging this ontology, they constructed a classification system grounded in the Open RAN architecture and its analytical objectives. Additionally, they elucidated their approach to conducting a survey of potential threats, which holds potential for subsequent evaluations of forthcoming Open RAN architecture iterations. Applying their assessment process, they conducted an exhaustive analysis of general threats to Open RAN. Their investigation encompassed past breaches across diverse domains, scrutinizing their relevance to the various risk facets within the Open RAN framework. They also correlated their operational and security consequences. Building upon the insights gleaned from their threat analysis, they pinpointed several promising avenues for future research aimed at bolstering the overall security of the Open RAN architecture. Toward the end of their work, it's important to note that while they delved into various aspects of the Open RAN architecture's security evaluation process, they did not explicitly address the security considerations of the RIC. Their focus primarily encompassed developing a comprehensive security evaluation process and conducting a general threat analysis of the Open RAN architecture~\cite{mimran2022evaluating}. 

The study proposed by Nguyen et al.~\cite{nguyen2021security} offers a comprehensive examination of the historical development of security structures and vulnerabilities within older network systems. By exploring the deficiencies in established standards and delving into technical flaws in the protocols of these networks, the necessary improvements for ensuring security and privacy in 6G networks are emphasized. Secondly, their survey presents a holistic perspective on security and privacy concerns, highlighting the need for adjustments in current solutions to meet the evolving requirements of 6G. Lastly, their discussions regarding the lessons gleaned from the deficiencies in existing security architecture and the ongoing technical challenges can assist researchers and developers in promptly identifying pertinent issues and establishing initial points of focus for future work~\cite{nguyen2021security}. It's important to highlight that the authors did not specifically address the RAN section of 6G in their work. Instead, they offered an overview of the broader 6G landscape, omitting discussions related to topics like the Open RAN, RIC architecture, security considerations, and associated vulnerabilities. This omission is significant because RANs constitute vital components in the evolving network paradigm, and their security and functionality play a pivotal role in shaping the next generation of networks.

The white paper by Boswell et al.~\cite{boswell2020security} presents a limited bibliographic survey on Open RAN security. It is noteworthy that this paper does not discuss the security aspects of the RIC. While it partially encompasses a brief review of RAN evolution, RAN virtualization, O-RAN security risks, areas of concern beyond open networks, and delineates security best practices, it additionally provides an overview of the security challenges prevalent within the domain of Open RAN architecture, particularly as applied to industrial use cases. It is important to acknowledge, however, that the scope of coverage pertaining to these subjects remains non-comprehensive.

The paper proposed by Habibi et al.~\cite{habibi2019comprehensive} delves into the details of the 5G communication system, discussing its service categories, proposed system architecture, advantages, applications, and implementation issues. A comprehensive review of the evolution of historical and current RAN architectures is presented. Various RAN implementations reported in the literature are discussed in depth, along with ongoing standardization activities and the need to redesign legacy RAN architectures to meet the requirements of next-generation mobile networks. The paper presents a comprehensive analysis and review of the literature on Cloud-RAN (C-RAN), Heterogeneous Cloud RAN (H-CRAN), Virtualized Cloud RAN (V-CRAN), and Fog RAN (F-RAN). It explores these RAN architectures from multiple perspectives: energy consumption, security, costs (CAPEX/OPEX), performance, spectrum utilization, mobility, resource allocation, and overall system architecture. A significant contribution of the paper is an extensive comparative analysis of C-RAN, H-CRAN, V-CRAN, and F-RAN. This analysis aims to study these RAN architectures from various angles, including energy consumption, system architecture, costs, and system-level performance. 
The paper provides an exclusive review and comprehensive survey of key enabling technologies for 5G RAN. It explores innovations that enhance spectrum efficiency, reduce costs, and meet end-user expectations and the requirements of vertical industries. The paper addresses existing challenges and outlines future research directions in the realm of 5G RANs and Radio Access Technologies (RATs). The intention is to stimulate the community to develop practical solutions with new technical advancements~\cite{habibi2019comprehensive}. The paper's contribution section does not specifically address the Open RAN concept or its security implications. Additionally, the authors do not discuss the RIC and its security considerations in the context of the paper's content. The focus remains on providing a detailed examination of 5G RAN.

The paper proposed by Abdou et al.~\cite{abdou2018comparative} identifies five crucial control functions necessary for a practical production network to deliver essential network services. A comprehensive analysis is conducted on threats targeting these five functions when implemented through conventional networks in Layer 2 (L2) and Layer 3 (L3) along with SDNs. The paper also explores defence mechanisms to counter these threats. The paper introduces a novel evaluation framework that enables an objective comparison of security between L2 networks, L3 networks, and SDNs. This framework employs two distinct threat models that define the attacker's position within the network~\cite{abdou2018comparative}. However, it should be emphasized that the paper does not delve into discussions about RANs, Open RAN, and the security aspects of RIC.

The paper proposed by Tian et al.~\cite{tian2017survey} introduces the C-RAN architecture and explores its primary application scenarios. It outlines the distinct characteristics that define C-RAN. The paper examines the security threats and vulnerabilities associated with C-RAN. It reviews existing studies in the literature that address these security concerns. Different solutions to mitigate security threats across various logic layers of C-RAN are introduced, along with a discussion of their advantages and disadvantages. The paper establishes security and trust requirements specific to C-RAN. These requirements serve as benchmarks to identify areas with unresolved issues. The paper uses these requirements to identify open problems within C-RAN security and trust. Notably, the paper does not discuss the Open RAN concept or its implications, focusing solely on C-RAN~\cite{tian2017survey}.

The paper proposed by Khan et al.~\cite{khan2016topology} offers a thorough understanding of SDN, covering its core concepts and discussing various threats that target the layered architecture of SDN. The paper emphasizes the significance of topology discovery and its role in both traditional networks and SDN environments. The paper introduces a comprehensive thematic taxonomy to systematically categorize topology discovery into distinct groups. These categories include objectives, controller platforms, dependent services, discovery entities, and controller services. The paper classifies threats related to topology discovery, detailing state-of-the-art security solutions. It covers aspects such as attack entities, vulnerabilities in controllers, types of attacks, and the occurrence of threats. The paper presents potential research directions for advancing topology discovery within SDN. It also offers recommendations for potential solutions to tackle the challenges posed by topology discovery~\cite{khan2016topology}. Notably, the paper does not encompass the RIC or Open RAN concepts, including their associated security aspects, within its purview.

Acknowledging the importance of the subject matter, it has come to our attention that the reviews discussed in this related work section do not explicitly address the intricate domain of 6G RIC and its associated security aspects. In light of this observation, we have taken the initiative to conduct an exhaustive survey, focusing explicitly on uncovering and elucidating the security risks, as well as identifying potential solutions and opportunities within the realm of 6G RIC.

\section{RAN evolution towards software-defined, open and intelligent RAN} \label{sec:RANEvolution}

To understand the security nuances of the Open RAN intelligent controller in the 6G era, we must first delve into the historical tapestry of cellular networks, spanning from 2G to 5G. This foundational exploration not only enriches our perspective but also equips us to navigate the intricate security challenges within the Open RAN framework. As we embark on this journey, section~\ref{sec:Standard} sheds light on the core tenets of cellular networks, emphasizing the pivotal role of standardization in molding the mobile communication landscape. Meanwhile, section~\ref{sec:LegacyRAN} offers a retrospective look at RAN's inception and evolution, charting its course from 2G through to 4G and the inherent challenges faced along the way. Transitioning to the more recent advancements, section~\ref{sec:NGRAN} captures the revolutionary essence of 5G RAN, highlighting its innovative strategies and features. As we set our sights on the future in section~\ref{sec:openIntRAN}, we delve into the anticipated world of 6G. Here, we encounter a realm marked by open architectures, intelligence-driven operations, and software-centric approaches. Our narrative culminates in section~\ref{sec:RANEVOLesson}, where we distill the pivotal moments and lessons from RAN's rich evolutionary history, reflecting on the innovations that have shaped this dynamic field.

\subsection{Cellular Network and Standardization} \label{sec:Standard}
The telecommunications mobile network consists of four primary parts, including User Equipment (UE), RAN, Transport Network (TN), and core network. The UE allows a user (e.g. smartphones, connected vehicles) access to mobile network services. The UE is subdivided into the Universal Subscriber Identity Module (USIM) or Universal Integrated Circuit Card (UICC) domain and the ME (Mobile Equipment) Domain. The core network processes multiple functionalities, including mobility management, call and session management, billing management, and security, providing the UE with requested services. The RAN is the access link to the UE to connect to the core network and the TN provides connectivity between the RAN and the core network. Different components of the mobile network in RAN, TN, and core network, communicate and exchange information with each other through a standardized method, which is referred to as an interface. Interfaces define the rules and protocols that govern how data is exchanged between different parts of the network, allowing them to work together seamlessly~\cite{parvez2018survey}. Mobile Network Operators (MNOs), as entities that provide mobile network services to users, rely on suppliers for assistance in operating and maintaining their own networks. These suppliers, including network equipment manufacturers, cloud service providers, security and maintenance contractors, and infrastructure providers, offer services and infrastructure to MNOs to support the construction and operation of their mobile networks. By working with suppliers, MNOs can acquire the necessary infrastructure and services to construct and maintain mobile networks, which enables them to provide mobile network services to users~\cite{coronado2022zero}.

The Third Generation Partnership Project (3GPP) is a global partnership comprising Standard Development Organisations (SDOs) from various continents, responsible for developing standards for mobile communications. It comprises seven Organisational Partners from different regions, including European Telecommunications Standards Institute (ETSI), the USA's Alliance for Telecommunications Industry Solutions (ATIS), China's Communications Standardization Association (CCSA), Japan's Association of Radio Industries and Businesses (ARIB) and Telecommunication Technology Committee (TTC), Korea's Telecommunications Technology Association(TTA), and India's  Telecommunications Standards Development Society, India (TSDSI). 3GPP's primary mandate is to establish and evolve the standards for mobile communications for all generations, ranging from 2G to 5G. These standards, which 3GPP designates as "Releases," encompass a vast array of novel nodes, interfaces, and functions that articulate each generation of mobile networks. By working with MNOs and suppliers, 3GPP can develop and establish these standards, which ensure interoperability and compatibility between different components of the mobile network industry~\cite{coronado2022zero}.

\subsection{Legacy RAN in 2G/3G/4G}\label{sec:LegacyRAN}
Legacy RAN refers to the first three generations of mobile networks, which include 2G, 3G, and early 4G networks. These networks played a pivotal role in transitioning from analog phone services to Internet Protocol (IP)-based networks, marking a transformative phase in telecommunications. These networks have a closed and monolithic architecture with tightly integrated hardware and software components. They were originally designed to support voice and basic data services and were not optimized for today's complex mobile applications. Additionally, traditional RAN uses proprietary protocols and closed systems, which limit interoperability. The standardization of these networks was led by the 3GPP, with specific releases dedicated to each generation, including 3GPP Release 99 for 2G, 3GPP Release 4-12 for 3G, and 3GPP Release 8-15 for early 4G. These releases laid the foundation for the initial mobile communication networks, paving the way for more advanced network architectures.

\subsubsection{GERAN (BSS) in 2G } The 2G wireless system, known as the Global System for Mobile Communications (GSM), was initially designed in the 1980s for direct and exclusive voice communication between two mobile stations~\cite{sempere1997overview}. However, the emergence of mobile Internet changed its purpose, leading to the development of data services and the introduction of the Enhanced Data Rates for the GSM Evolution (EDGE) network. This transition allowed wireless devices to connect to the Internet and Intranet, expanding the capabilities of the GSM network~\cite{sauter2010gsm}. A GSM/EDGE network comprises three subsystems: the Mobile Station (MS), the Base Station Subsystem (BSS) or the GSM/EDGE Radio Access Network (GERAN), and the Network Subsystem (NSS) or the Core Network (CN). The GERAN subsystem encompasses all the nodes and functionalities necessary for connecting MS to the CN. The network nodes within GERAN include Base Transceiver Stations (BTSs), also called base stations (BS), and the Base Station Controller (BSC)..

\begin{figure} 
    \centering
  \subfloat[Centralized BS (2G)\label{fig:RFBBU}]{%
       \includegraphics[width=0.43\textwidth]{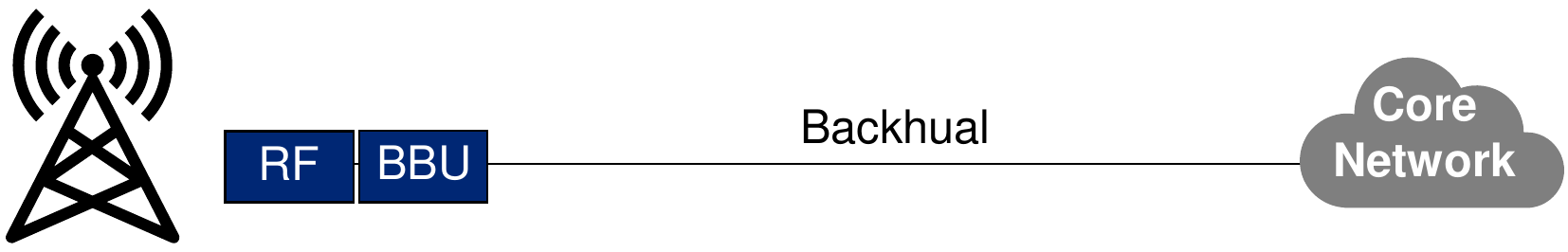}}
    \vspace{-4mm}
  \subfloat[RRH-BBU split (3G/4G)\label{fig:RF-BBU1}]{%
        \includegraphics[width=0.43\textwidth]{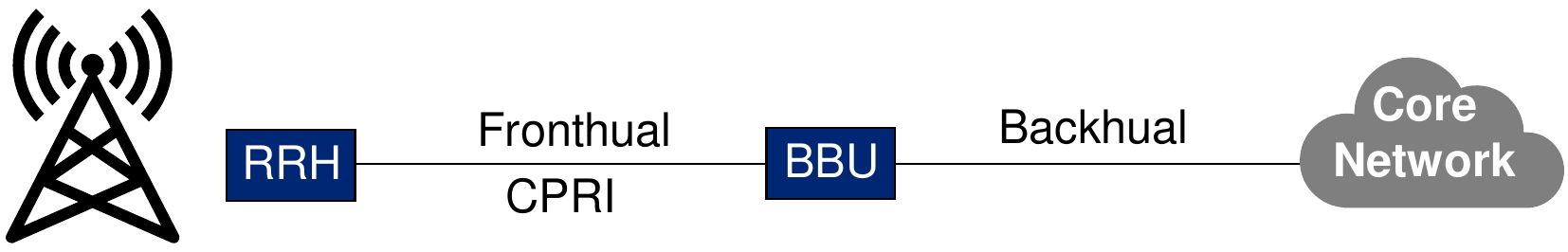}}
    \vspace{-4mm}
  \subfloat[RU-DU-CU split (5G)\label{fig:RF-BBU2}]{%
        \includegraphics[width=0.43\textwidth]{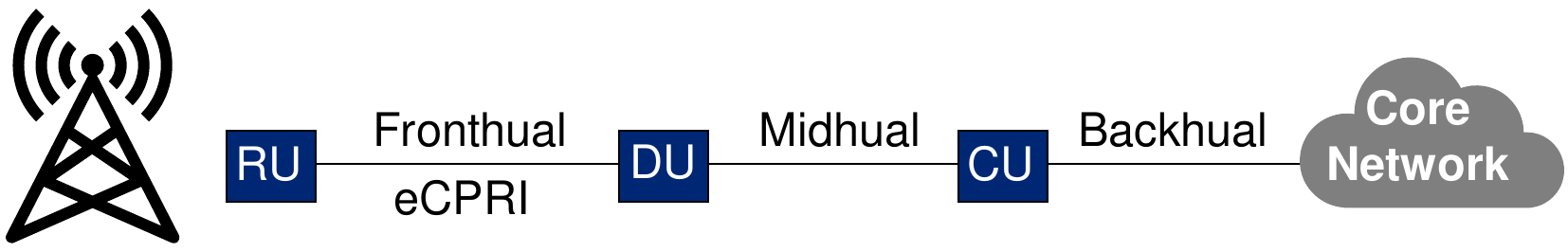}}

  \caption{Base station disaggregation from 2G to 5G}
  \label{fig:BScomponent} 
\end{figure}

The BTSs, which are the most numerous components of a mobile network, communicate with MS over the air interface or the \textit{Um} interface. They consist of five key components, namely antennas, Radio Frequency (RF) modules, digital units (also known as BaseBand Units or BBUs), transmission units, and control units~\cite{nahas2012base}. The antenna, which is the most visible part of a mobile network, can cover a specific area, referred to as a \textit{cell}. The RF module receives or transmits signals over the air interface, converting them to or from digital data. The BBU processes the encoded signal before sending or receiving it from or to the CN through the transmission unit. The coordination between these three functions is maintained by the control unit~\cite{simic2007evolution}. The RF and BBU units are centralized in a single location at the base of a tower. This arrangement allows coaxial copper (or RF) cables to connect the RF unit to the antennas~\cite{alimi2017toward}. This configuration is visually exemplified in Fig.~\ref{fig:RFBBU}, illustrating the centralized BS architecture in Legacy RAN.

The central device responsible for controlling BTSs is BSC which communicates through the Abis interface. The BSC's primary responsibilities include scheduling air interface resources, establishing, releasing, and maintaining all connections between the BTS and MS. The BSC also ensures the Quality of Service (QoS), such as minimum bandwidth, maximum throughput, latency, load balancing between different simultaneous radio bearers, mobility management, and interference management~\cite{sauter2010gsm}. The BSC interacts with the Circuit-Switched (CS) sections of the core network via the \textit{A} interface, which can manage voice services (see GERAN architecture in Fig.~\ref{fig:LRANArch}). All Packet-Switched (PS) services are routed to and from the CN through the \textit{Gb} interface, which often requires a connection to the Internet.
\subsubsection{UTRAN (RNS) in 3G } The 3G mobile network, also known as Universal Mobile Telecommunications System (UMTS), aimed to merge voice and high-speed data services into a single system. The UMTS Terrestrial Radio Access Network (UTRAN) was the foundation for this system. UMTS was primarily designed in the 1990s to offer fast packet data services for high-speed Internet access to both private and industrial customers. Initially, the 3G network started with bandwidths of 128 Kbps for mobile stations and 2 Mbps for fixed applications. However, as 3G evolved, it supported higher data rates and introduced new services like video telephony, streaming, Multimedia Message Services (MMS), mobile TV, and mobile Internet~\cite{garg2001wireless,patel2018comparative}.

UTRAN proposed a new architecture to replace the existing BTS and BSC with NodeB (NB) and Radio Network Controller (RNC), respectively (see UTRAN architecture in Fig.~\ref{fig:LRANArch}). The NB is responsible for transmitting and receiving information over the air interface, including channel coding, spreading, and power control. To improve the deployment of modules in the NB, the RF module is installed close to the antennas, and the BBU acts as an interface to the CN over a high-speed backhaul connection. This configuration reduces the length of costly coaxial copper cables to the antennas, known as \textit{Remote Radio Head (RRH)} or \textit{Remote Radio Unit (RRU)}~\cite{larsen2018survey}. The RRH contained only the radio functions, whereas the BBU contained all baseband processing functions. This deployment method can result in significant savings, especially if the antennas and the base station cabinet cannot be installed near each other. Every RRH and BBU pair was linked through a newly established network segment known as the fronthaul network, which typically constituted a point-to-point connection. The transmission of radio signals across this network employed protocols such as the Common Public Radio Interface (CPRI)\cite{CPRIspec}, the Open Base Station Architecture Initiative (OBSAI)\cite{OBSAspec}, or the Open Radio Interface (ORI)\cite{ORIspec}, as depicted in Fig.\ref{fig:RF-BBU1}. The \textit{Iub} interface connects the NBs to an RNC. The RNC establishes radio links to wireless devices through base stations, manages connections, ensures QoS, and facilitates handovers between BTSs. Compared to GSM, the \textit{Iub} interface has a higher capacity, and NBs provide high-speed data rate links. The RNC is linked to the core network's CS and PS sections through the \textit{Iu(cs)} and \textit{Iu(ps)} interfaces, respectively.
\subsubsection{E-UTRAN in 4G}
The 4G Long-Term Evolution (LTE) network offers significant improvements over previous radio networks, GERAN and UTRAN. E-UTRAN, or Evolved Universal Terrestrial Radio Access Network, is the RAN component of the 4G LTE network. E-UTRAN's base stations were redesigned in 2009 to provide improved user path latency and distributed management tasks. In contrast to previous generations of RANs, where a central device such as BSC or RNC controlled the base stations, the majority of functionalities that were formerly part of the radio network controller are now integrated into the base station itself. This base station is referred to as the eNodeB (eNB) which handles the management of the air interface, QoS, user management, load balancing, mobility, and interference control.

\begin{figure}[!htbp]
\centering
\includegraphics[width=0.47\textwidth]{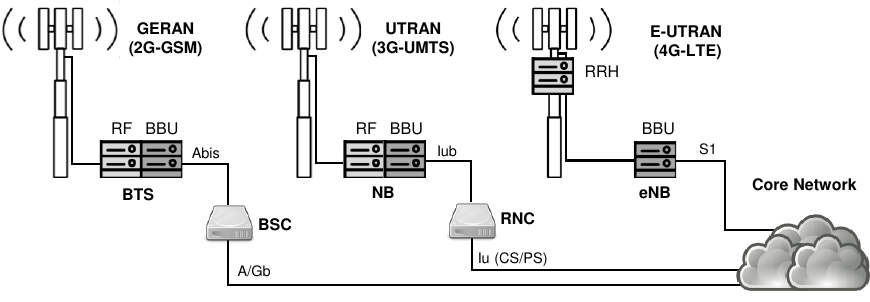}
\caption{\small Legacy RAN from 2G to 4G.}
\label{fig:LRANArch}
\end{figure}

The BBU in eNB acts as an interface to the CN over a high-speed backhaul connection, and the interface between the eNB and CN is referred to as the S1. The S1 interface is divided into two logical parts, where the S1 User Plane (S1-UP) transports user data, and the S1 Control Plane (S1-CP) transfers signaling data. The Fig.~\ref{fig:LRANArch} illustrates the evolution of legacy RAN from 2G to 4G cellular networks.


\subsection{Next Generation RAN in 5G} \label{sec:NGRAN}
 While Legacy RAN enabled mobile communications on a large scale, it also presented several limitations. These networks had tightly integrated hardware and software components, which made it difficult to replace or upgrade specific elements of the RAN without impacting the overall system. This resulted in high capital and operating costs, as well as limited flexibility in RAN deployment and management. Furthermore, the proprietary nature of RAN architectures created vendor lock-in, where mobile operators were dependent on a single vendor for their entire RAN infrastructure, which reduced competition and innovation in the market. 

To tackle these challenges effectively, the mobile industry has made significant strides towards next-generation RAN architectures in 5G. Within this 5G architecture, the term gNB refers to the next-generation NodeB, a base station that facilitates communication between the network and the user equipment. These architectures are designed to be highly flexible, scalable, and programmable. To achieve these goals, the industry has embraced several recent technologies in 5G, including SDN~\cite{kreutz2014software}, Network Functions Virtualization~(NFV)~\cite{mijumbi2015network}, and cloud-native principles advocated by the 3GPP~\cite{3GPPTS23.501}. By adopting these approaches, the industry has gained access to various features and benefits, ultimately leading to the realization of its objectives~\cite{shafi20175g}. Next-generation RANs are characterized by three key features: cloudification and virtualization of RAN functions, disaggregation of RAN software and hardware, and moving from specifically designed hardware to general-purpose hardware. In the following sections, we will explore the details of each category.
\subsubsection{Virtualization and Cloudification on RAN (C-RAN)}
Virtualization and cloudification are two technologies that work together to optimize the performance of RANs and allow network scaling based on demand. Virtualization creates an abstraction layer, known as a \textit{hypervisor}, on top of hardware, enabling software to operate. Through the use of NFV, Network Functions (NFs) and services can be moved from dedicated devices to generic servers and represented as Virtual Machines (VMs). Cloudification, on the other hand, involves relocating software from a local environment to a centralized cloud infrastructure, which provides a more efficient and flexible way to manage network resources~\cite{wu2015cloud}.

\begin{figure*}[!htbp]
\centering
\includegraphics[width=\textwidth]{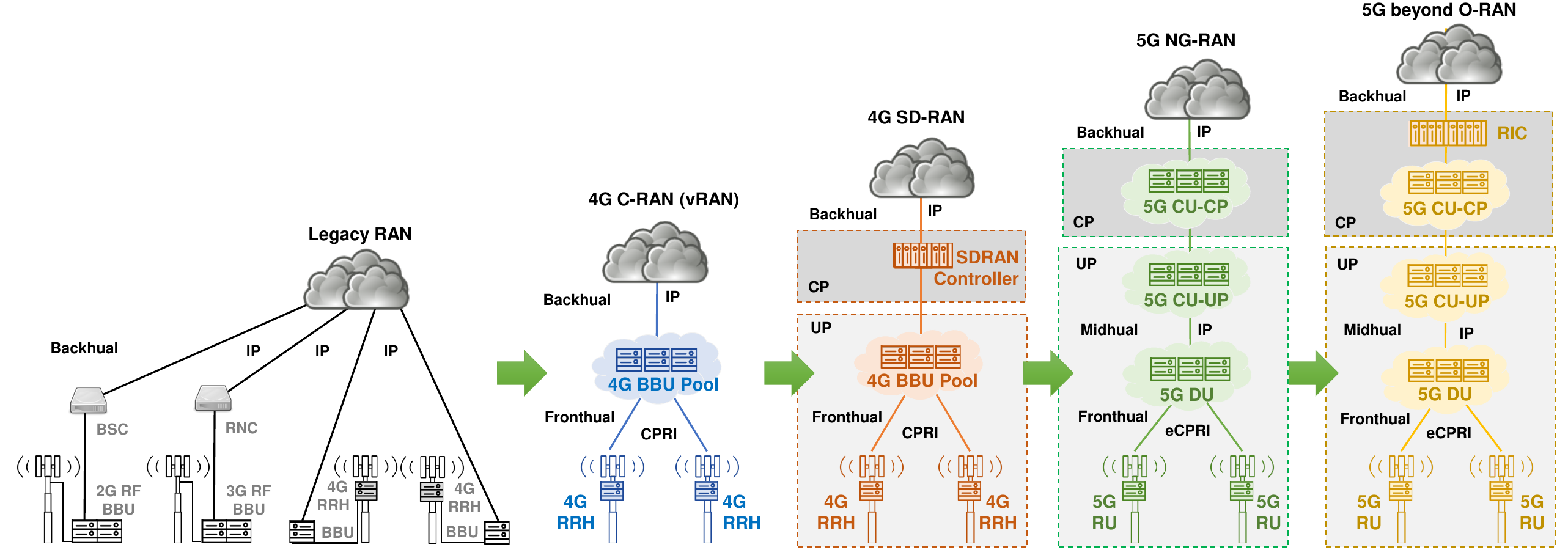}
\caption{\small RAN evolution toward software-defined open RAN.}
\label{fig:LRAN}
\end{figure*}

The combination of virtualization and cloudification in RAN, also C-RAN, utilizes centralized processing and virtualization to reduce the complexity and cost of RAN hardware while also enhancing network capacity and efficiency~\cite{checko2014cloud,peng2016recent}. In the C-RAN framework, BBUs execute the baseband processing tasks on a standardized and open software-defined radio platform. These BBUs are capable of handling multiple air interface standards and can readily adapt to upgraded wireless signal processing algorithms. Furthermore, the application of virtualization technology enhances the versatility of base stations, allowing those from various operators to collaborate effectively by sharing resources and computational capacity. C-RAN architectures employ a centralized BBU pool by consolidating the BBUs of a cluster of base stations, separating data processing from coverage (see C-RAN architecture in Fig.~\ref{fig:LRAN}). Hundreds and thousands of BBUs are placed centrally in a giant computer room. This enables the sharing of processing resources across different sites while maintaining support for low-latency services~\cite{parvez2018survey}. As a result, only antennas and a few active RF components, such as RRUs or RRHs, are left on the cell sites. The virtualization of the BBU pool introduced the concept of shared processing, which enables processing resources to be shared among various sites and provides additional processing capacity when required in different areas.

\subsubsection{Softwarization Technology on RAN (SD-RAN)}\label{sec:SDNarch}

The concept of SDN~\cite{cox2017advancing,casado2007ethane,casado2006sane} has been brought to the RAN domain under the generic term Software-Defined Radio Access Network (SD-RAN)~\cite{yang2013openran}. The SDN architecture contains three major layers, including the data plane (also known as the infrastructure layer), the control plane, and the management plane (also known as the application layer)~\cite{casado2006sane}. Specifically in SD-RAN architecture, BSs are considered reconfiguring devices in the data plane, which can be remotely controlled and programmed by the controller in the control plane~\cite{zaidi2018will,kitindi2017wireless}. The SD-RAN controller is the intelligent part and is in charge of translating the SDN application requirements in the management layer to a set of configurations to be applied on BSs in the data plane. For this communication, the SD-RAN controller uses the southbound interfaces through the management and control protocols, such as OpenFlow~\cite{mckeown2008openflow}, NETCONF~\cite{rfc6241} and OpFlex~\cite{smith-opflex-00}. Among these, OpenFlow is a well-known and open standard protocol designed and developed by Open Networking Foundation (ONF)~\cite{ONF}.

The SD-RAN controller communicates with the BSs in the data plane via the OpenFlow channel, delivering the necessary commands and messages. A software-defined BS is equipped with a flow-table pipeline, where each node hosts a series of flow tables beginning with table zero. The BS populates these tables with flow entries that contain instructions. When a data packet arrives at the BS, it triggers a table lookup sequence. During this sequence, the packet's header is matched against the flow entries' criteria. The packet progresses through the pipeline's tables until a match is found, at which point the actions specified in the corresponding flow entry are applied to the packet. If there is no match, a condition known as a \textit{table-miss}, the BS forwards the packet to the controller via a \texttt{PACKET-IN} message. The controller then assesses this message and decides the next steps, which may include discarding the packet or responding with a \texttt{FLOW-MOD} message to set up new flow entries in the BS for future packets of this kind. Alternatively, the controller may opt to simply forward the packet with a \texttt{PACKET-OUT} message without establishing any new rules at the BS~\cite{OpenFlowSpec}.

The application layer is designed to accommodate an array of sophisticated applications that serve as decision-makers for the management and enhancement of intricate RAN functions. Communication between the controllers and these applications is facilitated by a northbound interface, which also connects to the orchestration and automation layer. This layer encompasses critical systems including the Business Support System (BSS), Operations Support System (OSS), and Network Management System (NMS). Pioneering academic studies have contributed to the development of SD-RAN, with notable works on frameworks such as SoftRAN~\cite{gudipati2013softran} and FlexRAN~\cite{foukas2016flexran}. To further enhance the flexibility and efficiency of RANs, SD-RAN can be combined with Software Defined Radio (SDR) technology. SDR allows for the software-based control and management of radio resources, enabling the dynamic adjustment of parameters such as power, modulation, and frequency~\cite{marzouk2020energy}. By integrating SD-RAN and SDR technologies, network operators can create a highly flexible and programmable RAN capable of supporting a wide range of use cases and applications.

\begin{figure}[!htbp]
\centering
\includegraphics[width=0.48\textwidth]{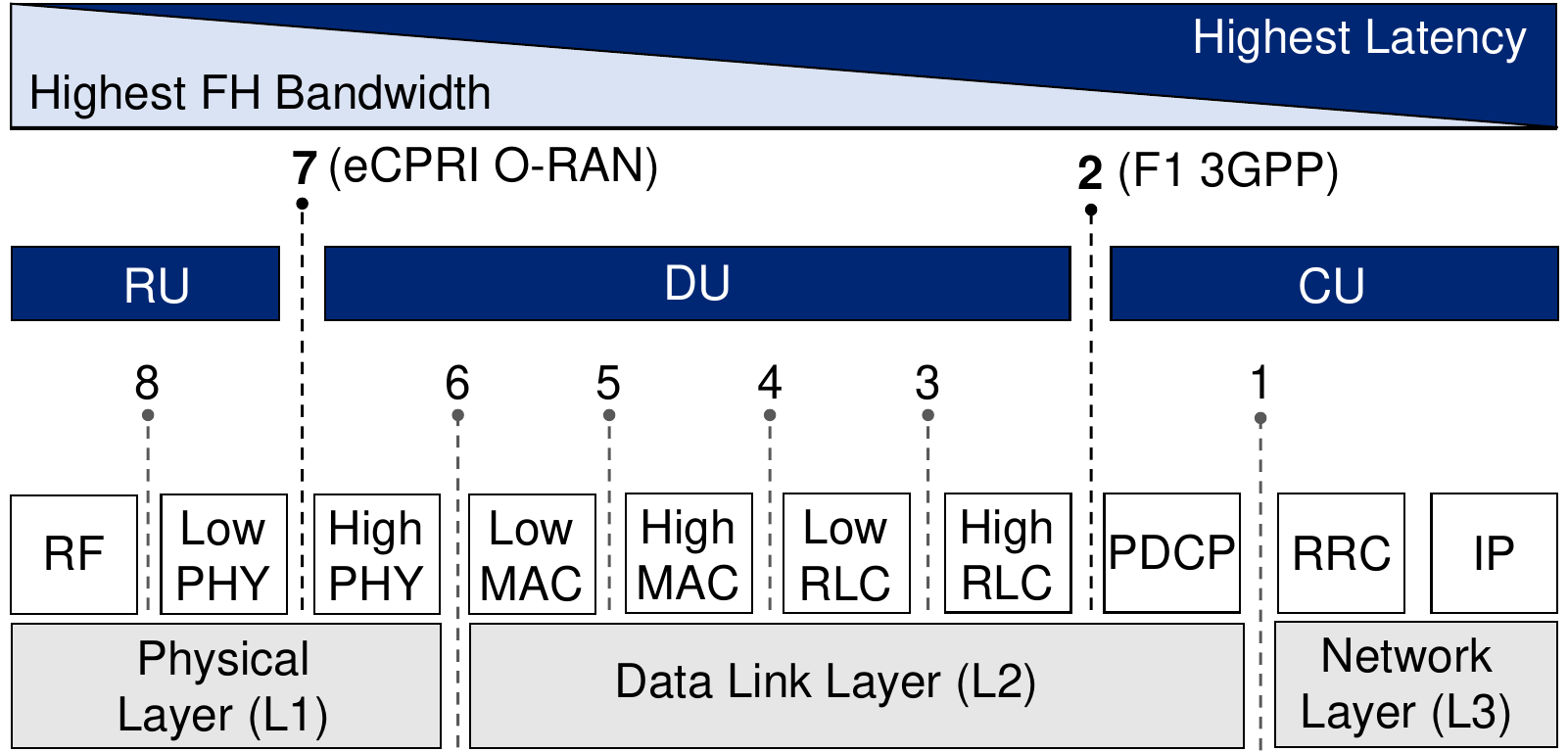}
\caption{\small BS protocol stack and different functional splits}
\label{fig:BSPS}
\end{figure}

\subsubsection{ Radio Functional Splits on RAN (NG-RAN)} 

In the context of 5G RANs, commonly referenced as New Radio (NR)~\cite{3GPPTS38.801}, there is a notable rise in data traffic, creating a challenge as the fronthaul bitrates needed for high capacity have become economically impracticals~\cite{sexton20175g,agiwal2016next}. This has spurred researchers to explore ways of reducing fronthaul bitrates while retaining the advantages of traditional RAN concepts. One avenue involves adding more functions to local sites and processing the signal thoroughly before transmission. To this end, the Next-Generation RAN (NG-RAN) represents a shift from the traditional RRH-BBU split to a more adaptable functional separation on the BS radio stack~\cite{schmidt2021flexric, TNS2020}. In this architecture, the gNB is split into a Radio Unit (RU), a Distributed Unit (DU) and a Central Unit (CU) as illustrated in Fig.~\ref{fig:RF-BBU2}. The gNB Distributed unit (gNB-DU) facilitates the management of the gNB Radio Unit (gNB-RU), while also potentially regulating bandwidth allocation to optimize network communication~\cite{polese2023understanding,larsen2018survey}. The gNB Control Unit (gNB-CU) serves as the central orchestrator for multiple gNB-DUs. This unit is divided into the gNB-CU Control Plane (gNB-CU CP) and the gNB-CU User Plane (gNB-CU UP) which communicate via the E1 interface~\cite{habibi2022mapping},\cite{habibi2021towards}. The NG-RAN functional split dictates which functions are retained locally at the antenna site and which are centralized in a high-processing data centre, optimizing function distribution for enhanced efficiency and performance~\cite{ojaghi2023benefits}.

In the NR, the whole functionality of the gNB in the protocol stack can be defined as a chain of NFs. The 3GPP proposes eight different functional split options for the allocation of these functions among gNB components, including gNB-RU, the gNB-DU, and gNB-CU~\cite{habibi2022mapping},~\cite{habibi2021towards}. This protocol stack comprises three layers, with the physical layer at the base, followed by the data link layer, and the network layer on top. The configuration of the BS protocol stacks is depicted in Fig.~\ref{fig:BSPS}. The primary role of the physical layer is to convert digital bits into outgoing radio waves, handling Downlink (DL) transmission by converting bits into waves and vice versa for Uplink (UL) communication. Within the data link layer, there are three sublayers: Packet Data Convergence Protocol (PDCP), Radio Link Control (RLC), and Media Access Control (MAC)~\cite{fischer2011user}. The data link layer receives radio bearers from the network layer in the DL direction and transmits transport blocks to the physical layer. The RLC subprotocol optimizes data transmission by resizing Protocol Data Units (PDUs) according to MAC requirements, managing acknowledgements, retransmissions, and error correction for reliability. Meanwhile, the PDCP oversees tasks like header compression and security functions like ciphering and handovers. In the network layer, the control plane features the Radio Resource Control (RRC) protocol, whereas the user plane operates on the Internet Protocol (IP). These layers, along with the RRC, interface with the datalink layer via radio bearers, facilitating pivotal functions like system information management, connection control, measurement configuration, and mobility~\cite{van2011control}.
\\\indent Functions retained in the gNB-DU remain in close proximity to end-users, being positioned at the antenna mast, while those in the gNB-CU gain from the aggregation of processing and robust computational resources within a data center, designated as the CU pool. Locating more functions in the gNB-DU means that a significant amount of processing is completed before data is transmitted across the fronthaul network, consequently reducing the bandwidth demand on this network. However, this can introduce a latency increase, given the gNB-DU's relatively lower processing capability compared to the robust power of the CU pool. A number of different functional splits are currently being investigated to be used for NR. For example, in one splitting function option, the gNB-DU hosts the RLC, MAC, and PHY layers of the gNB. On the other hand, the gNB-CU houses the RRC and PDCP protocols of the gNB. Option 2 is most often identified as the functional division between the gNB-CU and gNB-DU, utilizing the F1 interface as the communication channel.

\subsection{Open, intelligent and Software-defined RAN in 6G}\label{sec:openIntRAN}
\subsubsection{Open RAN}
Transitioning from legacy RAN to next-generation RAN brought significant improvements in network performance. However, it also presented challenges such as interoperability, standardization, and limited deployment options, creating a vendor lock-in scenario that constrained operator choice of RAN components. Open RAN utilizes open interfaces to create a more diverse and competitive ecosystem of RAN suppliers and operators~\cite{bonati2020open}. Open RAN allows operators to mix and match RAN components from different vendors, reducing vendor lock-in and enabling innovation and competition in the RAN market. In addition, new technologies, such as softwarization, cloudification, and virtualization, are used to allow for more interoperability between different network components in the RAN~\cite{nis2022euoran}~\cite{azariah2022survey}~\cite{garcia2021ran}~\cite{upadhyaya2022prototyping}. In addition, the Open RAN framework has garnered substantial attention as a potential foundational architecture for AI-enabled 6G mobile networks. Open RAN represents a revolutionary approach to cellular network architecture, aiming to enhance flexibility, interoperability, and innovation. The transition from 5G to 6G introduces new demands, including the integration of diverse technologies and the need for ultra-low latency, real-time edge processing, heterogeneous network management, dynamic resource allocation, and heightened security and privacy. These evolving requirements exceed the capabilities of the current RAN infrastructure, constrained by its inherent limitations. Open RAN, characterized by its flexibility and openness, emerges as a frontrunner for the next-generation RAN, poised to effectively address these diverse 6G requirements. By seamlessly integrating Open RAN with advanced AI technologies, 6G networks have the potential to deliver vastly improved performance, optimized resource allocation, and an enhanced user experience, thereby paving the way for a more capable and interconnected wireless ecosystem.

A variety of testbeds have been established to implement, deploy, and assess Open RAN networks, examining their performance across multiple use cases~\cite{bonati2022openran,upadhyaya2022prototyping,wang2021design,atalay2022scaling,niknam2022intelligent}. Initiated in 2016, the Telecom Infra Project (TIP) aims to expedite the advancement and implementation of open, disaggregated, and standards-compliant technological solutions. TIP contributes significantly to the global advocacy, education, and execution of Open RAN solutions. The O-RAN Alliance, created in August 2018, focuses on developing specifications for an open RAN architecture known as O-RAN. These specifications are designed to enhance the 3GPP 5G RAN standards by introducing new open interfaces, protocols, and components~\cite{coronado2022zero}.
\subsubsection{RAN Intelligent Controller (RIC)}

At the core of O-RAN architecture, the pivotal implementation of an intelligent controller, known as RIC, introduces a high degree of programmability and advanced management capabilities~\cite{bonati2021intelligence, polese2023understanding}. The RIC is divided into two distinct components: the Near-Real Time (nRT) RIC  and the Non-Real Time (non-RT) RIC. The nRT RIC focuses on real-time control and optimization and plays a critical role in making immediate decisions within RAN. It manages radio resources and handles handover control, mobility management, load balancing, and interference management, all with an emphasis on low latency and high responsiveness. The nRT RIC ensures that the RAN operates efficiently, dynamically adapting to changing network conditions, and maintaining a high level of QoS for network users through optimized resource allocation in real-time. 

The non-RT RIC is located in the Service Management and Orchestration (SMO) framework, which is responsible for RAN domain management, optimization, and orchestration and provides policy-based guidance and enrichment information to the nRT RIC. The non-RT RIC has two sub-functions, the non-RT RIC framework and non-RT RIC applications, known as \textit{rApp}, which work together to provide RAN optimization and other functions. The non-RT RIC framework exposes necessary services to the \textit{rApps} through its R1 interface, enabling \textit{rApps} to obtain information and trigger actions such as policies and re-configuration. The non-RT RIC engages in resource planning, capacity optimization, and spectrum allocation and provides intelligent decision support for long-term network planning, capacity expansion, and resource allocation decisions, contributing to the overall efficiency and adaptability of the O-RAN network~\cite{arnaz2022towards, alliance2022ran}.

\subsubsection{The nRT RIC as Next-Generation of SDN controllers}
 At a high level, nRT RIC is a software-based platform, similar in function to an SDN controller, overseeing the allocation of radio resources through the innovative E2 interface~\cite{balasubramanian2021ric}. This interface plays a crucial role in effectively segregating data and control plane functionalities~\cite{balasubramanian2021ric}. In the data plane, RAN components, namely E2 nodes comprising elements such as RUs, DUs, and CUs, oversee the routing of the subscriber traffic to the core network. In contrast, nRT RIC in the control plane manages radio resource management functions, such as handover control, mobility, and load balancing, intelligently. This architectural approach aligns O-RAN with the core principles of SDN, a paradigm that has transformed network management. SDN's innovation lies in separating the control plane from the data plane, enabling centralized control and dynamic resource allocation. O-RAN's integration of nRT RIC and the E2 interface mirrors these fundamental SDN principles, heralding a new era of adaptable and efficient network management. The O-RAN Alliance, through O-RAN Software
Community (OSC)~\cite{ORANOSC} deploys a nRT RIC consisting of microservices on a Kubernetes cluster for prototyping O-RAN solutions.
The ONF also proposed the exemplar platform for O-RAN architecture. They defined the SD-RAN open source project~\cite{sunay2020onf}, which is consistent with 3GPP standard~\cite{3GPPTS38.401} and O-RAN reference design specifications~\cite{alliance2022ran}. The project is designed in close coordination with SDN principles. It leverages the flexibility and scalability of SDN controllers and P4 programmable switches to implement the open-source components in control and data planes. In this implementation, the nRT RIC runs as an open-source SD-RAN controller which is implemented using a microservices-based ONOS ($\mu$ONOS) controller~\cite{microONOS}. The first generation of ONOS controllers was released by ONF in 2015, and the next generation, $\mu$ONOS, supports a microservice-based architecture adapted to Open RAN functionalities. Several other RIC implementations exist in the academic research domain such as FlexRIC~\cite{schmidt2021flexric}.

\begin{figure}[!htbp]
\centering
\includegraphics[width=0.47\textwidth]{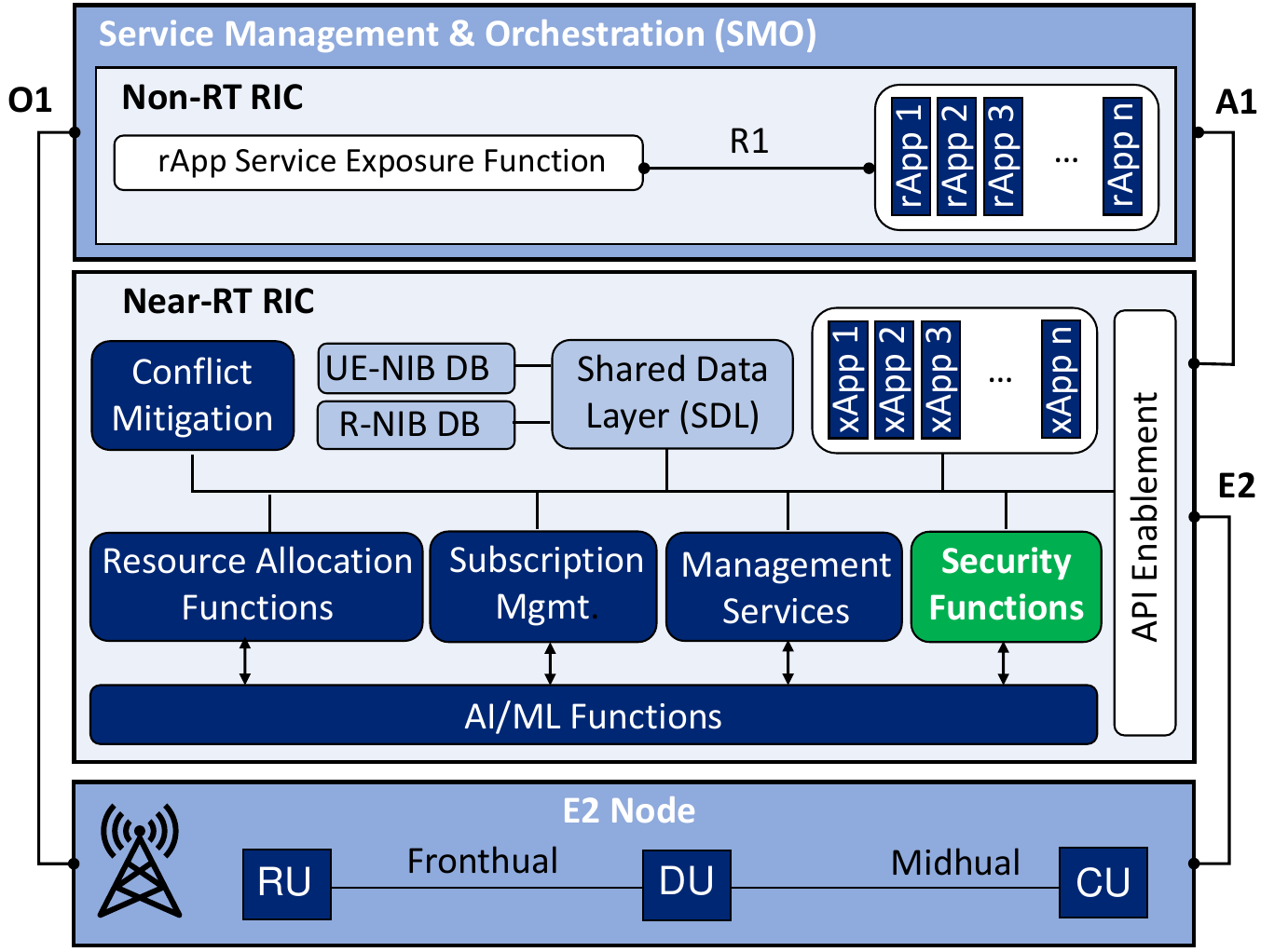}
\caption{\small nRT RIC reference architecute}
\label{fig:RIC}
\end{figure}

Fig.~\ref{fig:RIC} provides a visual representation of the nRT RIC's architectural layout. The nRT RIC incorporates a comprehensive set of interface termination functions, encompassing A1 termination, E2 termination, and O1 termination, each serving distinct roles within its architecture. The A1 interface establishes a vital connection between the nRT RIC function and the non-RT RIC function situated in the SMO layer. This interface plays a pivotal role by supporting three fundamental services: Policy Management, Enrichment Information, and ML Model Management. A1 policies, characterized by their temporality and non-critical nature to network traffic, hold a significant position in decision-making, taking precedence over configuration settings. It's worth noting that A1 policies are non-persistent, which implies they do not persist beyond a restart of the nRT RIC~\cite{polese2023understanding} \cite{alliance2022ran}. The nRT RIC enables the execution of services and the transmission of outcomes to E2 nodes via the E2 interface. In addition to its pivotal role in data exchange, the E2 interface supports essential API enablement functions tied to nRT RIC API operations, including authentication, service discovery, and generic event subscription. Additionally, it extends its capabilities by providing support for AI/ML operations, including data pipelining, training, and performance monitoring. The O1 interface serves as a crucial link connecting the O-RAN managed element to the management entity, facilitating the efficient management and integration of O-RAN components within the network infrastructure~\cite{alliance2022ran}.

Crucial software components known as \textit{xApps} collaborate closely with the nRT RIC platform to address specific and often specialized use cases. These \textit{xApps} play a pivotal role in enhancing the operational aspects of the RAN ecosystem. When the \textit{xApp} is registered with the nRT RIC platform, it imparts vital information regarding its Operations, Administration, and Maintenance (OAM) functions and control capabilities. This registration process is instrumental in enabling seamless coordination between the \textit{xApps} and the RIC platform, ensuring that the \textit{xApp} can effectively contribute to the RAN's overall functionality and performance, while also allowing for efficient management and control. The nRT RIC platform, in turn, takes on the responsibility of managing \textit{xApp} subscriptions. This management process streamlines data distribution to various \textit{xApps}, originating from diverse sources within the RAN infrastructure.

Furthermore, the nRT RIC platform excels in conflict mitigation, a crucial function that resolves potential conflicts arising from simultaneous and possibly conflicting requests made by multiple \textit{xApps}~\cite{alliance2023ricarch}. The RIC architecture incorporates Network Information Base (NIB) databases that serve to store critical information. Specifically, the RAN NIB (R-NIB) database stores data on the E2 nodes, while the UE-NIB database maintains entries for individual UEs and their respective identities. The R-NIB helps the RIC to have an abstract view of RAN resources and enables a smooth and efficient RAN operation. Moreover, R-NIB is crucial for the topologically dependent \textit{xApps}. UE-NIB allows for UE-specific control while simultaneously exposing sensitive information about the user. Overlayed on top of these databases is the shared data layer (SDL), which enables \textit{xApp} access. These databases can be queried by the \textit{xApps} through the SDL APIs. Furthermore, the messaging infrastructure enables interaction between the internal functions of nRT RIC, while the security function provides the necessary security measures for \textit{xApps}. The Near-Real-Time RIC provides management services including fault, configuration, and performance management to the SMO as a service producer. Its capabilities for logging, tracing, and metrics collection are designed to record, observe, and gather data on the internal status of the nRT RIC, which can then be relayed to an external system for additional analysis.

\subsection{Lessons Learned}\label{sec:RANEVOLesson}
The evolution of mobile network architecture from 2G to 6G has witnessed significant advancements and transformative changes. Few lessons are given below.

\begin{itemize}[leftmargin=*]
\item \textit{Adaptability Through Standardization:} The transition from 2G to 6G highlights the crucial role of standardization bodies like 3GPP and industry alliances such as O-RAN. Standardization fosters interoperability, accelerates innovation, and enables multi-vendor ecosystems. As networks evolve, standardization remains paramount to seamlessly integrating new technologies.

\item \textit{Disaggregation and Virtualization:} The shift from Legacy RAN to Next Generation RAN (5G) introduced the concepts of disaggregation and virtualization. These approaches decouple hardware from software, allowing for flexibility in network deployments. The success of C-RAN in 5G showcases the benefits of centralized processing and virtualization in optimizing network resources.

\item \textit{SDN Integration:} The integration of SDN principles into RAN architecture, as seen in SD-RAN, offers centralized control and dynamic resource allocation. SDN's separation of control and data planes enhances network adaptability and responsiveness. This integration aligns with the evolving demands of 5G and 6G networks.

\item \textit{Functional Split for Efficiency:} The concept of radio functional splits offers a solution to address rising data traffic efficiently. By distributing functions between local sites and centralized data centres, NG-RAN optimizes network resource allocation and minimizes fronthaul bitrates, essential for high-capacity networks.

\item \textit{Open RAN Ecosystem:} Open RAN emerges as a pivotal architecture, fostering diversity in vendor choices and promoting innovation. It counters vendor lock-in and enables operators to mix and match components. Open RAN, along with the integration of AI and ML technologies, holds promise for addressing the unique challenges posed by 6G networks.

\item \textit{Role of the RIC:} The RIC, comprising nRT and non-RT components, represents a crucial element in O-RAN architecture. It offers real-time control, optimization, and intelligent decision-making within the RAN. The RIC, resembling SDN controllers, enables adaptable and efficient network management.

\end{itemize}

\section{Security Evolution in RAN: Challenges and Progress } 
\label{sec:ORAN-SecurityRisks}

In the wake of exploring the transformative journey of RAN in the previous section, it becomes imperative to delve deeper into the security implications of this evolution. As RAN has transitioned from legacy systems to modern, open, and software-defined architectures, each phase has introduced its unique set of vulnerabilities and countermeasures. This Section offers a comprehensive exploration of this intricate security landscape. We commence with a reflection on the security mechanisms of 2G, 3G, and 4G networks in Section~\ref{sec:ORANSecLagacyRAN} emphasizing their inherent strengths and weaknesses. The narrative then shifts focus to the 5G paradigm in Section~\ref{sec:ORANSecNGRAN} shedding light on the challenges and solutions arising from the integration of transformative technologies like cloud computing, SDN, and NFV. Further, in Section~\ref{sec:ORANSecORAN} we dissect the risk landscape of O-RAN, drawing from the detailed assessments by the O-RAN Alliance Security Work Group (WG11). The section~\ref{sec:ORANSecLL} offering a reflective analysis of RAN's security evolution and emphasizing the significance of proactive measures in this ever-evolving domain.

\subsection{Security in Legacy RAN}\label{sec:ORANSecLagacyRAN}
In order to secure communication within GERAN, GSM provides three major mechanisms. Firstly, it utilizes shared-secret cryptography for UE authentication, employing a SIM card that stores a crypto variable and validates the identity of the mobile subscriber. Second, radio traffic encryption strategically safeguards voice and data circuits against unauthorised interceptions. Thirdly, for user anonymity, GSM introduces the concept of a Temporary Mobile Subscriber Identity (TMSI), a replacement for the International Mobile Subscriber Identity (IMSI). The IMSI acts as a unique identifier in mobile networks, enabling efficient communication routing and personalized services. The TMSI acts as a protective shield against identity-related attacks and unauthorised tracking. By employing the TMSI instead of the actual IMSI, communication security is strengthened, leading to improved network efficiency. Regular updates to the TMSI prevent consistent tracking and are crucial for maintaining user identity confidentiality within the GSM network~\cite{gindraux20022g}.

However, the 2G network exhibited several security limitations and weaknesses in the GERAN domain. Notably, its encryption algorithms were deemed insecure and susceptible to reverse-engineering, making them prone to attacks like eavesdropping~\cite{buttyan2007security}. For instance, the stream ciphers A5/1 and A5/2, utilized for call encryption, could be compromised through ciphertext-only attacks~\cite{hanser2014security}.  The absence of end-to-end encryption in communications served as the underlying cause for various eavesdropping-based exploits, such as Man-In-the-Middle (MITM) attacks~\cite{conti2016survey, cattaneo2013review}. Among the notable vulnerabilities is the IMSI-catcher attack, where attackers leveraged unencrypted IMSI information during authentication and paging procedures to trace mobile subscribers~\cite{dabrowski2014imsi}. Additionally, 2G networks lacked mutual authentication mechanisms between mobile phone subscribers and corresponding networks. While cellular networks authenticated UE unilaterally, UEs had no means to authenticate the network, thus enabling the attacker to interrupt the mobile communication with rogue BS attacks~\cite{conti2016survey,rappaport2002wireless}. Furthermore, GSM's lack of data integrity against channel hijacking in the absence of encryption rendered it vulnerable to both unauthorised access and Denial-of-Service (DoS) attacks~\cite{toorani2008solutions}.

In 3G network specifications~\cite{3GPPTS33.102}, 3GPP improved RAN security by addressing GERAN vulnerabilities and adding extra safeguards. UTRAN security ensures secure UE access to 3G services, guarding against potential radio link attacks~\cite{boman2002umts}. One example is mutual authentication, where both UE and network confirm identities for better privacy. Unlike GSM's one-sided authentication, UMTS employs mutual authentication, reducing rogue BS risks. 3GPP also set privacy rules for UTRAN, covering user identity, location, and traceability. User identity security, linked to IMSI, shields radio communication~\cite{3GPPTS33.102}. However, it's evident that UTRAN faces various attacks that mainly focus on subscriber identity and confidentiality, like IMSI paging~\cite{khan2014another}. Another concern is downgrade attacks~\cite{van2015defeating}, especially in the coexistence of 2G and 3G networks. In such attacks, perpetrators force victims to connect to less secure 2G networks without mutual authentication. Following a successful downgrade, attackers could execute MITM attacks~\cite{meyer2004man} to collect UE's IMSI for location tracking.

The 4G network introduced new cryptography methods and key structures distinct from 2G and 3G. It adopted algorithms like Evolved packet system Encryption Algorithms (EEA) and Evolved packet system Integrity Algorithms (EIA). In 4G, most keys are 256 bits compared to 128 bits in 3G. Moreover, 4G employs varied algorithms and key sizes for control and user plane traffic. Its main authentication mechanism is called the Authentication Key Agreement (AKA) protocol. Integrity and replay protection for 4G air interface traffic are ensured by Non-Access Stratum (NAS) and RRC signaling. Additionally, Global Unique Temporary Identifier (GUTI), a temporary identifier, is used to hide a subscriber’s long-term identity~\cite{3GPPTS33.401}. 
\\Although security features have been improved, the E-UTRAN architecture still has some weaknesses. The allocation pattern of GUTI, which identifies users, can be predicted and used for tracking~\cite{hong2018guti}. Two major vulnerabilities are MITM attacks and eavesdropping. These occur when attackers create fake base stations, posing as real ones~\cite{ferrag2018security}. Various solutions have been proposed, such as using cryptographic authentication protocols. Additionally, the eNB is susceptible to physical attacks, DoS attacks, and passive eavesdropping on long-term keys. To address these concerns, 3GPP has enforced strict requirements for the eNB, including secure setup, configuration of base station software, key management, and safeguarding user and control plane data~\cite{3GPPTS33.401}.

\subsection{Security in Next Generation RAN} \label{sec:ORANSecNGRAN}
As outlined in Section~\ref{sec:NGRAN}, the fundamental transformative technologies in 5G in contrast to prior generations encompass cloud computing, SDN, and NFV, all of which have been incorporated into RAN architectures. This section presents an examination of the security issues and solutions primarily within the context of their implementation in the RAN domain. 
\subsubsection{C-RAN security}

Virtualization and cloud technology play a crucial role in the creation and utilization of C-RAN, leading to considerable concerns about potential risks related to cloud resources and virtual platforms~\cite{xiao2012security,baroos2020enisa}. Cloud resources are highly distributed, heterogeneous, and virtualized, traditional security mechanisms
are no longer enough for the security of C-RAN~\cite{gonzalez2012quantitative,tian2017survey}. Within the array of challenges, below we emphasize the main security issues for C-RAN

Cloud servers, while entrusted with vast amounts of data, bring about the potential for malicious or accidental loss or modification of RAN control or user plane data. Furthermore, in scenarios involving BBUs from different operators, VMs might coexist on a single physical server (known as VM co-tenancy), which raised certain security issues, such as Cross-VM attacks and Malicious SysAdmin~\cite{xiao2012security}. While the streamlined processes of creating, deleting, and relocating BBU VMs across the resources offer advantages, they simultaneously amplify security concerns due to the increased complexity of tracking malicious VMs~\cite{ahmad20175g}. The dynamic capability for migrating VNF /VM to different resources brings about a new level of vulnerability. A BBU VNF that needs to conform to a high-security level might be maliciously moved to another physical server which may offer lower
security. The process of service chaining different NFVs will further add complexity to the analysis of the underlying reasons behind security threats~\cite{vaughan2008virtualization}. The other security challenge is related to the hypervisor. The hypervisor can be targeted for
a number of attacks such as exploiting the host operating system to DoS attack on VMs, and VM hopping attacks~\cite{huang2012security}.  An insider attack, involving service provider personnel who have access to the physical servers containing user data, poses a significant threat to the concept of C-RAN if these insiders misuse or mishandle user data and information.

\subsubsection{SD-RAN security}

The centralized control plane of SD-RAN makes it a highly targeted point for compromising the network or carrying out malicious activities in the network due to its pivotal role in decision-making~\cite{chica2020security,farris2018survey}. Furthermore, the implementation of the SD-RAN control plane in OpenFlow magnifies these security repercussions, as illustrated by the following instances. 

Initially, a variety of connected entities, such as Mobile Edge Computing (MEC) servers~\cite{alwarafy2020survey, ranaweera2021survey}, monitoring platforms, and other systems, can easily send packets to the controller by provoking a \textit{table-miss} event at the BS and triggering a \texttt{PACKET-IN} message~\cite{dhawan2015sphinx}. This mechanism could be exploited by malicious entities to deluge the BS with non-conforming packets, potentially leading to a DoS attack on the BS and the controller, causing network disruptions and outages~\cite{abdulkarem2020ddos,xue2018linkscope}.
Next, the controller's requirement for a complete overview of the RAN topology, encompassing all connected hosts, BSs, and linkages, is imperative. Applications and services dependent on topology, such as routing, mobility, load, and congestion management, consult the SDN controller for this information to inform their decision-making. Malicious entities could exploit this by distorting the controller's view of the network layout, leading to a topology poisoning attack~\cite{hussain2020indirect,shrivastava2021topology,kim2021bottlenet}, which in turn could alter traffic routes and lay the groundwork for MitM attacks~\cite{pakzad2014efficient}.
Moreover, the controller issues traffic routing directives as flow rules to BSs within the data plane, which could be manipulated by a compromised BS to redirect or intercept traffic, altering the prescribed flow rules~\cite{hua2020flow} or leaking confidential information about the flow rules~\cite{marin2019depth, jero2017identifier}. Lastly, the network is vulnerable to the injection of counterfeit or spoofed messages by malicious hosts or BSs, especially since the controller lacks inherent mechanisms for detecting such security breaches~\cite{alimohammadifar2018stealthy}. 

Furthermore, within the framework of SD-RAN architecture, the majority of RAN functions are implemented as software applications. This configuration presents a concerning scenario: if a malicious application is granted access, it possesses the potential to sow chaos throughout the RAN domain, and in more severe instances, could disrupt the entire cellular network. The north-bound interface compounds this challenge, especially for remote applications, primarily due to the absence of standardized interfaces~\cite{ahmad2021scalability}. The security predicaments stemming from applications can be attributed to several factors. Notably, the presence of open APIs in network equipment creates vulnerabilities. Additionally, the lack of robust trust mechanisms between applications and controllers further exacerbates the situation. Moreover, the absence of adequate authentication and authorisation techniques for applications, particularly those originating from third parties, adds to the complexity of the security landscape~\cite {kreutz2013towards}.

To provide a secured communication channel, the south-bound interface uses Transport Layer Security (TLS)~\cite{rfc5246} and Datagram Transport Layer Security (DTLS)~\cite{rfc6347}. However, due to the configuration complexity in the use of TLS and DTLS, their use is left optional. SDN networks combat malicious application access through rigorous pre-verification. Solutions like \textit{PermOF}~\cite{kreutz2013towards} establish strict boundaries, confining apps within defined privileges. Trusted access to the control plane is enforced by~\cite{scott2014operationcheckpoint}, while \textit{FortNOX}~\cite{porras2012security} employs role-based authorisation. Similarly, \textit{ROSEMARY}~\cite{porras2012security} proposes a permission system to guard against malicious and faulty applications, collectively enhancing SDN security. 

To address scalability challenges of the controller and enhance defences against DoS attacks, \textit{AVANT-GUARD}~\cite{shin2013avant} employs a connection migration tool to restrict the number of flow requests (unsuccessful TCP sessions) reaching the control plane. Various approaches have been explored to enhance controller security against topology attacks. Notable defences include \textit{TopoGuard} utilizing port labeling for Link Fabrication Attacks (LFAs) detection~\cite{hong2015poisoning}, and an advanced version, \textit{TopoGuard+}, incorporating a Link Latency Inspector~\cite{skowyra2018effective}. Further strategies encompass threshold-based techniques~\cite{marin2019depth,smyth2017detecting}, Stealthy Probing Verification (SPV)~\cite{alimohammadifar2018stealthy}. ML-based Link Guard (MLLG)~\cite{soltani2021link} and Real-time Link Verification (RLV)~\cite{soltani2021link,soltani2023real } address LFA and Link Latency Attack (LLA), in larger-scale SDNs. Additionally, deep reinforcement learning bolsters topology poisoning defence \cite{wang2020topology,wang2021location}. The Hybrid-Shield approach counters Multi-Hop Link fabrication in hybrid SDN networks \cite{shrivastava2021topology}.


\subsubsection{NG-RAN security}
In 5G NG-RAN, each subscriber is assigned a globally unique Subscription Permanent Identifier (SUPI). This SUPI comprises the UE IMSI along with specific home network identifiers. Importantly, the SUPI remains concealed during the initial attachment request and connection establishment phases. Instead, a temporary one-time user identifier known as the Subscription Concealed Identifier (SUCI) is employed, generated by the UE. The SUCI serves to encrypt and hide the subscriber's actual information. Notably, this design prevents malicious entities such as fake base stations (commonly referred to as IMSI catchers) from ascertaining the subscriber's identity, as the SUPI is only unveiled once a full connection has been established, involving interaction with the core network~\cite{cao2019survey}. To enhance security further, the NG-RAN incorporates a critical feature known as the Subscriber Identifier De-concealing Function (SIDF) within the core network. The SIDF exclusively entertains requests originating from network functions within the subscriber's home network, and its primary role is to de-conceal the SUPI from the SUCI. This security measure represents a significant improvement over earlier generations of networks, as detailed in a study by Mao et al.~\cite{mao2023security}. Additionally, 5G NG-RAN offers robust protection against bidding-down attacks. In the 5G ecosystem, malicious actors would need to intercept and capture the entire set of coded and protected NAS messages to impersonate a base station. This heightened security makes it considerably more challenging for IMSI catchers to succeed in compromising the network.

NG-RAN security encompasses the principle of unified authentication, requiring both 3GPP (5G NG-RAN) and non-3GPP access networks (like WiFi) to implement consistent authentication methods. This consistency ensures that UE and networks are capable of supporting both EAP-AKA~\cite{rfc5448} and 5G AKA authentication, regardless of the access network type.
Moreover, the security measures extend to all RAN functions, which robustly employ TLS with client and server-side certificates. Operators are strongly encouraged to activate integrity protection for user data, with IPSec Encapsulating Security Payload (ESP), DTLS, and Internet Key Exchange (IKE)-v2 security certificate-based authentication being utilized between the gNB-CU and gNB-DUs across both control and user planes. Furthermore, NG-RAN enhances security by allowing selective integrity protection for user plane traffic on a per-radio bearer basis. This ensures that signaling, including NAS and RRC signaling, is subjected to mandatory integrity protection. Additionally, the interfaces between gNB components, such as F1 and E1, are required to guarantee and support confidentiality, integrity, and replay protection, adding an extra layer of security to the network's operations.

\subsection{Security in open, intelligent and software-defined RAN} \label{sec:ORANSecORAN}
O-RAN Alliance Security WG11 conducted a risk analysis of O-RAN architecture and investigated additional security requirements to protect O-RAN~\cite{O-RAN.SFGthreat, O-RAN.SFGthreatProtocol, O-RAN.SFGReq, O-RAN.SFGSecTest} and the end-to-end security test specifications have also
been defined~\cite{O-RAN.SFGtest}. The WG11 also comes up with a risk assessment on the severity and likelihood of existing threats. The risks are analyzed by considering a Zero Trust Architecture (ZTA) defined by the National Institute of Standards and Technology (NIST)~\cite{rose2020zero} and using the International Organization for Standardization (ISO) 27005 methodology~\cite{ISOIEC27005}. The O-RAN Alliance WG11 has identified seven distinct threat categories that delve into the security risks and vulnerabilities associated with O-RAN systems, shedding light on the multifaceted challenges that must be addressed to safeguard the integrity and reliability of these innovative network architectures.
\begin{enumerate}[leftmargin=*]
\item Security risks to O-RAN systems:
The O-RAN architecture introduces new elements and interfaces, expanding the threat surface. These include unauthorised access to disaggregated RAN components, malicious \textit{xApps}, and attacks compromising data integrity and confidentiality. Weak authorisation mechanisms can allow network penetration, potentially leading to data breaches and infrastructure compromise.
\item Vulnerabilities in O-Cloud infrastructure:
O-Cloud, as a virtualization environment, is not immune to threats. Potential attacks involve compromising virtual network functions, exploiting interfaces like O2, and abusing VMs/containers. Security concerns extend to spoofing, network compromise, and the misuse of privileged VMs/containers within virtualized environments.
\item Security challenges in open source code projects:
The softwarization of RAN and O-RAN components introduces vulnerabilities. Threats include backdoors intentionally introduced by trusted developers or vulnerabilities in upstream libraries not controlled by O-RAN developers. These code-based risks may compromise the overall system's security.
\item Risks to 5G radio network security:
Attacks on the radio network can lead to performance degradation. Threats range from jamming data and synchronization signals to denials of service. Combining equipment from different vendors may also pose a risk if configurations or supported functionalities are mismatched.
\item Threats to ML systems:
O-RAN systems incorporate AI/ML models for inference and control. Threats include poisoning attacks, where attackers inject altered data into datasets used for offline AI/ML training. Adversaries may also gain control over O-RAN nodes to manipulate real-time data, potentially leading to wrong predictions, control decisions, and performance issues.

\item Security concerns in protocol stack layers:
Attacks can target protocol stack layers, including injection, cross-site scripting, DoS, and unauthorised exposure of objects identifiers through REST APIs, JSON, or HTTP exploits. These vulnerabilities can compromise the integrity and reliability of the protocol stack.
\item Security challenges in open fronthaul interface:
The Open fronthaul interface presents vulnerabilities, particularly due to its lack of encryption on the control plane. This opens the door to man-in-the-middle attacks, impersonation of elements like DU or RU, and potential compromises of user data or configurations. Attacks on the synchronization infrastructure can also lead to performance degradation.
\end{enumerate}

Table~\ref{table:risk} shows their results as a risk assessment matrix where a total of 68 threats associated with O-RAN component vulnerabilities are identified~\cite{soltani2022can}. Each threat within the system is allocated an impact level, delineated as low, medium, or high. The determination of a threat's severity hinges on two primary factors: (1) the extent of affected RU and DU, and (2) the degree of influence on key security principles such as privacy, confidentiality, integrity, and availability~\cite{O-RAN.SFGthreat}. Additionally, the probability of a threat occurring is evaluated based on the presence of security measures within the O-RAN framework. Risk levels are then established by assessing both the potential impact and the likelihood of threats, which allows for categorization into various risk tiers. A matrix that cross-references severity and likelihood is utilized to graphically represent the risk associated with each threat. This method arranges threats in order of priority: high, medium, or low. The O-RAN Alliance Working Group 11 has identified 52 threats as high-risk, 9 as medium-risk, and 7 as low-risk based on their impact and probability of occurrence. Given that approximately 76\% of threats fall into the high-risk category, adopting a security-by-design philosophy in O-RAN becomes critically essential.

\begin{table}  [!htbp]
    \caption{\small Risk assessment result on threats against Open RAN} \label{table:risk}
	\centering
	\footnotesize
	\begin{tabular}{|l|p{1.43cm}|p{1.7cm}|p{1.75cm}|p{1.7cm}|}
    \hline
	    \multicolumn{2}{|c|}{}&\multicolumn{3}{c|}{\textbf{Severity Leve}l}\\
	    \cline{3-5}
        \multicolumn{2}{|c|}{}&  
        \textbf{Low} & 
       \textbf{ Medium}& 
       \textbf{ High}\\

        \multicolumn{2}{|p{2cm}|}{Total Threats: 68}&  
        \tiny (One DU is affected with one served RU) & 
        \tiny (One DU is affected with several served RUs) & 
        \tiny (Several DUs and RUs are affected)\\
	    \hline %
	    \multirow{8}{*}{\begin{turn}{+90}\textbf{Likelihood Level}\end{turn}}
	    
	    &\textbf{High}&\multirow{3}{*}{\cellcolor{yellow!25}}&\multirow{3}{*}{\cellcolor{red!25}}&\multirow{3}{*}{\cellcolor{red!25}}\\
	     & \tiny (Lack of Open RAN security controls)&\cellcolor{yellow!25}0&\cellcolor{red!25}2&\cellcolor{red!25}21\\
	     \cline{2-5}
	     
	    &\textbf{Medium}&\multirow{3}{*}{\cellcolor{green!25}}&\multirow{3}{*}{\cellcolor{yellow!25}}&\multirow{3}{*}{\cellcolor{red!25}}\\
	     & \tiny (Inadequate Open RAN security controls)&\cellcolor{green!25}1&\cellcolor{yellow!25}4&\cellcolor{red!25}29\\
	     \cline{2-5}

	    &\textbf{Low}&\multirow{3}{*}{\cellcolor{green!25}}&\multirow{3}{*}{\cellcolor{green!25}}&\multirow{3}{*}{\cellcolor{yellow!25}}\\
	     & \tiny(Inefficient Open RAN security controls)&\cellcolor{green!25}3&\cellcolor{green!25}3&\cellcolor{yellow!25}5\\
	     \hline

	\end{tabular}
\end{table}

It's noteworthy that security measures in O-RAN networks extend beyond software solutions and have the potential to be integrated into hardware components. In the study by Haas et al.~\cite{haas2022trustworthy}, they propose a comprehensive solution to bolster the security and speed of O-RAN, crucial for critical tasks and resource-sharing scenarios. The paper introduces the concept of \textit{capabilities} as a crucial mechanism to regulate and safeguard O-RAN communication. These capabilities, enforced by hardware mechanisms, strictly control authorised communication, preserving network integrity and efficiency. The paper presents two systems, \textit{M3} for smaller segments and \textit{FractOS} for larger deployments like data centres, both relying on capabilities to manage communication access rights. In \textit{M3}, individual tiles include a special component called a TCU, responsible for checking permissions. In contrast, \textit{FractOS} extends these principles to larger environments and utilizes SmartNICs to safeguard communication. This approach seeks to strike a balance between network security and performance, ensuring O-RAN networks remain secure and responsive.

Numerous security threats targeting the O-RAN architecture have been pinpointed and explored~\cite{O-RAN.SFGthreat,habler2022adversarial}. Among these, the most critical are the ones that compromise the central control element of O-RAN, specifically the RIC. If such an attack prevails, it holds the potential to dominate the entire network, leading to information breaches or enabling other malevolent actions. Whenever the attacker gets control of the RIC thoroughly, it leads to a network service failure and subsequently affects the entire network. The O-RAN Alliance WG11 has published a technical specification~\cite{O-RAN.SFGRICXAPP} that offers a foundational understanding of security concerns in this context, raising awareness of potential risks associated with RIC systems. However, it provides a high-level overview and does not delve into the intricate specifics of these threats. Notably, the WG11 report does not explore the intricate relationship between RIC and SDN controllers or incorporate the latest research findings in the field. In contrast, this study aims to surpass this foundational groundwork. We dive deeper into the nuanced aspects of RIC security, recognizing that comprehensive security understanding requires meticulous examination. We have thoroughly investigated often-overlooked threats stemming from the interaction between RIC and SDN controllers. Additionally, our research remains dynamic, staying updated with the latest developments in RIC security. Our ultimate goal is not just to identify vulnerabilities but to provide actionable, technically rigorous countermeasures, empowering stakeholders to proactively enhance RIC security in the evolving telecommunications landscape. The rest of the paper is presented with detailed descriptions of the threat landscape and security countermeasures on RIC as it is related to the topic of this survey. 

\subsection{Lessons Learned}\label{sec:ORANSecLL}

The evolution of RAN from its traditional form to the modern architecture underscores pivotal security lessons. As RAN technologies continue to evolve, there's a pressing need to persistently refine and bolster security protocols. Historical vulnerabilities in legacy RAN systems, notably weak encryption and the lack of mutual authentication, emphasize the necessity of proactively enhancing security in response to technological advancements. In contemporary RANs, a comprehensive approach to security is paramount, encompassing both control and user plane traffic. This demands the deployment of robust authentication, encryption, and diverse threat protection mechanisms. Transitioning to O-RAN architectures unveils distinct challenges tied to disaggregated components, virtualized ecosystems, and open-source initiatives, all of which mandate a forward-thinking, security-by-design strategy. Central to the O-RAN architecture is the security of the RIC. Any compromise to the RIC can precipitate grave repercussions, including potential full-scale network breaches by adversaries. A meticulous examination of the interplay between RIC and SDN controllers is essential, accompanied by the development of technically sound countermeasures to fortify RIC security in the ever-evolving telecommunications domain.
\section{Security Challenges in RIC} \label{sec:RIC-SecurityRisks}

The RIC adopts the SDN controller's principles to centralize control over radio functions and enable programmability within communication networks. Yet, these groundbreaking features also potentially open doors to network intrusions and exploitations. Centralized control, for instance, could become a prime target for DoS attacks, and the exposure of essential APIs to unauthorized software could lead to the collapse of the network. While RIC allows for the integration of smart security detection algorithms via security-focused \textit{xApps}, there currently isn't an established framework for incorporating security services into Open RAN architecture. Moreover, this new technology introduces significant security challenges; novel infrastructure, systems, and devices bring along fresh threats to the network's integrity. To draw attention to the criticality of RIC security, Section \ref{sec:ORAN-Sec-Vul} delves into the key vulnerabilities inherent to RIC's role in the radio network. Additionally, RIC designs don't directly address the emerging security concerns of SDN controllers. Thus, Section\ref{sec:RICSecuritySDN} explores the vulnerabilities of SDN controllers and how they may be affected or altered within the RIC framework. Finally, Section~\ref{sec:ORAN-Sec-attEx} outlines potential attack strategies that could leverage these identified weaknesses in RIC, with two practical use case examples illustrating how these attacks could unfold.

\subsection{RIC Vulnerabilities}\label{sec:ORAN-Sec-Vul}
\vspace{1mm}
Our examination of current O-RAN proposals and their associated applications has led us to categorize RIC vulnerabilities into six separate groups. We present an overview of these vulnerability categories in this section. Subsequently, in Section \ref{sec:ORAN-Sec-attEx}, we delve into a comprehensive analysis of the attacks that could take advantage of these weaknesses, using two concrete examples of attacks sourced from the latest research on RIC security. Therefore, we now turn our attention to a methodical and detailed investigation of RIC vulnerabilities.
\subsubsection{Uncertified access to the radio resource} \label{sec:UncertifiedAccess} 
The deployment of Open RAN technology necessitates a broader selection of radio resources sourced from multiple suppliers. This expansion in supplier diversity, in conjunction with the evolving supplier landscape, has led to an increased demand for third-party access to these radio resources. The RIC plays a crucial role, providing access to various \textit{xApps} and facilitating communication through APIs. The inclusion of multiple \textit{xApps} from diverse suppliers introduces a significant challenge in ensuring the seamless integration of reliable and proficient applications and APIs throughout network deployment, operation, and maintenance~\cite{nis2022euoran}. Moreover, while the strategic adoption of open-source software within this framework offers advantages such as flexibility, community support, and cost-effectiveness, it simultaneously necessitates a vigilant approach to risk management. Vulnerabilities like insecure \textit{xApps}, API exploitation, and open-source software vulnerabilities have the potential to compromise the security of the RIC, which, in turn, could result in unauthorised access to these vital radio resources. 

\textit{Insecure xApps}. Developing applications demands careful measures to safeguard radio network operations from potential application malfunctions and software flaws. \textit{xApps}, which execute on the RIC, are responsible for a vast array of radio network functions and are often crafted by entities other than the RIC providers. These applications acquire privileged access to both radio network resources and sensitive information. Moreover, they frequently have the ability to alter RAN functions without robust security protocols to defend against hostile acts. If an attacker impersonated the \textit{xApp}, it could gain access to radio network resources and manipulate the network operation. For instance, a malicious \textit{xApp} could exploit these vulnerabilities to disrupt services for subscribers or specific network areas~\cite{boswell2020security}. Additionally, \textit{xApps} that enable increased automation within the RAN, if compromised, may gain access to sensitive attributes. Similarly, poorly designed or buggy \textit{xApps} can inadvertently introduce vulnerabilities into the system, with known bugs potentially exploited by attackers to compromise the application's integrity. Moreover, the presence of nested \textit{xApp} applications, where one application uses an instance of another, can create indirect communication pathways with the RIC, potentially serving as gateways for unauthorised access to E2 nodes.

\textit{API exploitation}. There are various APIs offered by the RIC platform. The \textit{xApps} can utilize SDL APIs to query R-NIB and UE-NIB databases, and there are APIs available for \textit{xApp} registration, discovery, and subscription. Additionally, the E2 interface facilitates control plane functionalities like mobility and handover through APIs, while the A1 interface supports RAN optimization and management with APIs for tasks such as load balancing and interference management. However, despite their intended benefits, these APIs are vulnerable to a range of security threats that could compromise the confidentiality, integrity, and availability of the RAN system. One such threat is API exploitation, which involves exploiting vulnerabilities in the APIs to gain unauthorised access to the system's sensitive data or functions. For instance, in the context of SDL APIs, an adversary could potentially exploit a vulnerability to access the sensitive RAN information stored in the R-NIB database in RIC. If successful, this could allow them to intercept network state history, E2 node configurations, cell information,  flows, and their associated mappings.

\textit{Open-source software vulnerabilities.} 
Implementation of RIC and \textit{xApps} rely on open-source software due to their various benefits such as flexibility, community support, and cost-effectiveness. Software development processes frequently incorporate the use of prebuilt, reusable open-source software components. However, it is important to acknowledge the potential risks associated with open-source software, which need to be effectively managed. There are several risks that may arise from the use of open-source components in RIC and \textit{xApps} software, such as the opportunity for attackers to exploit known vulnerabilities, unintentional insertion of vulnerable code, intentional backdoors inserted by malicious developers, and the potential for attackers to identify vulnerabilities through code review. In addition, non-compliance with open-source licenses could result in legal action and financial damage to consumers.

\subsubsection{Unauthorised access to UE identifications} \label{sec:AccessControlE2} 
As the UE identity is a key and sensitive piece of information in the RIC that enables UE-specific control, its exposure could lead to the compromise of sensitive information on users. This is a significant security concern as it can compromise the privacy and confidentiality of the users' data. For example, the UE identifiers, such as SUPI, will be exchanged frequently over the E2 and A1 interfaces.  If an attacker gains access to the E2 or A1 interface, they could potentially use a \textit{xApp} as a "sniffer" to identify specific UEs~\cite{boswell2020security}. Moreover, a vulnerability such as SDL API exploitation could potentially be exploited by the adversary to access the sensitive UE information stored within the UE-NIB database in RIC, which contains entries for UEs and their identities. Such data could include real-time location information of users connected to the network, potentially exposing sensitive personal information to the adversary.

\subsubsection{Conflict and inconsistency in radio access policies}\label{sec:lackOfclearFunctionalSplit}RIC enables mobile and wireless networks to be easily programmed through \textit{xApp}. However, these advantages may also lead to security vulnerabilities. With configuration changes from multiple \textit{xApp} or decisions from multiple components, inconsistencies might happen in the network. Lack of functionality in RIC to support preventing conflict to avoid mandatory network policies from being bypassed. The following vulnerabilities are two potential conflict decisions arising between different Open RAN components that the adversary could exploit.
\begin{itemize}[leftmargin=*]
\item \textit{RIC and gNB}. The division of responsibilities between the RIC and the gNB is not distinctly defined; it varies with the \textit{xApps} in use and the functionalities made available by the gNB. Such ambiguity can result in discrepancies between the RIC's actions and those of the gNB, potentially causing network instability and opening up exploitable vulnerabilities. For instance, a malicious actor could deploy a rogue \textit{xApp} designed to initiate radio resource management (RRM) actions that conflict with the gNB's own decisions, thereby orchestrating a DoS attack.

\item \textit{Multiple \textit{xApps}}. A multitude of interoperability limitations and policy collisions may be caused by the variety of third-party \textit{xApp} applications developed in various independent development environments with the help of many programming algorithms. For instance, the \textit{xApp} applications from multiple sources may make conflicting decisions to alter the same parameter in time for the same user. These types of conflicts lead to inconsistency in the network and give opportunities to the attacker to exploit the vulnerability.

\end{itemize}
\subsubsection{Network misconfiguration} \label{sec:misconfiguration}
RIC is responsible for managing and controlling the configuration of various network elements in data plane. It receives information from various sources in RAN and core network and then makes decisions on how the network should be configured to optimize performance and meet user requirements. The configuration of the network is constantly changing due to the dynamic nature of the network. RIC is responsible for ensuring that the network is configured correctly at all times. However, the network complexity can make it challenging for RIC to manage and control the network configuration effectively, increasing the likelihood of misconfiguration and faults. Therefore, it is essential to minimize the complexity of network configuration and avoid potential risks~\cite{nis2022euoran}. Here are some factors which cause network complexity in O-RAN.
\begin{itemize}[leftmargin=*]
    \item \textit{Integrating multiple suppliers.} The incorporation of multiple components from various suppliers in the data plane of O-RAN networks leads to an increase in complexity of network configuration. This complexity is managed and controlled by RIC, which may be prone to errors during the integration process. As a result, there is an elevated risk of network misconfiguration and faults. The impact of such misconfiguration and faults can be significant, as it can lead to network outages and security vulnerabilities, thereby posing a threat to the reliability and security of the network. 
    \item \textit{Decoupling hardware and software.} When hardware and software are decoupled, and various software components from different suppliers are used, there is a potential additional risk of misconfiguration of networks. This is because some software may offer security features that rely on a hardware-implemented secure element or support from other software components. As there are multiple combinations of software and hardware, or combinations of software components, integrating these security features may require extra effort. It could lead to situations where security features are not properly utilized. This presents a significant vulnerability as it may result in security breaches and network misconfiguration. Hence, it is important to carefully select and integrate software and hardware components to ensure that security features are properly implemented and utilized in the networks.
\end{itemize}

\subsubsection{Poisoned AI/ML RAN functions} \label{sec:PoisionedAI/ML} AI has the potential to enhance network efficiency and security substantially. Nonetheless, inadequately designed AI frameworks and algorithms may broaden the attack surface and reveal additional security weaknesses~\cite{masur2022artificial}. In O-RAN architecture, the RIC provides a software platform for deploying and managing AI/ML-enabled RAN functions. This means that the RIC can receive data from the data plane and use AI/ML algorithms to optimize network performance and functionality~\cite{lee2020hosting}. However, the application of AI/ML increases the risks of security attacks and privacy leakage. More specifically, placing the trained ML models in RIC can become new security attack surfaces and introduce the possibility of injecting poisoned models into RIC which could then be used to manipulate RAN functions in unexpected and harmful ways. This is a particular concern in O-RAN, where there is a focus on open interfaces and multi-vendor deployments, which could increase the risk of vulnerabilities in the RIC software. 

Open interfaces in Open RAN allow external sources to provide data to intelligent RAN functions and configure the network intelligently. However, the interfaces have not been further defined. This lack of standardization and clear definition for the open interfaces creates a higher risk for poisoning AI/ML RAN functions in O-RAN, as malicious actors can potentially inject data into the network through these interfaces, which could lead to a reduction in accuracy rate. Moreover, open interfaces allow for the integration of multiple components from different suppliers, which may not have undergone proper security testing and validation. These components can introduce new vulnerabilities that attackers can exploit to poison the AI/ML models used in RAN functions~\cite{rahman2021network}. Therefore, the risk of poisoned AI/ML RAN functions is primarily applicable in O-RAN architecture because of the unique features of the RIC and the use of AI/ML for RAN functions. It is crucial for O-RAN implementers and operators to be aware of this risk and to take appropriate measures to ensure the security and integrity of the RIC software and AI/ML models.

\subsubsection{Incomplete or inadequate hardening of the RIC}\label{sec:hardening}
Every function and module within the RIC needs to be fortified to minimize their vulnerability exposure. It is essential that hardening protocols ascertain that all default settings, encompassing the Operating System (OS) software, firmware, and applications, are correctly configured~\cite{baroos2020enisa}. RIC functions might be vulnerable if they miss appropriate security hardening. An attacker could, in such case, either inject malware and/or manipulate existing software, harm the Open RAN components, gain unauthorised access to the network or disrupt its operations. Following items are a few instances of weaknesses that may occur as a result of inadequate hardening of the RIC.

\begin{itemize}[leftmargin=*]
    \item \textit{Operating system (OS) vulnerabilities}. The OS supporting RIC components should provide a safe and stable environment for RAN functions. An attacker may exploit OS vulnerabilities such as default passwords, back-door accounts, open ports, unprotected services, and insecure protocols.
    \item \textit{Software vulnerabilities}.The RIC functions as a software platform, meaning that common software vulnerabilities translate directly to vulnerabilities within the RIC controller. Such vulnerabilities, whether they are flaws, defects, weaknesses, or errors in the software's design, can be exploited by attackers to disrupt the normal operations of the RIC or reconfigure the entire network, facilitating additional attacks.
    \item \textit{Improper cryptographic key management}. The RIC system needs to provide cryptographically separated secure environments for different applications. Improper mechanisms to manage cryptographic keys, including the use of weak algorithms and undermine trust in integrity and confidentiality protection mechanisms, can compromise sensitive functions and information on RIC.
    \item \textit{Lack of patch management process}. Lack of security patches and updates is a common security vulnerability that attackers can exploit to gain unauthorised access to RIC components. Security patches and updates are released by software and hardware vendors to address known security vulnerabilities, bugs, or other issues that attackers could exploit to compromise the system or network. 
    \item \textit{Improper log and audit mechanisms}. Security event logs need to be sent or uploaded to a central location or external systems, and these logs need to be securely stored and transferred to prevent unauthorised access. Failure to adequately protect these logs could result in delays, inaccurate audit outcomes, and prolonged security risks. Inadequate monitoring can also lead to undetected attacks or failures that are not addressed.
\end{itemize}

\subsection{RIC Vulnerabilities Inherited from SDN Controller}\label{sec:RICSecuritySDN}
In the previous section, we specifically discussed vulnerabilities and attacks which are newly emerging in the context of RIC in O-RAN architecture. We here remark that RIC originates from the main idea of using a centralized control plane, similar to SDN controller, to control and manage nodes in data plane. For this reason, RIC naturally inherits most of SDN controller vulnerabilities, in particular the ones generated by an inconsistent and inaccurate global view of the network. Maintaining a complete and accurate information of the network states is utmost important and a prerequisite for various RAN management tasks including monitoring, diagnosing, and resource management. 
Once the attack succeeds in poisoning this global view of the RIC, it could cause significant impact on the whole network. In what follows, we provide a summary of the vulnerabilities extracted from SDN security studies~\cite{hong2015poisoning, smyth2017detecting}, which could also impact RIC security.

\subsubsection{Relaying LLDP Frames} 
\begin{figure}[!htbp]
\centering
\includegraphics[width=0.47\textwidth]{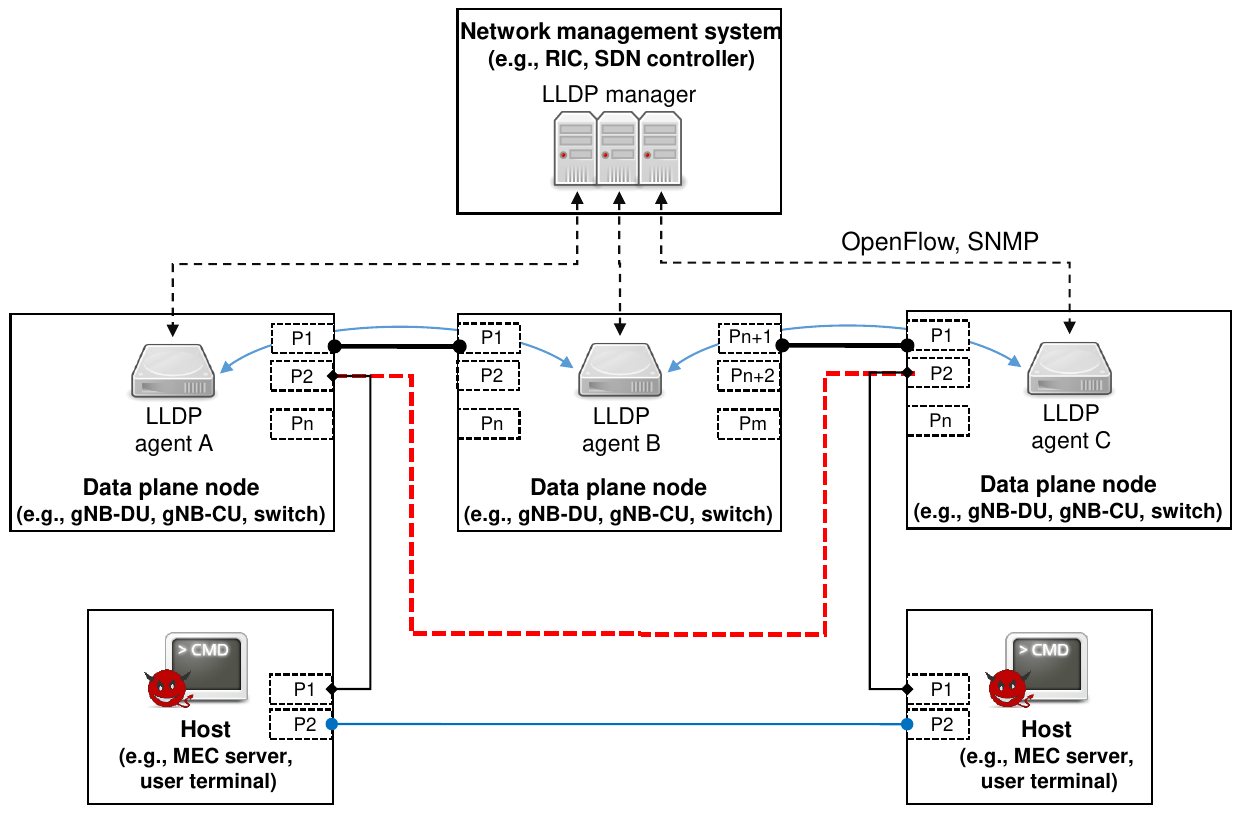}
\caption{\small Relaying LLDP frames which mislead the SDN controller and RIC view}
\label{fig:LLDPframe}
\end{figure}

The centralized control plane necessitates a comprehensive understanding of the entire network topology, encompassing all nodes and connections, to effectively strategize and enhance end-to-end communication routes. The R-NIB database, housed in the RIC, retains information regarding radio network states, RAN nodes (such as DUs, CUs, and UEs), and RAN links. To discern the interconnections between nodes in the data plane, the Link Layer Discovery Protocol (LLDP), as standardized in IEEE 802.1AB~\cite{1490127}, is employed. Fig.\ref{fig:LLDPframe} illustrates an attack scenario where a malicious actor relays LLDP information between two compromised hosts. When an LLDP agent (like a DU, CU, or OpenFlow switch) in the data plane receives an LLDP frame from its manager (such as the RIC or SDN controller), it disseminates the frame across all its ports. Any host or additional LLDP agent linked to this agent would consequently receive the frame. Under normal circumstances, hosts dismiss the incoming LLDP frame, resulting in no subsequent action. However, during an attack, a malicious actor controlling two hosts doesn't disregard the LLDP frame and instead channels the frame to a second LLDP agent, potentially using an out-of-band method. The attacker might also employ an in-band approach or a dual-home host to transmit the received LLDP. When the second agent receives this frame, it relays it to the LLDP manager. This deceptive act can mislead the network management system, be it an SDN controller or RIC, into falsely recognizing a connection between two nodes that aren't directly linked\cite{hong2015poisoning}. Within the O-RAN structure, DUs share LLDP data with CUs via a robust Midhaul network, typically facilitated through an Ethernet connection.

The challenge with relaying LLDP frames is that during link discovery, hosts may intercept these frames via their connected port to the LLDP agent. To confine the LLDP's reach solely to agents, a solution involving real-time port verification has been suggested in the framework presented in~\cite{hong2015poisoning}. This method employs a port classification system where each agent's port is designated based on the initial traffic it observes. A port will be categorized as SWITCH (or DU or CU in the context of Open RAN) if it transmits LLDP, while it's considered a HOST if it originates host traffic. Ports with no connected devices are classified as ANY. According to this tactic, the LLDP manager dismisses any LLDP frames emanating from a port identified as HOST. The limitation of this approach is that an intruder could compromise a host and mimic a SWITCH by transmitting LLDP messages. Additionally, the research in~\cite{smyth2017detecting} introduces a threshold-based defensive mechanism against such attacks, using statistical analysis of the distribution of link latency. When a new link is established, it's closely observed for a set vetting period before allowing normal traffic to flow through it. This vetting period is sustained until a sufficient number of link latency measurements are gathered. After the vetting period concludes, these latency figures are evaluated against a preset threshold, which represents the typical range of latency values under normal conditions.

\subsubsection{Compromised LLDP Frame}
When a compromised host receives the LLDP frame from the agent, it manipulates the field of agent DPID and Port ID inside the frame. By sending the modified LLDP back toward the agent, the controller announces a non-existing link between two agents. The reason is that neither LLDP integration nor authentication is guaranteed during the LLDP process. The work in~\cite{hong2015poisoning} provides this assurance by adding a key-Hash Message Authentication Code (HMAC)~\cite{krawczyk1997hmac} as a signed TLV inside each LLDP packet. The HMAC is computed using eq.~\eqref{eq:calHMAC}.
\begin{equation}
HMAC(K,m) = h((K \oplus opad)|h((K \oplus ipad)|m))\label{eq:calHMAC}
\end{equation}
where $m$ represents TLV fields in LLDP, i.e, the agent DPID and PortID. Parameters of $h$ and $K$ indicate hash function and secret key, respectively. Sign $|$ and $\oplus$ represent concatenation and XOR function. Also $opad$ and $ipad$ are considered 
constant values~\cite{krawczyk1997hmac,alharbi2015security}. Particularly, the author utilizes a static secret key and SHA-256 hash function. Another approach is to select a random dynamic secret key, i.e, $Ki,j$, where $i$ is the LLDP frame number and $j$ indicates the topology discovery round~\cite{alharbi2015security}. By using the dynamic secret key, each LLDP has a unique HMAC value. Hence, the adversary fails to use one instance of LLDP to create other fake LLDPs. Upon receiving the LLDP, the controller verifies the genuinity of the frame by checking the HMAC TLV.

\subsubsection{Port Amnesia} \label{sec:PAA}
Port amnesia attack~\cite{skowyra2018effective} can bypass the port classification module. In this attack, when the compromised host receives the LLDP frame, it relays the frame toward the peer compromised host.
Before sending LLDP toward the agent, the adversary disconnects and then connects the connection to the LLDP agent. This action triggers a \textit{Port-down} message toward the controller, causing the resetting label of the port from HOST to ANY. When the controller receives the LLDP packet from the port labelled with ANY, verifies LLDP path legitimacy and announces the new link. In the case of using an in-band channel to relay the LLDP frame, both compromised hosts need frequent context-switching between HOST and SWITCH labels which imposes an extra complexity on the attack. This is because compromised hosts need to have a HOST label when they forward data-plane traffic while their label must set SWITCH in case of using the fabricated link.

The work in~\textit{TopoGuard+}~\cite{skowyra2018effective} mitigates the risk of port amnesia attack by adding two modules to the controller, namely Link Latency Inspector (LLI) and Control Message Monitor (CMM). The former module detects the attack if the adversary uses the out-of-band channels between compromised hosts, while the latter protects the network against in-band LFA. The strategy of \textit{LLI} to detect a fake link is the extra latency that the out-of-band channel imposes on LLDP propagation duration. The controller periodically issues probe packets to measure the Round Trip Time (RTT) between the controller and LLDP agent. Upon receiving the packet, each switch provides a suitable response to that packet. Assume that we have two LLDP agents, namely, $s_1$ and $s_2$, that are connected to a controller. Also, suppose that $T_{p_1}$ and $T_{p_2}$ are the corresponding link latency of the probe packets sent to $s_1$ and $s_2$. The \textit{LLI} computes the link latency $T_l$ using eq.~\eqref{eq:calLinkDelay}.

\begin{equation}
T_l=T_{LLDP}-T_{p_1}-T_{p_2},\label{eq:calLinkDelay}
\end{equation}
where $T_{LLDP}$ indicates the propagation delay of the LLDP frame. To calculate the $T_{LLDP}$, the controller adds a times-tamp to the issued LLDP frame toward the LLDP agents and takes the difference when receiving it. Moreover, LLI stores the values of link latency based on received LLDP frames and calculates a latency threshold as shown in eq.~\eqref{eq:LLIThreshold}.
\begin{equation}
T_h=q_3+3*(q_3-q_1),\label{eq:LLIThreshold}
\end{equation}
where $q_1$ and $q_3$ indicate the lower and upper quartiles on stored latency values, respectively. By comparing latency $T_l$ and threshold $T_h$, LLI verifies the validity of the link and raises a security alarm in case of suspicious delay (i.e, $T_l > T_h$). However, the LLI module could be bypassed if the adversary gradually increases the latency of the threshold, i.e, $T_h$. For this purpose, the adversary overloads one LLDP agent in the network during a long period to slightly grow the latency of the LLDP packet in each propagation round~\cite{marin2019depth}. Eventually, this approach leads to a high value for the threshold which is greater than the latency of the out-of-band channel. However, it needs hours of preparation.

\renewcommand{\tabcolsep}{2pt}
\begin{table*}[htbp]
	\caption{An exhaustive view on vulnerabilities of RIC in Open RAN architecture.} \label{tbl:RICThreats}
    \centering
	\scriptsize
	\begin{tabular}{l>{\raggedright}m{1.8cm}>{\raggedright}m{1.7cm}>{\raggedright}m{2.9cm}>{\raggedright}m{4.6cm}>{\raggedright}m{3.3cm}>{\raggedright}m{1cm}m{1.1cm}}
            \hline
		\textbf{ID}
		&\textbf{RIC Vulnerability} 
		&\textbf{Source of Vulnerability}
		&\textbf{Affected Assets  }
		&\textbf{Root Cause} 
		&\textbf{Potential Impact} 
		&\textbf{Severity Level} 
		&\textbf{Likelihood Level}
		\\
		\hline
  	   \hyperref[sec:UncertifiedAccess]{V-01} & Uncertified access to the radio resource &Exclusive to Open RAN
	    &\begin{itemize}[leftmargin=0.13in]
		    \item Subscribers’ data
		    \item Network data
		    \item gNB-DUs
		    \item gNB-CUs\vspace{-2mm}
		\end{itemize}
	  	&\begin{itemize}[leftmargin=0.13in]
		    \item Compromised \textit{xApp}
		    \item Poorly designed \textit{xApp}
		    \item Malicious nested \textit{xApp}
		    \item Insecure APIs
		    \item Open-source software \vspace{-2mm}
		\end{itemize}
	    &\begin{itemize}[leftmargin=0.1in]
		    \item Service outage
		    \item Performance degradation\vspace{-2mm}
		\end{itemize}
		&High
		&High
		\\
			    \hline
		\hyperref[sec:AccessControlE2]{V-02} &
		Unauthorised access to UE identifications
		&Exclusive to Open RAN
		&\begin{itemize}[leftmargin=0.13in]
		    \item Subscribers’ data
		    \item Subscriber geo locations\vspace{-3mm}
		\end{itemize}

		& \begin{itemize}[leftmargin=0.13in]
		    \item Sniffing through a malicious \textit{xApp}
		    \item Compromised RIC\vspace{-3mm}
		\end{itemize}
		&\begin{itemize}[leftmargin=0.13in]
		    \item Information integrity
		    \item Information confidentiality\vspace{-3mm}
		\end{itemize}
		&High
		&Medium
	    \\
	    \hline
	    \hyperref[sec:lackOfclearFunctionalSplit]{V-03}  &
	    Conflict and inconsistency in radio access
        policies 
                &Exclusive to Open RAN
        &\begin{itemize}[leftmargin=0.13in]
            \item Network policies
		    \item gNB-DUs
		    \item gNB-CUs\vspace{-3mm}
		\end{itemize}
        &
        \begin{itemize}[leftmargin=0.13in]
                \item Conflict between RIC and O-gNB
                \item Conflicts arising from multiple \textit{xApp}\vspace{-3mm}
        \end{itemize}
	    &\begin{itemize}[leftmargin=0.13in]
		    \item Service outage
		    \item Performance degradation\vspace{-3mm}
		\end{itemize}
		&High&Medium
	    \\
		\hline
		\hyperref[sec:misconfiguration]{V-04} & Network misconfiguration &Exclusive to Open RAN
		    &\begin{itemize}[ leftmargin=0.13in]
		    \item Data traffic
		    \item gNB \vspace{-2mm}
		\end{itemize}
		& \begin{itemize}[leftmargin=0.13in]
		    \item Integrating multiple suppliers
		    \item Decoupling hardware and software \vspace{-2mm}
		\end{itemize}
		& \begin{itemize}[leftmargin=0.13in]
		    \item Service outage 
		    \item Performance degradation \vspace{-2mm}
		\end{itemize}
		&Medium&Medium
	    \\
		\hline
		\hyperref[sec:PoisionedAI/ML]{V-05}  &Poisoned AI/ML RAN functions &Exclusive to Open RAN
		    &\begin{itemize}[leftmargin=0.13in]
		    \item Network data
		    \item gNB
		    \item Data traffic\vspace{-2mm}
		\end{itemize}

		& \begin{itemize}[leftmargin=0.13in]
		    \item Open Interface
		    \item Multi-vendor deployments\vspace{-2mm}
		\end{itemize}
		& \begin{itemize}[leftmargin=0.13in]
		    \item Service outage 
		    \item Performance degradation
		    \item Information confidentiality\vspace{-2mm}
		\end{itemize}
		&High&Medium
	    \\
		\hline
	\hyperref[sec:hardening]{V-06} &
			Incomplete or inadequate hardening of the RIC
		    &Exclusive to Open RAN
		    &\begin{itemize}[leftmargin=0.13in]
		    \item Data traffic
		    \item gNB\vspace{-2mm}
		\end{itemize}

		& \begin{itemize}[leftmargin=0.13in]
		    \item Operating system vulnerabilities
		    \item Software vulnerabilities
		    \item Improper cryptographic key management
		    \item Lack of patch management process
		    \item Improper log and audit mechanisms\vspace{-2mm}
		\end{itemize}
		& \begin{itemize}[leftmargin=0.13in]
		    \item Service outage 
		    \item Information confidentiality
		    \item Information integrity
		    \item Performance degradation \vspace{-2mm}
		\end{itemize}
		&Medium&High
	    \\
		\hline
  		\hyperref[sec:RICSecuritySDN]{V-07} & Inaccurate topology discovery of data plane &Inherit from SDN
	 		&\begin{itemize}[leftmargin=0.13in]
		    \item Data traffic
		    \item gNB-DUs
		    \item gNB-CU-UPs\vspace{-2mm}
		\end{itemize}

		& \begin{itemize}[leftmargin=0.13in]
		    \item Unauthorised relaying of LLDP
		    \item Compromised LLDP 
		    \item Port amnesia technique
		    \item Miscalculated link latency \vspace{-2mm}
		\end{itemize}
		& \begin{itemize}[leftmargin=0.13in]
		    \item Service outage
		    \item Information integrity
		    \item Performance issues \vspace{-2mm}
		\end{itemize}
	    &High&Medium\\
     		\hline
	
	\end{tabular}
	\vspace{-5mm}
\end{table*}

In in-band port amnesia, the adversary frequently resets the adversary port to forward both LLDP traffic and host-generated traffic without raising labelling violations. This means that before sending host traffic, the port label should be set as HOST, while during sending the LLDP packet, it should be set as SWITCH. Hence, the CMM module monitors the frequency of port down and up during the LLDP  forwarding process and in case of any abnormal port connection and disconnection, it generates security alerts.

\subsubsection{Link Latency Attack}\label{sec:LLA}
The adversary launches Link Latency Attack (LLA) in two phases, including the overload phase and relay phases~\cite{soltanilink}.
In the overload phase, two compromised hosts inject ARP flooding traffic toward two agents $s_1$ and $s_2$. The traffic causes two negative consequences. First, it generates a huge number of \textit{Packet\_In} messages directed to the controller. Second, it imposes a high workload on the agent which leads to growing the RTT of the probing packet. In the relay phase, host $h_1$ listens for the LLDP packet and upon receiving the LLDP, it relays the packet to host $h_2$ through an out-of-band channel. Host $h_2$ directs the LLDP to the LLDP agent and it has to forward this packet to the controller. The controller receives the LLDP response packet, finds a change in the network topology, and updates it. To do so, it performs a check on the threshold and receives LLDP packet latency using eq.~\ref{eq:calLinkDelay} and eq.~\ref{eq:LLIThreshold}. Here, the values of $T_{p_1}$ and $T_{p_2}$ are high compared to the normal LLDP packets since they experience high latency in the \textit{overload phase}. However, by applying eq.~\eqref{eq:calLinkDelay}, the latency of the extra link between switches $s_1$ and $s_2$, i.e., $T_{l}$, stays in the valid range from the controller point of view, i.e, $T_l \leq T_h$. Even in some cases, the calculation shows a negative value for $T_l$. Hence, the LLI module in \textit{TopoGuard+} fails to detect the LLA, and finally, the controller updates its view of the network topology by adding an extra link between switches $s_1$ and $s_2$.

Having discussed the intricacies of RIC vulnerabilities in the O-RAN architecture, a summarized overview is provided for clarity. Table~\ref{tbl:RICThreats} presents a structured breakdown of the six vulnerabilities, detailing their source, affected assets, root cause, potential impact, severity, and likelihood. Readers are encouraged to refer to this table for a comprehensive overview of RIC vulnerabilities in the evolving Open RAN context.

\subsection{Attack Examples on RIC}\label{sec:ORAN-Sec-attEx}

In this section, we provide attack examples to the existing RIC schemes, exploiting the vulnerabilities highlighted in Section~\ref{sec:ORAN-Sec-Vul}.

\subsubsection{ Bearer Migration Poisoning (BMP)}\label{sec:BMP-attack}
The author in~\cite{soltani2022poisoning} introduces a novel attack, named \textit{Bearer Migration Poisoning (BMP)}, which misleads the nRT RIC to changes the user plane traffic path and causes signalling overhead. BMP has a remarkable feature that even a weak adversary with only two compromised hosts could launch the attack without compromising the RIC, RAN components, or applications. The attack works based on two following procedures which RIC initiates.
\begin{itemize}[leftmargin=*]
    \item \textit{Bearer context migration procedure: }A bearer context is a collection of signalling data transmitted via the E1 interface, which connects the CU-CP and CU-UP~\cite{O-RAN.CPlanProfile}. The establishment of a bearer context organizes the necessary resources and information to facilitate the transfer of user plane services among the CU-UP, the corresponding DU, and the UE~\cite{3GPPTS38.401}. For this purpose, the CU-CP employs bearer context management operations, which the nRT RIC platform can instigate~\cite{alliance2022E2GAP}. During this process, the CU-CP dispatches a \texttt{BEARER CONTEXT SETUP} message to form a new bearer context between the designated CU-UP and DU~\cite{masini2021guide}. It then signals the DU with a \texttt{F1 BEARER MODIFICATION} message to modify the configuration of the F1 interface. Ultimately, the CU-CP issues a \texttt{BEARER CONTEXT RELEASE} message to discard the previous bearer context. This results in the CU-CP transitioning the bearer context from the originating CU-UP to the intended CU-UP for a given DU. However, this functionality exposes \mbox{Open RAN} to an increased spectrum of security risks.
    \item \textit{Link discovery procedure: }The RIC is capable of managing the network topology for RAN elements on the data plane through the E2 interface, which includes carrying out the link discovery protocol. Initially, the controller formulates and dispatches LLDP messages to every node in the data plane. Subsequently, as each node receives the LLDP packets, it propagates them from all of its ports. Next, the node that receives these packets relays the LLDP back to the RIC. Consequently, the controller is able to detect all the current links among the nodes in the data plane. This process is methodically repeated at consistent, scheduled intervals.
\end{itemize}

\noindent By leveraging the bearer context migration procedure and the link discovery procedure, adversaries can strategically mislead the RIC, initiating malicious activities that threaten the network's integrity. In the following, we delve deeper to describe this threat in greater detail.

\noindent\textit{Threat: }The malicious actor's objective is to deceive the RIC into thinking that it's necessary to initiate a bearer context migration procedure. We outline the BMP attack using a streamlined O-RAN structure, which includes a single DU, two CU-UPs, a pair of MEC servers functioning as hosts, and several routers within the mid-haul network, as depicted in Fig.~\ref{fig:BCA}. It's assumed that the adversary has taken control of two MEC hosts, specifically h$_1$ and h$_2$, which are connected to DU$_1$ and CU-UP$_2$ respectively. In this particular attack setup:

\begin{figure}[!htbp]
\centering
\includegraphics[width=0.47\textwidth]{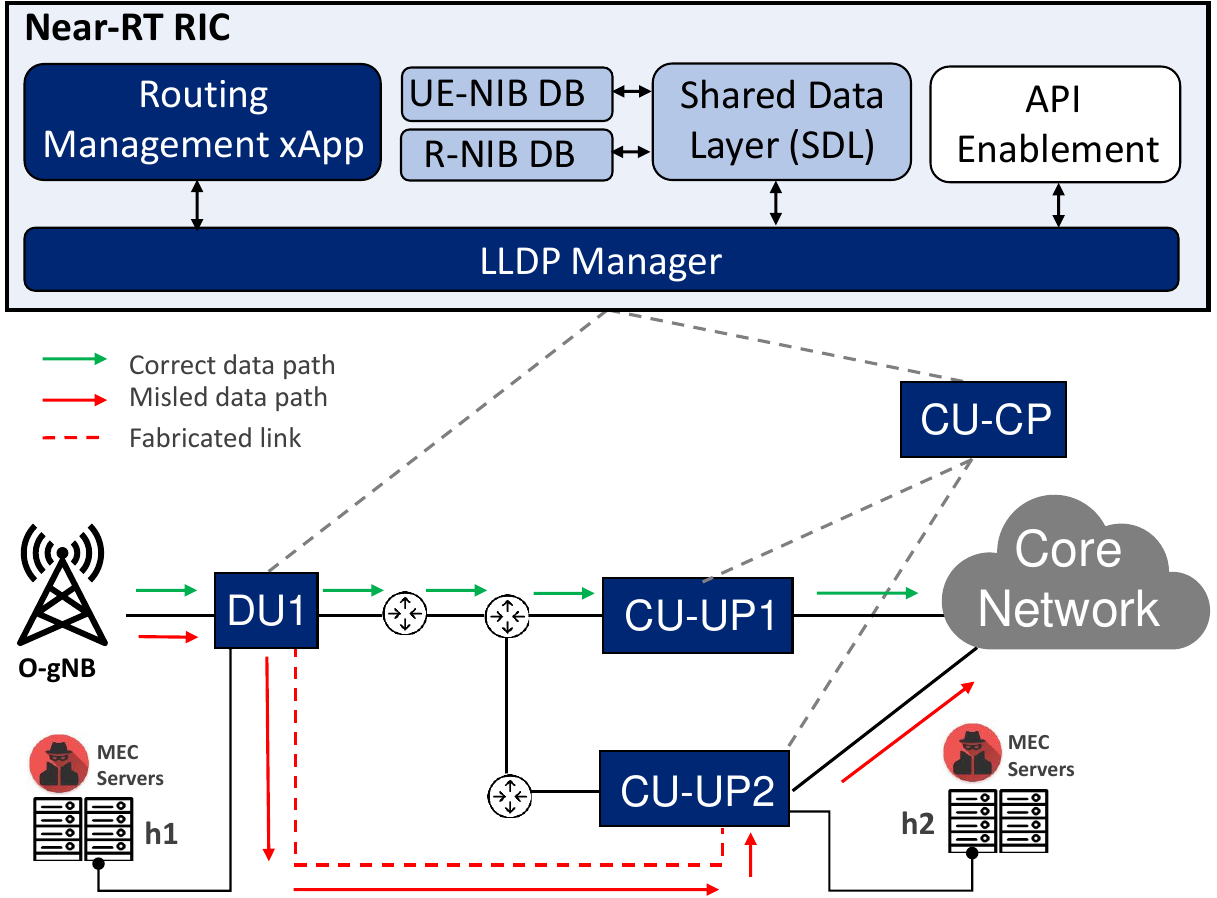}
\caption{\small Bearer migration poisoning attack in Open RAN}
\label{fig:BCA}
\end{figure}

\begin{enumerate}[label=(\roman*)]
    \item The malicious attacker at \( h_1 \) actively monitors the port linked to \( DU_1 \), designated as port 1 for this illustration. In the course of the link discovery process, the RIC dispatches an LLDP packet in the direction of \( DU_1 \). Once \( DU_1 \) acquires the LLDP packet, it disseminates it across its ports. This packet is then intercepted by the adversary at \( h_1 \), who redirects it towards \( h_2 \) through an alternative communication channel. Subsequently, the compromised \( h_2 \) relays this LLDP packet to \( CU-UP_2 \), which in turn sends it back to the RIC. The RIC, upon receipt of the LLDP packet from \( CU-UP_2 \), is misled into cataloguing a non-existent link between \( DU_1 \) and \( CU-UP_2 \), illustrated by the dashed red line in Fig.~\ref{fig:BCA}. This manoeuvre is predicated on the link fabrication tactic detailed in~\cite{skowyra2018effective,soltani2021link}.

    \item Given that the routing \textit{xApp} is configured to receive updates on the network's radio topology, it is alerted by the RIC with a \texttt{Topology Update Report}, which details the addition of the spurious link. The routing algorithm, typically reliant on determining the shortest path, opts for this new link due to its perceived directness between \( DU_1 \) and \( CU-UP_2 \). Consequently, the \textit{xApp} dispatches a \texttt{Path Update Request} to the RIC to apply the updated routing configuration.
    \item In response to the path update notification, the RIC issues a \texttt{RIC Control Request} to the CU-CP to instigate the bearer context modification. Following this, the CU-CP commences the switch of CU-UP upon receipt of the request from the RIC, terminating the existing bearer context with \( DU_1 \) and \( CU-UP_1 \), and forming a new bearer context with \( DU_1 \) and \( CU-UP_2 \).

\end{enumerate}

The reconfiguration of the bearer context, which diverts user traffic through the fake created link between \( CU-UP_1 \) and \( CU-UP_2 \), adversely affects the cell performance managed by the RU, as depicted by red arrows in Fig.~\ref{fig:BCA}. As demonstrated by empirical data and analyses in \cite{soltani2022poisoning}, such a BMP attack can precipitate two major repercussions. Initially, it can critically diminish the throughput of both downlink and uplink to almost \( 0 \, \text{Mbps} \), severely undermining service quality and user satisfaction, potentially leading to a loss of customers and revenue for the operators. Secondly, it can lead to a substantial surge in signalling overhead, thus inflating network latency and squandering valuable radio spectrum resources, with the signalling cost experiencing an approximately tenfold increase due to the BMP attack.

\subsubsection{ Signaling Storm Attack (SSA)} 
The advancement of 6G technology implies that a vast number of Internet of Things (IoT) devices, will utilize the Open RAN network. Therefore, the IoT may escalate the danger of Open RAN network congestion due to Distributed Denial of Service (DDoS) attacks~\cite{soltani2022can},~\cite{O-RAN.SFGthreat}. A DDoS attack involves numerous compromised machines, or bots, working together to target the RIC controller and deny its service to users. These attacks have various negative impacts on RIC controllers including service downtime, financial losses, and both short-term and long-term consequences. DDoS attacks come in different forms, with varying objectives, methods, and scales. DDoS attacks can fall into two categories: brute force and semantic. In brute-force attacks, also known as flooding or high-rate DDoS attacks, attackers inundate the targeted cloud server with a large number of malicious requests, with the goal of overwhelming its network bandwidth. However, these attacks are easy to detect due to their high traffic rate.
In contrast, semantic attacks, also known as vulnerability attacks, exploit weaknesses in protocols instead of using up network bandwidth or cloud computing resources. Attackers use a lower volume of malicious traffic to target specific protocols or applications, which makes these attacks more challenging to identify as they look similar to legitimate traffic. These attacks are referred to as low-rate DDoS attacks \cite{agrawal2019defense}. In this section, we plan to devise one potential attack scenario known as the Signaling Storm Attack (SSA), where an adversary exploits standard network CP mechanisms to initiate a DDoS attack. For example, they may flood the network CP with invalid or repeated registration requests. Even if these registration requests are declined, they still consume core network resources in the CP that are crucial during the authorisation process.

In this scenario, attackers are using a specific type of malware that is designed to infect IoT thermometers. This malware is called "remote-reboot" malware because it allows attackers to remotely control the IoT thermometer and reboot it on command. The attackers first need to find a vulnerability in the IoT thermometer's security, such as a weak password or outdated software, that they can exploit to gain access to it. Once they have access, they can install the remote-reboot malware onto the thermometer without the owner's knowledge. The malware allows the attackers to take control of the IoT thermometer and add it to a botnet army. A botnet is a network of infected devices that can be controlled by a single entity, in this case, the attackers. By infecting a large number of IoT thermometers, the attackers can create a powerful botnet army that they can use to carry out the DDoS attack in order to target the RIC controller. The attackers can instruct the remote-reboot malware to perform specific actions, such as rebooting all the infected IoT thermometers at the same time. Overall, by infecting a large number of IoT thermometers with remote-reboot malware, the attackers can create a botnet army that they can control and use to carry out a DDoS attack. 

\begin{figure}[!htbp]
\centering
\includegraphics[width=0.47\textwidth]{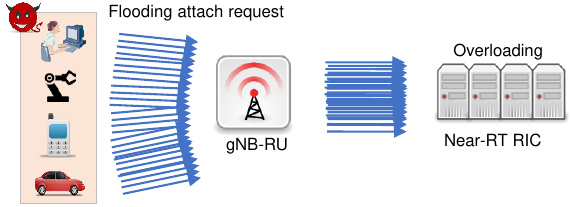}
\caption{\small Signaling storm attack in Open RAN}
\label{fig:ddos}
\end{figure}

The botnet army floods the controller with a massive volume of \texttt{RRC} connection request signalling messages, overwhelming its capacity and rendering it unable to process legitimate requests from the network (see  Fig.~\ref{fig:ddos}). As the attack continues, the botnet army may change its tactics and start targeting specific vulnerabilities in the RIC controller's software. The attacker may also exploit vulnerabilities to gain unauthorised access to the controller and to execute arbitrary code. Therefore, the attacker can take control of the network functions managed by the RIC controller. They may be able to modify traffic routing rules, reconfigure network slices, or even shut down critical network services~\cite{O-RAN.SFGthreat},~\cite{liao2022development}. After the RIC natively identifies a DDoS attack, some form of mitigation measure is necessary. The RIC is the most efficient Open RAN component for dealing with this type of attack since it manages RRC connections, making it ideal for preventing excessive malicious Attach Requests. Detecting and mitigating the attack in this manner would showcase built-in closed-loop automation \cite{5GAmericas}. 

\subsection{Lessons Learned}
The analysis of vulnerabilities within the RIC in O-RAN technology has revealed critical lessons for network security. Firstly, it emphasizes the need for a meticulous assessment of third-party applications and APIs due to Open RAN's integration of diverse radio resources and third-party \textit{xApps}. Rigorous security evaluations are essential to minimize vulnerabilities, and securing APIs must be prioritized, necessitating robust authentication, authorisation, and encryption mechanisms to safeguard sensitive data and functions. Secondly, this analysis underscores the importance of careful management of open-source software within Open RAN. While open-source software offers flexibility and cost-effectiveness, it can introduce risks such as known vulnerabilities and malicious code. Adequate open-source software management, including vulnerability monitoring and patching, is crucial. Furthermore, protecting user data, simplifying network configuration, implementing robust conflict resolution mechanisms, securing AI/ML deployments, and prioritizing security hardening are all integral aspects of securing Open RAN networks.

Additionally, the technical lessons extend to addressing vulnerabilities inherited from SDN controllers in the context of the RIC and Open RAN architecture. RIC's centralized control plane approach shares conceptual similarities with SDN controllers, leading to the inheritance of vulnerabilities related to maintaining an accurate global view of the network. To mitigate these risks, several key lessons emerge. First, the vulnerability associated with relaying LLDP frames highlights the necessity of real-time port verification mechanisms to prevent unauthorised manipulation of LLDP frames, ensuring only authorised agents handle them. Second, compromised LLDP frames can be addressed through the implementation of techniques like HMAC to sign LLDP packets, enhancing security and thwarting attempts to announce non-existent links. Lastly, the Port Amnesia Attack, wherein compromised hosts manipulate port states to bypass security measures, underscores the importance of detection and mitigation. Modules like LLI and CMM
play a crucial role in monitoring link latency and abnormal port connections to identify and respond to such threats effectively. In conclusion, these integrated technical lessons emphasize a comprehensive approach to network security in O-RAN technology, including robust assessments, open-source software management, and protection against vulnerabilities inherited from SDN controllers, all supported by security measures like real-time verification, cryptographic authentication, and vigilant monitoring.

\section {Resilient RIC Security: Advancements for 5G and 6G}\label{sec:RICSecurity5G6G}
In Section \ref{sec:RIC-SecurityRisks}, we identified the potential threats to the RIC security. In this section, we will delve into the security solutions and practices that can help enhance the resilience of RIC security in the face of these threats. Specifically, we will outline the RIC requirements that must be addressed in 5G and 6G networks in order to effectively mitigate these identified threats.
The O-RAN Alliance WG11 has published a number of technical specifications that mandate authentication and encryption procedures for RIC and \textit{xApps}~\cite{O-RAN.SFGRICXAPP}. The inclusion of third-party \textit{xApp} in the RIC can provide attackers with the exploit of taking control of different nodes of the network. Security controls are required to prevent malicious \textit{xApps} from leaking sensitive RAN data or from affecting the RAN performance. The security of RAN architecture must not be compromised as it becomes more software-driven, open, and intelligent. As Open RAN architecture is being defined and developed, security controls need to be embedded from the beginning to ensure the security-by-design principles.

The decoupling of both software and hardware in Open RAN architecture demands a reliable process. One of the encouraging algorithms in order to suppress security hazards in the Open RAN ecosystem is the ZTA which is proposed in~\cite{gilman2017zero,ramezanpour2022intelligent }. The Open RAN attack surface
must be protected through the ZTA approach, built upon a secure-by-specification and design foundation. Based on ZTA, the network is in a perilous environment at all times, and many internal and external hazards exist in it. Physical location is not adequate to specify the reliability of a network. All users, entities, and network traffic inside and outside a network are untrustworthy unless authenticated and verified. Also, security guidelines should be dynamic and computed from as many information origins as possible. 
The ZTA decreases malicious access and attacks by applying the least privilege guidelines and firmly executing access control guidelines and policies. It recognizes and logs all network traffic and constantly traces user's actions. It ensures secured access to Open RAN and 6G networks, regardless of whether the access location is from an internal or external network. The concept of “zero-trust” meets the security necessities of a 5G network that organizes many linked gadgets and sessions with unclear risks. Therefore, the zero trust algorithm is efficient for addressing the network threats and Open RAN security limitations. A ZTA module, with real-time monitoring and risk evaluation and the objectives of maximum availability and least privileges, seems a necessary and promising solution to protect the beyond 5G infrastructure from emerging security threats. The O-RAN Alliance WG11 recommended the intended countermeasures and best practices to mitigate the identified risks in high-level and general forms. Many research and development directions exist that need to be discovered. These include the mutual authentication for validating access to Open RAN structure and prohibiting malicious elements and applications, procedures enabling reliable implementation of \textit{xApps} and artificial intelligence algorithms, and cryptography with protected key management, consisting of key generation, storage, rotation, and revocation.

The purpose of this section is to present a comprehensive framework for understanding and enhancing RIC security, reflecting on the lessons learned and looking ahead to future developments. This section will unfold in four interconnected subsections, each addressing a different facet of RIC security within the context of current and next-generation networks. Section~\ref{sec:RICSEC5G} will scrutinize the security measures currently deemed essential for the protection and robust performance of RIC in the 5G infrastructure.Section~\ref{sec:RICSEC6G} will pivot to the anticipated advancements in 6G, exploring the innovative security paradigms designed to leverage the advanced capabilities of RICs. Section~\ref{sec:RICSECUsecase} will illustrate practical scenarios where RIC security measures have been effectively implemented. Section~\ref{sec:RICSECLL} aims to encapsulate the key insights and takeaways from the current implementation of security measures in RIC systems. Together, these sections aim to provide a strategic blueprint for the evolution of RIC security, detailing an approach that is as proactive as it is responsive, within the rapidly evolving landscapes of 5G and 6G networks.

\subsection{Enhancing Resilient RIC Security in 5G}\label{sec:RICSEC5G}

Recognizing the imperative of RIC security is the first step toward safeguarding the 5G infrastructure. The RIC not only holds the key to network efficiency and service innovation but also bears the responsibility for protecting the network against a host of security threats. These threats range from data breaches and unauthorised access to service disruptions and attacks on network integrity. In light of these challenges, ensuring the security of the RIC is not just a technical necessity but a fundamental requirement to maintain trust and reliability in 5G networks. In the discussion that follows, we will outline a series of critical security controls essential for maintaining a secure RIC. We will explore the multifaceted strategies and mechanisms necessary to ensure that every facet of the RIC's operation, from \textit{xApp} lifecycle management to interface communications, is fortified against the spectrum of cyber risks. By instituting these comprehensive security controls, we aim to establish a robust framework that not only protects the RIC but also enhances the resilience and integrity of the entire 5G network.

\subsubsection{Ensuring secure \textit{xApp} registration}

It is imperative to incorporate effective, adaptable, and verifiable technical safeguards to ensure the secure and reliable deployment of \textit{xApps} on the nRT RIC platform. Additionally, it is crucial to prioritize the validation of \textit{xApp} integrity during the registration phase, thereby confirming its authenticity as provided by a reputable source. The \textit{xApps} are applications that undergo an onboarding process onto the SMO system. Subsequently, they are deployed and officially registered within the nRT RIC.

\begin{figure}[!htbp]
\centering
\includegraphics[width=0.47\textwidth]{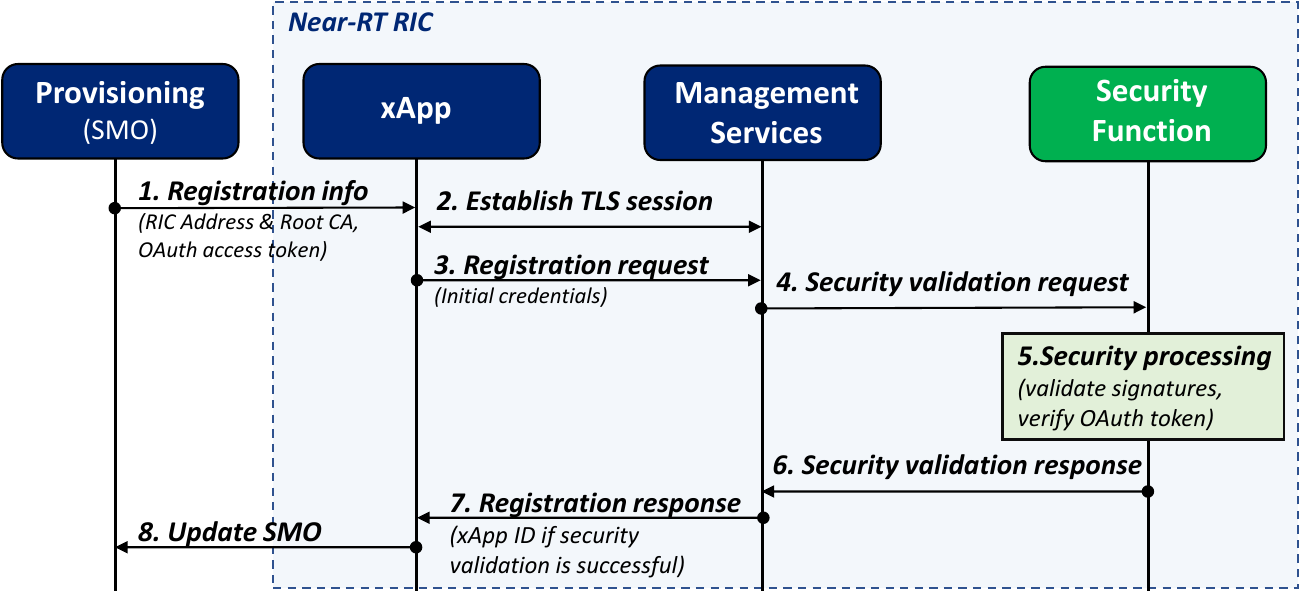}
\caption{\small Secure \textit{xApp} registration on nRT RIC}
\label{fig:xAppRegistration}
\end{figure}  

The O-RAN Alliance WG11 advocates for the mandatory inclusion of authentication and authorisation within the \textit{xApp} onboarding, provisioning, and registration processes~\cite{O-RAN.SFGRICXAPP}. 
\begin{enumerate}[label=(\roman*)]

    \item \textit{Secure onboarding.} The application developer generates a digital signature to sign the \textit{xApp}, which is subsequently transmitted to the network provider. The network provider performs validation of the digital signature and, if it passes, adds its signature to the package. This process ensures that the \textit{xApp} is verified to have come from a legitimate developer and has not been tampered with in transit. 
    \item \textit{Secure provisioning.} As a pre-requisite to the registration procedure, the \textit{xApp} obtains information from a provisioning system in SMO. This information is used to authenticate and establish a secure TLS communication with the nRT RIC platform during the registration process. The information encompasses specific attributes of the nRT RIC platform, including its address and the Root Certificate Authority (CA) certificate. The Root CA certificate, acting as a cornerstone in digital security, establishes trust in the RIC's identity. Additionally, this information covers an initial registration credential, exemplified by the OAuth 2.0 access token, which plays a pivotal role in securing access to the RIC, as we will discuss later in Section~\ref{sec:APIaccess}. 
    \item \textit{Secure registration.} The registration process for \textit{xApps} starts with establishing a secure session between the \textit{xApp} and the security function in the nRT RIC platform using TLS as illustrated in Fig.~\ref{fig:xAppRegistration}. Subsequently, the \textit{xApp} sends a registration request message along with the initial registration credential pre-provisioned during the provisioning process (see message 3). The security function in the nRT RIC platform validates the initial registration credential and if the validation is successful, responds with a \textit{xApp} registration response message that includes the \textit{xApp} certificate and xApp ID (see message 7). This \textit{xApp} ID plays a crucial role in facilitating communication between \textit{xApps} and the nRT RIC platform, as it serves as a means for the platform to unequivocally identify each \textit{xApp} when processing API request messages.  
\end{enumerate}

\subsubsection{Ensuring strict API access controls}\label{sec:APIaccess} It's critical to differentiate \textit{xApps} based on their security profiles before they are permitted to interact with network data and radio assets. This stratification is pivotal to ensure that \textit{xApps} from service providers, third-parties, and platform vendors are properly authenticated and authorized, coupled with rigorous application auditing. Each \textit{xApp}, with its unique operational demands from the controller and network, should meet specific security standards. For instance, \textit{xApps} like those for load balancing may require access to network metrics such as byte or packet counts to function, whereas \textit{xApps} for intrusion detection might necessitate packet header inspection. Moreover, \textit{xApps} from service providers and third parties should have distinct levels of access to network details and radio resources, necessitating the enforcement of stringent network access control on API interactions.

For API authentication, the implementation of mutual authentication, with support for mutual TLS (mTLS 1.2 or higher) or IPsec can be utilized~\cite{O-RAN.SFGRICXAPP}. The mTLS requires both the \textit{xApp} (as API consumer) and the nRT RIC platform (as API producer) to have their certificates and verify each other's identity using their public and private keys. On the other hand, IPsec can provide secure communication by encrypting network traffic between two endpoints. To maintain secure and limited access to APIs in compliance with established authorisation standards, 3GPP recommends the use of Open-authorisation 2.0 (OAuth 2.0) protocol~\cite{3GPPTS33.210,3GPPTS33.310}. This approach ensures that third-party \textit{xApps} are only permitted to access resources for which they have explicit authorisation. OAuth 2.0 is a well-known authorisation framework, defined in RFC 6749~\cite{hardt2012oauth}, which enables third-party \textit{xApps} to request limited access to a particular resource. The main actors in OAuth 2.0 are the resource server, i.e., API producer, the resource requester, i.e., client or API consumer, and the authorisation server. The O-RAN Alliance WG11 also advocates for the use of OAuth 2.0 as a token-based authorisation mechanism to securely restrict \textit{xApp} access to APIs, such SDL APIs~\cite{O-RAN.SFGRICXAPP}.

\begin{figure}[!htbp]
\centering
\includegraphics[width=0.50\textwidth]{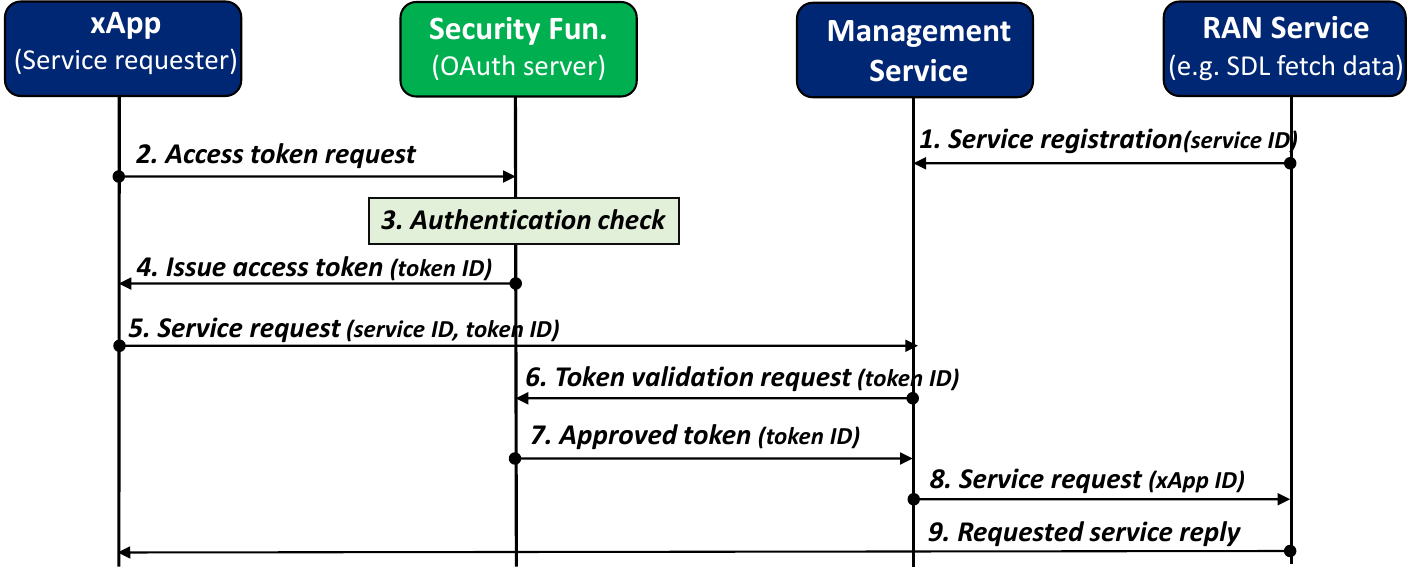}
\caption{\small RIC OAuth 2.0 authorisation flow to protect API access to the SDL data.}
\label{fig:OAuth}
\end{figure}

Fig.~\ref{fig:OAuth} illustrates an example of OAuth 2.0 authorisation flow to restrict API access to the SDL data. In this procedure, the \textit{xApp} requests access to the radio resource of SDL data, through an exposed API. The resource server is the nRT RIC that offers various RAN services via APIs. All offered services are initially registered in the management service module located in nRT RIC (see message 1). The resource requester or API consumer is the \textit{xApp}, and the nRT RIC security function serves as the authorisation server. The flow involves the \textit{xApp} first sending a  request for an access token (e.g., JSON access token) to the RIC security function for an SDL data service (see message 2). Then the security function authenticates the request and generates an access token, which the \textit{xApp} uses to access the API (see messages 3-5). Next, the management service module validates this request through the security function and, if the token is valid, forwards the request to the service provider module (see messages 6-8). Finally, the RAN service provider as the API producer provides the requested resource or service in response to the original request (see message 9). By implementing these measures, the industry can maintain a high level of security while allowing for the necessary access to resources.

\subsubsection{Ensuring Granular Access Control for R-NIB and UE-NIB}

The \textit{xApps} can utilize the SDL APIs to access both the R-NIB and UE-NIB databases. This access provides \textit{xApps} with the ability to leverage the sensitive information stored in these databases to perform various network management functions. As described in Section~\ref{sec:APIaccess}, the OAuth 2.0 mechanism is used to authorise \textit{xApps} to access SDL APIs. However, in order to guarantee fine-grained access controls for the R-NIB and UE-NIB databases, a Role-Based Access Control (RBAC) solution is also necessary. To this end, each \textit{xApp} is assigned an RBAC role, which determines what operations the \textit{xApp} can perform on what database. During \textit{xApp} onboarding, the \textit{xApp} solution provider produces a database token that details the required role, access type, and operations for the \textit{xApp}. Subsequently, the \textit{xApp} sends a database access request to the SDL for approval. The SDL forwards a client authorisation request to the Security Function, which compares the \textit{xApp's} database token with the one validated by the Service Provider to authorise access. Once authenticated, the Security Function assigns the \textit{xApp} a role based on the token and communicates the response to SDL.
\subsubsection{ Ensuring Secure Software Development} Open RAN suppliers need to implement secure software development best practices and cannot rely exclusively upon the open-source community to build secure software.
To mitigate these risks, it is recommended that security best practices are implemented, such as maintaining an inventory of open-source components, tracking and analyzing vulnerabilities, remediating open-source vulnerabilities, and continuously monitoring for new risks. By following these best practices, businesses can effectively manage the risks associated with the use of open-source software and libraries. Furthermore, it is necessary to implement appropriate tools and procedures that maintain software integrity by accurately monitoring modifications and the state of updates, especially during the application of software upgrades and security patches on the nRT RIC platform~\cite{O-RAN.SFGRICXAPP}.

\begin{table*}[htbp]
    \caption{Resilient RIC Security Measures and Vulnerabilities Mitigated$\CIRCLE$: Mitigated. $\LEFTcircle$: Partial Mitigated. $\Circle$: Not Mitigated}
    \label{tbl:RICSecurityFeaturesReversed}
    \centering
    \scriptsize
    \begin{tabular}{m{5.5cm}m{9.8cm}*{7}{>{\centering\arraybackslash}m{0.23cm}}}
         \hline
        \textbf{RIC Security Measures} & \textbf{Security Controls} &  \textbf{V-01} & \textbf{V-02} & \textbf{V-03} & \textbf{V-04} & \textbf{V-05} & \textbf{V-06} &\textbf{V-07} \\
        \hline
        \multirow{2}{=}{Ensuring strict API access controls} & 
        \begin{itemize}[leftmargin=*,nosep]\vspace{1mm}
            \item Mutual authentication (mutual TLS or IPsec).\vspace{-2mm}
            \end{itemize} 
        & $\CIRCLE$ & $\CIRCLE$ & $\Circle$ & $\Circle$ & $\CIRCLE$ & $\CIRCLE$ & $\Circle$ \\
        & 
        \begin{itemize}[leftmargin=*,nosep]
            \item Utilization of OAuth 2.0 for authorisation..\vspace{-2mm}
            \end{itemize} 
        & $\CIRCLE$ & $\CIRCLE$ & $\CIRCLE$ & $\Circle$ & $\Circle$ & $\Circle$ & $\Circle$ \\
        \hline
        \multirow{4}{=}{Ensuring secure \textit{xApp} deployment} & 
        \begin{itemize}[leftmargin=*,nosep]\vspace{1mm} 
            \item Digital signature generation and validation.\vspace{-2mm}
            \end{itemize} 
        & $\CIRCLE$ & $\CIRCLE$ & $\Circle$ & $\Circle$ & $\Circle$ & $\Circle$ & $\Circle$ \\
        & 
        \begin{itemize}[leftmargin=*,nosep] 
            \item Secure TLS communication during provisioning.\vspace{-2mm}
            \end{itemize} 
        & $\CIRCLE$ & $\CIRCLE$ & $\Circle$ & $\Circle$ & $\Circle$ & $\Circle$ & $\Circle$ \\
        & 
        \begin{itemize}[leftmargin=*,nosep] 
            \item Authentication in onboarding, provisioning, and registration.\vspace{-2mm}
            \end{itemize} 
        & $\CIRCLE$ & $\CIRCLE$ & $\Circle$ & $\Circle$ & $\Circle$ & $\Circle$ & $\Circle$ \\
        \hline
        \multirow{2}{=}{Ensuring granular access control for R/UE-NIB} & 
        \begin{itemize}[leftmargin=*,nosep]\vspace{1mm}
            \item Role-Based Access Control (RBAC).\vspace{-2mm}
            \end{itemize} 
        & $\CIRCLE$ & $\CIRCLE$ & $\CIRCLE$ & $\Circle$ & $\CIRCLE$& $\Circle$& $\CIRCLE$  \\
        & 
        \begin{itemize}[leftmargin=*,nosep]
            \item Database token generation and validation.\vspace{-2mm}
            \end{itemize} 
       & $\CIRCLE$ & $\CIRCLE$ & $\CIRCLE$ & $\Circle$ & $\CIRCLE$ & $\Circle$ & $\CIRCLE$ \\
        \hline
        \multirow{2}{=}{Ensuring secure software development} & 
        \begin{itemize}[leftmargin=*,nosep]\vspace{1mm}
            \item Maintain an inventory of open-source components and track vulnerabilities.\vspace{-2mm}
            \end{itemize} 
        & $\CIRCLE$ & $\Circle$ & $\Circle$ & $\CIRCLE$ & $\Circle$ & $\CIRCLE$ & $\Circle$  \\
        & 
        \begin{itemize}[leftmargin=*,nosep]
            \item Continuous monitoring for new risks and vulnerabilities.\vspace{-2mm}
            \end{itemize} 
        & $\CIRCLE$ & $\Circle$ & $\Circle$ & $\CIRCLE$ & $\Circle$ & $\CIRCLE$ & $\Circle$ \\
        \hline
        \multirow{2}{=}{Ensuring secure RIC interface communications} & 
        \begin{itemize}[leftmargin=*,nosep]\vspace{1mm}
            \item Utilize secure protocols like IPsec, TLS, and DTLS for data confidentiality and integrity.\vspace{-2mm}
            \end{itemize} 
        & $\CIRCLE$ & $\CIRCLE$ & $\Circle$ & $\Circle$ & $\CIRCLE$ & $\Circle$& $\Circle$  \\
        & 
        \begin{itemize}[leftmargin=*,nosep]
            \item Safeguard against eavesdropping and unauthorised O\&M connectivity.\vspace{-2mm}
            \end{itemize} 
        & $\Circle$ & $\Circle$ & $\Circle$ & $\CIRCLE$ & $\Circle$ & $\CIRCLE$ & $\Circle$ \\
        \hline
        \multirow{1}{=}{Protecting against conflicting \textit{xApps}} & 
        \begin{itemize}[leftmargin=*,nosep]\vspace{1mm}
            \item Function initiated by authorised \textit{xApps}, Subscription Management, or the function itself.\vspace{-2mm}
            \end{itemize} 
         & $\Circle$ & $\Circle$ & $\CIRCLE$ & $\CIRCLE$ & $\Circle$ & $\Circle$ & $\CIRCLE$  \\
        \hline
    \end{tabular}
\end{table*}

\subsubsection{Ensuring Secure RIC Interface Communications} The connection of RIC to external components must utilize a secure access protocol, such as IPSec or TLS, to safeguard against unauthorized interception of network data. Additionally, it is advised to enforce mutual authentication for O\&M connections to ensure secure access.
\begin{itemize}[leftmargin=*]
    \item \textit{Secure E2 interface.} One effective security control that can be implemented on the E2 interface is to adopt existing 3GPP recommendations for secure communication protocols. For instance, the use of IPsec (Internet Protocol Security) in tunnel mode, DTLS (Datagram Transport Layer Security), and TLS (Transport Layer Security) protocols can help ensure the confidentiality, integrity, and authenticity of data transmitted over the E2 interface. IPsec provides end-to-end security by encrypting and authenticating IP packets, while DTLS and TLS provide security for User Datagram Protocol (UDP) and Transmission Control Protocol (TCP) data, respectively.
    \item \textit{ Secure A1 interface.} The non-RT RIC sends policies to the nRT RIC via the A1 interface, which is secure and has measures to ensure confidentiality, integrity, and mutual authentication. However, the nRT RIC should not automatically trust the policies it receives and must implement its own security measures based on a zero-trust architecture. This means that the nRT RIC should not rely solely on perimeter security to protect against internal threats. The security measures that can be applied to the nRT-RIC's A1 Termination interface are to ensure policies comply with the specified schema, verify the correctness and range of policy values, use rate limiting to prevent resource depletion and Dos attacks, and set up security logging to record any failures. 
    \item \textit{ Secure O1 interface.}In certain network setups, vulnerabilities may be present in the O1 functions of an Open RAN architecture. To mitigate these risks, it is essential to apply security measures that adhere to zero trust principles, which are at the core of Open RAN's security criteria. To ensure this, the O1 interface will employ encryption to provide confidentiality, integrity, and authenticity, as well as to enforce least privilege access control through the use of a network configuration access control model. Inadequate security measures could result in adverse effects on network operations from unauthorized write operations like 'edit-config' and expose sensitive data through unrestricted read operations. Network operators must use a standards-based method that works with NETCONF for role-based access management to securely administer Open RAN functions. The Network Configuration Access Control Model (NACM) restricts access to certain NETCONF protocol operations and contents, aligns authentication and authorization with centralized access management systems, protects the running configuration against unauthorized changes or deletions, and supports change management protocols for network function configurations and NACM adjustments within a network function. For confidentiality, integrity, and authenticity, O1 will utilize TLS 1.2 or later versions, and employ NACM to implement least privilege access principles \cite{O-RAN.SFGRICXAPP}.
\end{itemize}

\subsubsection{Protecting Against Conflicting \textit{xApps}} The Conflict Mitigation function is designed to resolve conflicts that may arise from multiple \textit{xApps}making overlapping or conflicting requests. There are four different actors who can initiate this function at different stages in the life cycle of an \textit{xApp}, including authorised \textit{xApps}, the Subscription Management function, message monitoring by the Conflict Mitigation function, and the Conflict Mitigation function itself. Open RAN Alliance assumes that the Conflict Mitigation function has sufficient information to identify potential conflicts and decide on the best way to resolve them for these procedures. The proposed solution suggests that the first Conflict Mitigation guidance procedure, initiated by an authorised \textit{xApp}, should be mandatory as part of the E2 Subscription process performed at the conclusion of \textit{xApp} Registration. By requesting this guidance during \textit{xApp} Registration, conflicts that may arise from new \textit{xApps} being deployed can be reduced. However, it is important to note that conflicts may still occur after deployment, and additional Conflict Mitigation messages and procedures will need to be developed by Open RAN Alliance to address these situations~\cite{O-RAN.SFGRICXAPP}.

Table~\ref{tbl:RICSecurityFeaturesReversed} succinctly aligns proposed RIC security measures with the vulnerabilities V-01 through V-07 identified in Section~\ref{sec:RIC-SecurityRisks} and detailed in Table~\ref{tbl:RICThreats}. This tabulation highlights the relative effectiveness of each security control in mitigating identified threats, providing a concise reference to assess the defensive strategies explored for RIC system.

\subsection{Advancing Resilient RIC Security for 6G }\label{sec:RICSEC6G}
Next-generation mobile networks, embracing 6G, are pivoting towards Open RAN architecture to enable a new wave of mobile communications and services. This strategic choice unlocks potent avenues for reinforcing RAN security by effectively utilizing the distinct strengths of non-RT and nRT RICs. The non-RT RICs are designed to manage intensive security tasks within sub-second intervals, engaging in secure RAN analytics, policy crafting, and the provision of AI/ML-powered defence models to bolster the nRT RIC operations. In parallel, nRT RICs respond to critical security directives within milliseconds, applying AI/ML inference to ensure rapid and lightweight threat detection. The collaboration between these two types of RICs facilitates the creation and optimization of sophisticated AI/ML-driven security algorithms for RAN, transcending the limitations of single-vendor solutions and fostering a more dynamic and competitive telecommunications sector. Following this, we will delve into five key opportunities that RICs introduce, leveraging Open RAN to elevate the security framework of 6G networks comprehensively.

 \subsubsection{Shifting security capabilities to the edge of the network} The incorporation of open interfaces within the RAN architecture offers network operators the ability to distribute security analytics throughout the network, effectively shifting RAN monitoring to the network edge. This approach presents a unique opportunity to develop analytics that are specifically tailored to edge-based security threats, allowing for the rapid detection and prevention of network attacks, threats, and vulnerabilities. By facilitating closed-loop actions at the RAN level, malicious traffic can be effectively blocked from reaching the core network, ensuring the secure and efficient delivery of mobility services, particularly IoT services, by preventing DDoS attacks on the RAN by rogue mobile devices. In addition, distributed security analytics provide network operators with the ability to share insights between the RAN and core, as well as between different RAN locations. These insights can be used to safeguard radio units adjacent to a unit under attack or to utilize core insights to protect potentially vulnerable RAN units. Overall, this distributed approach to security analytics allows network operators to adopt a proactive approach to network security, identifying and addressing potential security risks before they can escalate into more significant issues~\cite{kang2013crossfire}.

\subsubsection{Seamless integration of industry-leading platforms} The Open RAN technology provides network operators with the opportunity to utilize industry-standard security protocols to integrate the best available security platforms, resulting in improved security for their networks. This is a significant advantage as it enables network operators to select the most effective and secure security solutions without having to consider compatibility issues that may arise. For instance, if a network operator seeks to improve their network's security, they may come across a security platform that they believe will work well but employs a different security protocol than the one they currently use. With Open RAN technology, the network operator can easily integrate this security platform without making significant changes to their network or incurring high costs associated with purchasing custom adaptors. Furthermore, Open RAN facilitates regular protocol updates by network function vendors to remain up-to-date with industry best practices. This enables network operators to keep their networks secure without incurring any additional costs or effort.

\subsubsection{Automation and zero-touch management}Open RAN technology has the potential to significantly accelerate the automation of network management, which in turn can enhance security by eliminating inherent risks associated with human access to network functions. These risks stem from the possibility of unintended alterations to the security posture of network functions or the deliberate exploitation of credentials, configuration changes, or the introduction of malicious software into the network. In addition, automation facilitates rapid response to changes in the network and improves the ability to manage security incidents through closed-loop management. For example, the use of an open management interface can provide network operators with an efficient means of verifying the security posture of network functions and promptly identifying and resolving degraded configurations or suspicious network activity. This benefit underscores the potential of Open RAN to mitigate security risks associated with human error and malicious activity, as well as to provide effective security incident management capabilities.


\subsubsection{Improved network visibility and auditing} 
The utilization of open standards for the interfaces between RAN components in Open RAN allows for a standardized connection between different components from various suppliers, which enhances visibility and transparency. This can aid auditors and security testers in comprehending the functionality of the RAN implementation and verifying its accuracy, facilitating the identification of potential security issues and vulnerabilities. The amplified use of open-source components in Open RAN can further improve transparency by providing greater visibility into how the components operate internally, aiding in the identification of potential security issues and facilitating a more secure implementation of the RAN~\cite{nis2022euoran}. Additionally, the disaggregation of components and the use of open interfaces in Open RAN provide network operators with direct access to a greater amount of data on network performance, enabling them to detect potential security issues more quickly. The use of open management interfaces also provides easier access to data about the running state of network functions, which can be combined with security log data for root cause analysis~\cite{coalition2021policy}.
However the disaggregation of the RAN into multiple components is seen by some as opening up the RAN to security vulnerabilities by introducing a greater attack surface.
\subsubsection{Interoperability enhances security} Open RAN systems, with a higher number of components that have open interfaces, provide greater interoperability, which can enhance security by enabling components from different vendors to work seamlessly together. In contrast to a traditional RAN, where all components are from a single vendor and replacing one component can be challenging, Open RAN allows for easy replacement of a vulnerable component without affecting the entire network. This can improve network security by reducing the time to fix vulnerabilities and lowering the risk of a single point of failure.

\subsection{RIC Security Use Cases} \label{sec:RICSECUsecase}
In this section, we illustrate the practical application of security mechanisms within the O-RAN architecture, as identified through our survey analysis. By exploring critical use cases, we highlight how advanced technologies can be strategically integrated into the RIC to enhance network defence and resilience. Each case underscores the potential of these solutions to fortify the RIC against emerging threats.

\subsubsection{Real-time link verification}
\begin{figure}[!htbp]
\centering
\includegraphics[width=0.32\textwidth]{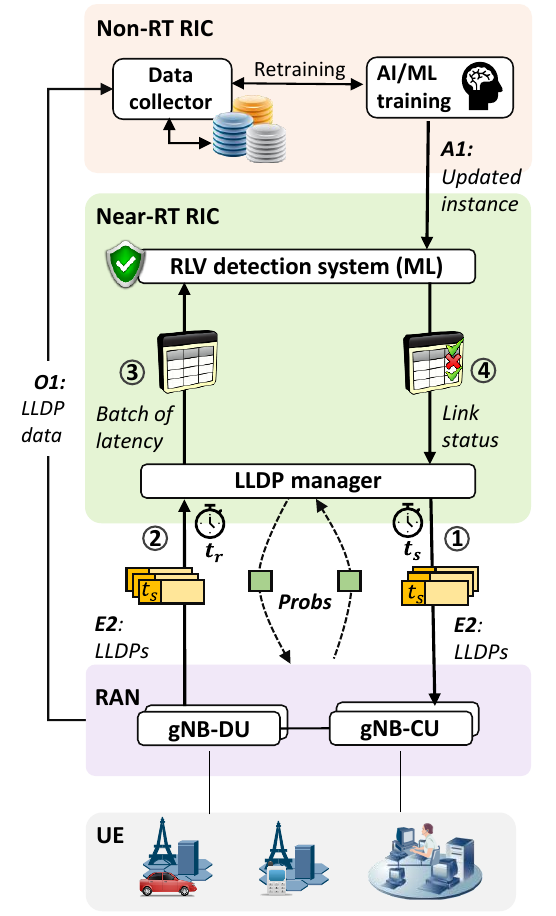}
\caption{\small RLV defence system deployed in Open RAN architecture}
\label{fig:RICLLDPframe}
\end{figure}

The work in~\cite{soltani2023real} have pioneered a defence mechanism in the form of a Real-time Link Verification (RLV) system, originally conceived for SDN controllers. This innovative technique leverages ML algorithms to detect and counteract LLA and LFA attacks, key concerns in SDN security. In our survey, we have meticulously analyzed and recognized the exceptional potential of RLV to reinforce the security landscape of O-RAN architectures. 
By adapting this methodology, the robustness of the RIC can be significantly enhanced, benefiting from RLV's real-time monitoring and verification capabilities. Our proposed architecture in Fig.~\ref{fig:RICLLDPframe} delineates the practical implementation of RLV within the O-RAN architecture, demonstrating its seamless integration and the role it plays in safeguarding the network. The illustration provides a visual representation of how RLV operates in conjunction with RIC, thereby offering a tangible example of how SDN-based defence algorithms can be transposed to fortify O-RAN security frameworks.

The proposed RLV system architectures are structured into three tiers: the data plane, which encompasses RAN elements like gNB-RU, gNB-DU, gNB-CU, along with both the nRT and non-RT RICs. Within the RIC, an LLDP manager is tasked with creating and dispatching LLDP and probe packets at set intervals to the data plane elements. Upon receipt, the gNB-DU and gNB-CU nodes return each LLDP packet to the RIC. These packets, once gathered by the LLDP manager, are processed to extract relevant metrics such as latency which are then assembled into batches for analysis. Utilizing the RLV, this data is subjected to an ML classification model within the non-RT RIC, which evaluates the information and relays the outcomes to the nRT RIC. To keep pace with network dynamics, the ML model is continuously refined through retraining with the latest LLDP data via the O1 interface, ensuring the nRT RIC is updated with the latest model iteration for optimal performance. Based on the outcome of the classifier, the RIC either drops the LLDP packet or updates the topology database.

The adaptation of RLV for the O-RAN architecture not only exemplifies the versatility of SDN-centric solutions but also highlights the potential for cross-pollination of technologies between different network paradigms. By tailoring this SDN-oriented defence mechanism to the requirements of the RIC, we can address specific security challenges inherent to the O-RAN ecosystem, thereby enhancing overall network integrity and reliability.

\subsubsection{Protection against signalling storms} 
In SSA, the adversary may flood the network control plane with invalid or repeated registration requests. Even if these registration requests are declined, they still consume core network resources in the control plane that are crucial during the authorisation process. It would be advantageous to somehow identify adversary devices early in the registration process within RAN to safeguard core network resources. Achieving this can be challenging in state-of-the-art mobile networks, where both hardware and software are typically supplied by a single vendor with limited configuration options. Conversely, the Open RAN concept enables interaction with the RAN through dedicated interfaces and the interception of RAN protocol messages. The O-RAN Alliance WG11 has recognized SSA detection as a critical issue to address through the development of a dedicated \textit{xApp}. In the proposed paper by Hofmann et al.~\cite{hoffmann2023signaling}, they introduced an \textit{xApp} designed to detect abnormal activity such as signaling storm in Industrial Internet of Things (IIoT) devices right at the start of their registration procedure. This \textit{xApp} leverages O-RAN interfaces to intercept control plane messages and gather the necessary long-term network statistics. These statistics are then utilized to identify abnormal activity associated with adversaries~\cite{hoffmann2023signaling}.

The primary defence algorithm upon attacks coming from various existing devices in the network is based on the configuration of the devices themselves and trust that the devices will simply fulfill to constraints specified by mobility standards. Back-off timer is a defence algorithm that limits the quantity of frequent device registrations, therefore, it prohibits devices from overloading the network with attach requests. If the mentioned reliance is disaffirmed, there will be no extra choices for defending the existing network rather than refusing service randomly to both malicious and benign gadgets. This incident is practically equipollent for DDoS attack. Regrettably, a few hundreds gadget categories exist in today's modern networks that randomly disaffirm this reliance and give permission to the gadgets to aggressively attach to the network at a rate of a thousand times per hour. The attacker is capable of causing an attach storm that would cause a long outage of large parts of the network by finding a solution to deal with a huge number of these vulnerable devices. On the other hand, the attacker is able to carry on this attack for many hours, and every time choosing thousands of devices from a large set of devices accessible; the network carrier cannot cease the attack without efficient controls to take action upon a specified structure of behaviour. Since the behavior of these chosen devices is very abnormal and different from the other devices in the existing network, the detection and recognition of these attacking devices is feasible. 

Applying dynamic constraints over these devices to cease them from overloading the control plane is essential for the network. These constraints should be efficient and well-organized to authorise benign devices to register to the network with no discontinuity. Ceasing these types of attacks without overloading various sections of the network in the core is feasible by having an intelligent security and defence system at the RAN. Building an \textit{xApp} with two major responsibilities including a DDoS recognition and mitigation ability is a very effective approach in order to defending the network against UE-originated signaling storms. The DDoS recognition ability contains two sections: near real-time recognition, which operates in the nRT RIC\textit{ xApp}, and non-real-time recognition, which operates at the non-RT RIC and is highly dependant on enrichment data originated in the external system. On the other hand, the DDoS mitigation ability \textit{xApp} is able to implement an E2 control loop where the \textit{xApp} can decide for each attach request if it should be accepted or denied, or it may update an appropriate E2 policy when a UE is defined to be suspect. An exemplary E2 policy in this use case may allow suspected UEs to be denied thoroughly at the E2 node. In this case, the Open RAN architecture allows implementing protection and defence system to the network at the edge, hence lowering the trace of these signaling storm DDoS attacks on network infrastructures. The integration of near real-time logic at the nRT RIC for fast detection, with slower scale analysis and input from the Non RT RIC supplies an efficient combination of advanced and quick detection schemes. It is also possible to apply lighter detection proceedings as \textit{xApps} while more pioneering heavy processing ML methods can work externally and send input to RIC~\cite{O-RAN.SFGusecase}.

\subsection{Lesssons Learned} \label{sec:RICSECLL}
In the context of 5G and the upcoming 6G networks, securing the RIC is paramount. Part I underscores the critical aspects of achieving this security. Secure \textit{xApp} management is foundational, emphasizing secure practices during\textit{ xApp} onboarding, provisioning, and registration. Digital signatures play a pivotal role in ensuring the authenticity of \textit{xApps}, preventing tampering, and maintaining data integrity. Furthermore, OAuth 2.0 authentication is recommended to enforce precise API access control, ensuring that only authorised \textit{xApps} gain access to specific resources. Granular access control is another vital lesson, urging network operators to segregate \textit{xApps} based on their security implications. Implementing mutual authentication, such as mTLS or IPsec, bolsters API access control, ensuring that both \textit{xApps} and the RIC platform verify each other's identity using public and private keys. The incorporation of OAuth 2.0 enhances security by permitting third-party \textit{xApps} to access only the resources they are explicitly authorised for. The lesson further underscores the necessity of secure software practices, including maintaining an inventory of open-source components, addressing vulnerabilities, and ensuring software integrity during updates. Lastly, secure RIC interface communications are critical, with protocols like IPSec and TLS recommended to protect RIC connectivity from eavesdropping, and mutual authentication should be enforced before allowing O\&M connectivity.

Part II dives into the advancements for resilient RIC security in 6G networks. It highlights a paradigm shift, where security moves to the network edge. Open RAN architecture allows the distribution of security analytics throughout the network, enabling the rapid detection and prevention of network attacks. By securing the edge, malicious traffic is efficiently blocked before reaching the core network, ensuring the secure delivery of services, particularly for IoT applications. Moreover, seamless integration of industry-standard security platforms is vital, allowing network operators to select the most effective security solutions without compatibility concerns. Regular protocol updates by network function vendors keep security aligned with industry best practices. Automation and zero-touch management are emphasized as open RAN technology accelerates network management automation, reducing the risks associated with human errors and malicious activities. This automation facilitates swift responses to network changes and improves security incident management through closed-loop management. Context-aware networks are another 6G security lesson, where security mechanisms leverage both current and predicted security contexts to identify real-time threats. User context acquisition respects privacy, ensuring a balance between security and user rights. Lastly, improved network visibility and auditing are critical, achieved through open standards for interfaces, enhancing transparency, and aiding auditors and security testers in comprehending network functionality and identifying potential security issues. Overall, these lessons ensure a resilient, adaptable, and secure wireless ecosystem for the future of mobile communication networks.

\section{Securing the Future: RIC Integration with Emerging Technologies} \label{sec:lessonlearn}

Currently, it is still too early to determine definitively which technologies will be integrated with open RAN in the context of 6G. However, there are several highly promising technologies that should be considered as potential candidates. Three notable contenders in this regard are Reconfigurable Intelligent Surface (RIS), Satellite, and Digital Twin (DT). These technologies have garnered significant attention and warrant further investigation and research within the academic community. Adhering to the principle of security by design, it is crucial to carefully evaluate the security risks and opportunities that these technologies may present for 6G, especially when integrated with Open RAN, with a specific focus on the RIC. In the subsequent sections, our aim is to illuminate the security aspect of integrating RIC with other technologies in the future. We seek to draw the attention of researchers towards the potential implications and associated security considerations that arise when RIC is combined with diverse technological frameworks. By considering these aspects, we can ensure a thorough understanding of the security landscape and foster informed discussions in the field.
\subsubsection{Secure interface between RIC \& RIS controller} 
\begin{figure}[!htbp]
\centering
\includegraphics[width=0.38\textwidth]{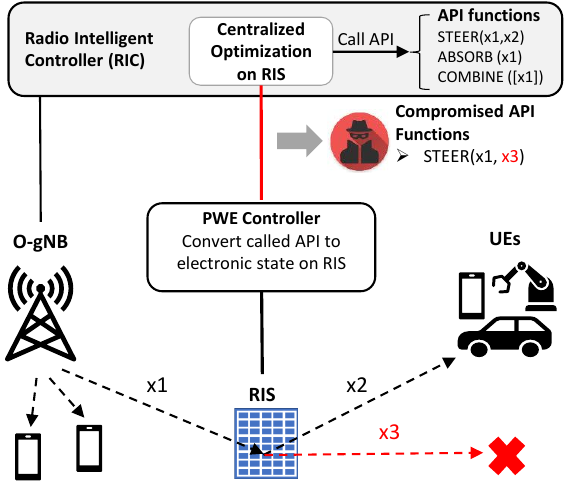}
\caption{\small Risk of integration of PWE management with RIC}
\label{fig:RISRIC}
\end{figure}

One possible future work scenario is integrating a Programmable Wireless Environment (PWE) controller into RIC~\cite{liaskos2022software}. The PWE comprises a set of software-defined RISs connecting to the controller for receiving Electro-Magnetic (EM) manipulation commands via the EM API. RIS refers to a technology that employs intelligent surfaces with the ability to adaptively manipulate wireless signals. These surfaces can enhance wireless communication performance by intelligently reflecting or refracting signals, enabling improved coverage, capacity, and energy efficiency in wireless networks~\cite{liu2021reconfigurable}. Recent advancements in research have unveiled a promising possibility of establishing a direct interface connecting the nRT RIC with the PWE controller~\cite{strinati2021reconfigurable} (see Fig.~\ref{fig:RISRIC}). This development holds significant potential as it allows for simultaneous control over the configurations of RIS and the beamforming of gNB. This approach allows different stakeholders, including network tenants and meta-surface providers, to coexist within the system. 


In the context of the aforementioned scenario, where the integration of the PWE control system into Open RAN infrastructures is taking place, ensuring the security of the RIC becomes of paramount importance. The secure interface between the RIC and the PWE controller is particularly crucial. Furthermore, within the integrated PWE control system framework, any security vulnerabilities present in the RIC could potentially be exploited to manipulate or compromise the functionality of the RIS. This could lead to unauthorised changes in custom air paths, interference issues, or other malicious activities that may significantly impact the performance and stability of the network. For example, Fig.~\ref{fig:RISRIC} illustrates a potential risk associated with the integration of RIS management with RIC, highlighting the possibility of compromised APIs. This risk underscores the need to address security concerns in order to safeguard the integrity and functionality of the RIS within the integrated PWE control system framework. 


Network operators and stakeholders should closely collaborate with security experts to ensure that the design, implementation, and maintenance of the RIC and the integrated PWE control system adhere to the highest level of security standards. Therefore, it is worth considering the exploration of security aspects concerning the RIC and its integration within the PWE control system in RIS as a prospective future research direction.

\subsubsection{Secure integration of satellite with RIC}

In the future, a potential work scenario within Open RAN 6G involves implementing a communication system based on Non-Terrestrial Networks. This system would utilize multiple Low Earth Orbit (LEO) satellites to serve a diverse range of users with varying needs. One scenario for using satellites in O-RAN 6G entails gathering network information, such as channel conditions, network traffic, and Quality of Service (QoS) requirements, through these satellites. Subsequently, this data is transmitted to a cloud server, specifically to thE RIC, using a gateway. At the RIC, a Deep Neural Network (DNN)-based learning model is implemented. This model utilizes trained algorithms to make resource allocation decisions. The training unit continuously provides decisions and collects data to improve the accuracy of the training process. This iterative approach allows the model to adapt to network changes and provide more effective responses~\cite{al2023digital}.

\begin{figure}[!htbp]
\centering
\includegraphics[width=0.34\textwidth]{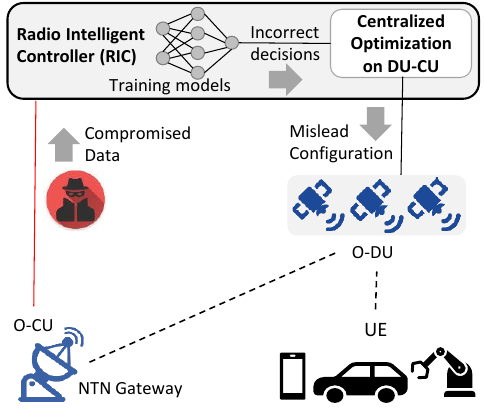}
\caption{\small Securing the integration of Satellite management with RIC}
\label{fig:SATRIC}
\end{figure}

The integration of satellites with the RIC in future advancements introduces significant security considerations. Given the central role of the RIC, there are concerns about potential vulnerabilities, particularly regarding the Gateway-to-RIC interface. Fig.~\ref{fig:SATRIC} illustrates a threat in this architecture, where the attacker compromises the collected data from the DU and CU, leading to misleading configurations on DU and CU. To address these concerns, a security-centric approach is crucial. It should include robust measures such as advanced encryption, stringent authentication, access controls, and comprehensive monitoring systems. By implementing these measures, the system can ensure the integrity and confidentiality of the data, safeguard against unauthorised access, and detect any potential security breaches.

\subsubsection{Secure integration of RIC with SON, MEC, and network slicing}


O-RAN can be effectively integrated with other important technologies such as Self-Organizing Network (SON), and Multi-Access Edge Computing (MEC). The MEC is a new concept introduced by the ETSI to address the challenges arising from the complexity of rapidly advancing mobile and wireless communication networks. The core idea behind MEC is to expand the capabilities of Cloud Computing (CC) to the edges of mobile networks, which helps alleviate the limitations associated with conventional cloud infrastructure. In simpler terms, MEC serves as a complement to business data centres, offering computing, storage, networking, and data analysis resources closer to where the data is generated~\cite{ranaweera2021survey}. On the other hand, A SON is an automated and intelligent system used in telecommunications and wireless network management. Its primary purpose is to optimize and enhance the performance of mobile and wireless networks autonomously, without requiring extensive manual intervention.

The rise of the 6G network era is anticipated to bring a plethora of opportunities and challenges, especially in terms of security. As we transition from 5G to 6G, the integration of SON and RIC becomes even more pivotal. SON's inherent ability to autonomously optimize and enhance network performance could be further augmented by RIC's capabilities~\cite{Intelligent2023optimization}. However, this integration is not without its concerns. The sophisticated and interconnected nature of 6G networks, combined with the dynamic functionalities of SON and RIC, can introduce potential vulnerabilities. Threat actors could exploit these to gain unauthorised access, manipulate network configurations, or disrupt essential services. RIC, in particular, with its real-time decision-making abilities, could become a prime target. As it interacts with various components and layers in a 6G network, it's essential to ensure its robustness against potential attacks. The integration of SON, with its automated network management features, can offer an additional layer of protection. By leveraging SON's intelligence, anomalous behaviors can be quickly detected, and appropriate countermeasures can be activated. Nevertheless, as the 6G ecosystem continues to evolve, continuous research and development efforts are vital to ensure that the integration of SON and RIC remains secure, resilient, and efficient.

In a study conducted by Kuklinski et al~\cite{kuklinski2020ran}, they have discussed O-RAN, SON, MEC, and network slicing technologies, highlighting their collaborative potential while also pinpointing areas where their architectural elements overlap. Based on their analysis, they've introduced Integrated nRT RIC (I-nRT RIC). Their paper demonstrates that adopting the ORAN-centric approach proves advantageous and addresses certain issues that O-RAN alone hasn't fully tackled. They illustrate that through integration, some components from these contributing technologies can either be eliminated or repurposed. With the introduction of I-nRT RIC as a key player in this integration, it's imperative to explore both the advantages it offers and the security implications it brings. One aspect to consider is the increased attack surface resulting from the integration of multiple technologies. 

As the I-nRT RIC interacts with various components and interfaces across these technologies, it potentially opens up new vectors for cyberattacks. Security measures need to be implemented to safeguard against threats such as unauthorised access, data breaches, and malicious control attempts. Additionally, the integration concept presents an opportunity to enhance security. By consolidating control and monitoring functions under the I-nRT RIC, it becomes possible to implement more centralized and robust security measures. This can include advanced threat detection, anomaly detection, and security policy enforcement. The integration may also enable better coordination among different security mechanisms across the integrated technologies. Furthermore, as the concept is presented at a high level and the nature of these systems is not yet fully defined, security considerations should be incorporated into the design and development phases. This proactive approach can help identify and address security vulnerabilities early in the integration process, reducing the risks associated with deploying such a complex and interdependent network ecosystem.

\subsubsection{Integration of DTs into RIC}


The integration of DTs into Open RAN architectures holds significant promise for enhancing the security of RIC. DTs are highly accurate digital replicas of physical objects, providing valuable data and virtual prototypes for various purposes. Operators rely on DTs to enhance preventive maintenance, drive innovation in business models, accelerate product development, and optimize sustainability and efficiency in real-world scenarios~\cite{nguyen2021digital}. DTs provide a valuable platform for proactive security testing, assessment, and risk mitigation. Security professionals can simulate various attack scenarios within the DT environment, enabling the identification of vulnerabilities and the development of robust security measures before implementing them in the physical network. 

By integrating DTs with the RIC, organizations can further enhance the security of the RIC itself. The RIC can leverage the simulation and prediction capabilities of the DT to analyze network behavior, detect security vulnerabilities, and take proactive security measures. Centralized monitoring and management of security events within the DT environment enable organizations to gain a comprehensive view of the network and respond swiftly to security incidents affecting the RIC. Moreover, DTs offer real-time monitoring and analytics capabilities~\cite{thiruvasagam2023open}, allowing for the detection of network anomalies and security threats. By leveraging ML and analytics within the DT environment, security professionals can gain valuable insights into network behavior, identify potential security risks, and respond promptly to mitigate them. This proactive approach to security, enabled by DTs, helps protect the integrity and confidentiality of the RIC and the overall Open RAN infrastructure. Further investigation and research are crucial to fully explore the potential of integrating DTs into the RIC and to address the specific security considerations in this context.

\vspace{-10px}
\section{Conclusion}
Open RAN has emerged as a disruptive force in the telecommunications industry, promising increased interoperability, reduced vendor lock-in, and enhanced network optimization. At the heart of this transformative paradigm, RICs are the orchestrators, enablers, and intelligence hubs that empower network operators to harness the full potential of disaggregated RAN components.
In this paper, we have conducted an extensive survey focusing on the security aspects of 6G RIC within the context of O-RAN. Our specific emphasis has been on addressing the challenges associated with securing RIC infrastructure. We have categorized various types of attacks and vulnerabilities and thoroughly reviewed them within the context of RIC, along with a review of relevant solutions. Additionally, we have presented security implications and outlined future research challenges, aiming to provide the community with valuable insights into the security landscape of RIC.

Finally, we have provided a summary of the challenges, limitations, open issues, and potential future research paths in the realm of RIC security. In conclusion, our analysis suggests that RIC has the potential to bolster the security of O-RAN networks by enabling proactive security measures and dynamic adaptations. However, it's crucial to acknowledge that the integration of RIC may also introduce novel vulnerabilities into the ecosystem. These vulnerabilities can arise from increased complexity, potential misconfigurations, or unforeseen interactions between RIC components and existing network elements. Therefore, while RIC offers promising security enhancements, its deployment should be approached with careful consideration and a comprehensive risk assessment to mitigate these new challenges effectively. 
The hope is that this in-depth survey will inspire and drive progress in both theoretical and experimental research, ultimately contributing to the future success of the RAN Intelligent Controller while enhancing its security aspects.

\bibliographystyle{IEEEtran}
\bibliography{references}

\vspace{-15 mm}
\begin{IEEEbiography}[{\includegraphics[width=1in,height=1.25in,clip,keepaspectratio]{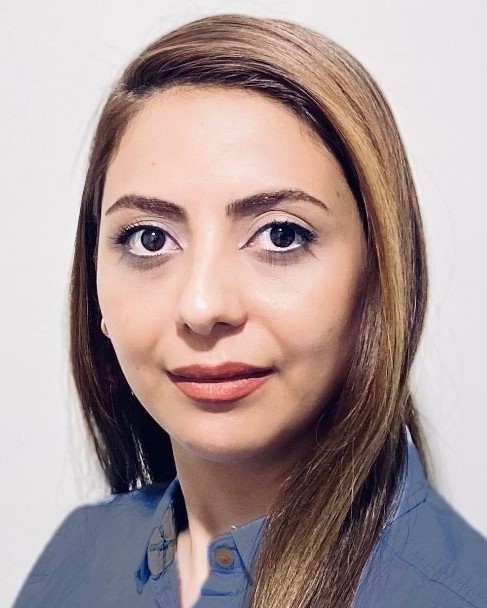}}]{Sanaz Soltani} is a PhD researcher in Information and Communication Systems (ICS), working in the 5GIC \& 6GIC Innovation Centre at the University of Surrey, UK. Before, she was a network specialist in Huawei and MTN telecommunication companies involved in 4G and LTE projects. She received her master's degree in software engineering from the Amirkabir University of Tech. (Tehran Polytechnic) Iran, 2014. Her research interests include Network Softwarization, Open RAN, Network Security, and Privacy.
\end{IEEEbiography}

\vspace{-15 mm}
\begin{IEEEbiography}[{\includegraphics[width=1in,height=1.25in,clip,keepaspectratio]{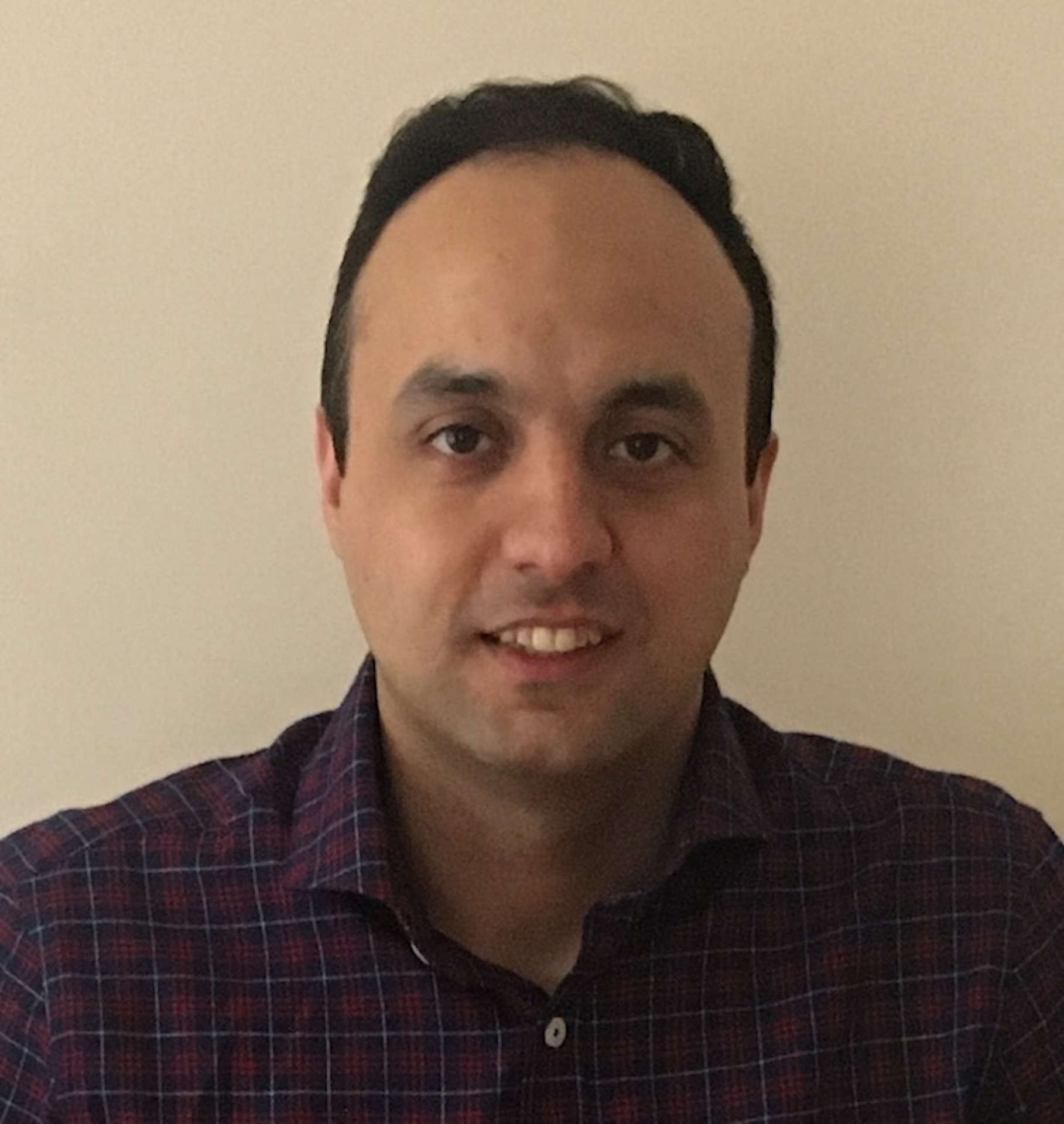}}]{Mohammad Shojafar} \textbf{(M'17-SM'19)} is a Senior Lecturer (Associate Professor) in network security and an Intel Innovator, a Professional ACM member and ACM Distinguished Speaker, a Fellow of the Higher Education Academy, and a Marie Curie Alumni, working in the 5G \& 6G Innovation Centre (5GIC \& 6GIC), Institute for Communication Systems (ICS), at the University of Surrey, UK. Dr Mohammad secured around $\pounds$1.7M as PI in various EU/UK projects, including D-XPERT (funded by I-UK/UK;2024), 5G Mode (funded by DSIT/UK;2023), 5G ONE4HDD (funded by DSIT/UK;2023), TRACE-V2X (funded by EU/MSCA-SE;2023), AUTOTRUST (funded by ESA/EU;2021), PRISENODE (funded by EU/MSCA-IF:2019), and SDN-Sec (funded by Italian Government:2018). He was also COI of various UK/EU projects like HiPER-RAN (funded by DSIT/UK;2023), APTd5G project (funded by EPSRC/UKI-FNI:2022), ESKMARALD (funded by UK/NCSC;2022), GAUChO, S2C and SAMMClouds (funded by Italian Government;2016-2018). He received his PhD in ICT from Sapienza University of Rome, Rome, Italy, in 2016 with an ``Excellent'' degree. He is an Associate Editor in \textit{IEEE Transactions on Network and Service Management}, \textit{IEEE Transactions on Intelligent Transportation Systems}, and Computer Networks. 
\end{IEEEbiography}

\vspace{-15 mm}
\begin{IEEEbiography}[{\includegraphics[width=1in,height=1.25in,clip,keepaspectratio]{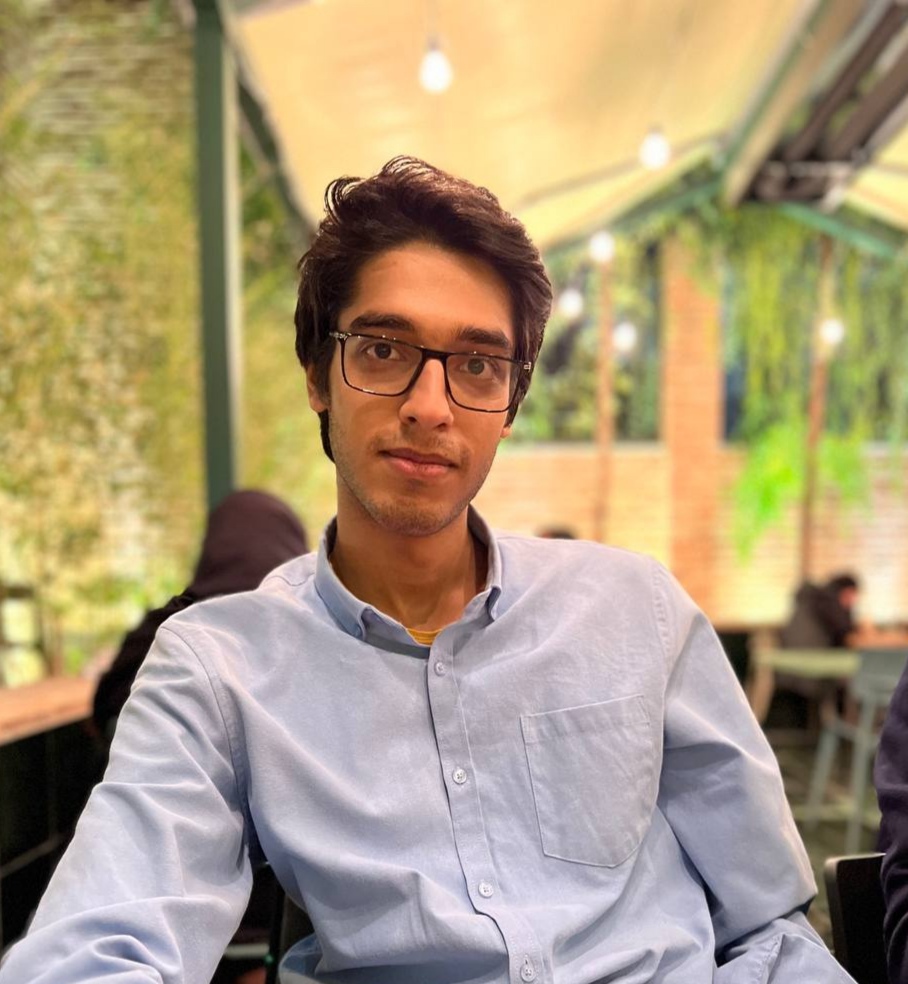}}]{Ali Amanlou} is a visiting researcher at the 5GIC \& 6GIC Innovation Centre at the University of Surrey, UK. He earned his B.Sc. degree (Hons.) in electrical engineering, specializing in telecommunications. 
He has authored numerous publications in peer-reviewed journals. He frequently serves as a paper reviewer for esteemed journals such as \textit{IEEE Transactions on Neural Networks \& Learning Systems}, \textit{IEEE Transactions on Multimedia}, \textit{IEEE Transactions on Image Processing}, and \textit{IEEE Access}. His research interests encompass artificial intelligence, deep learning, 5G/6G technology, network security, and Open RAN.
\end{IEEEbiography}

\vspace{-10 mm}
\begin{IEEEbiography}[{\includegraphics[width=1in,height=1.25in,clip,keepaspectratio]{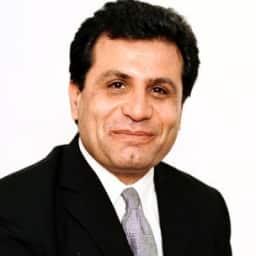}}]{Rahim Tafazolli (SM'09)} is the Regius Professor and Professor of Mobile and Satellite Communications. He is the Director of ICS and the founder and Director of world's first 5G Innovation Centre at the University of Surrey, UK. Many governments regularly invite him for advice on mobile communications and, in particular, 5G technologies. Regius Tafazolli has given many interviews to International media in the form of television, radio interviews and articles in the international press.
\end{IEEEbiography}
\balance
\end{document}